\documentclass[a4paper,fleqn,usenatbib]{mnras}

\usepackage{newtxtext,newtxmath}

\usepackage[T1]{fontenc}
\usepackage{ae,aecompl}

\usepackage{graphicx}	
\usepackage{amsmath}	
\usepackage{amssymb}	
\usepackage{cases}
\usepackage{tabularx}
\usepackage{ulem} 

\title[Compact planetary system formation]{Formation of compact systems of super-Earths via dynamical instabilities and giant impacts}

\author[Poon et al.]{
Sanson T. S. Poon,$^{1,2}$\thanks{E-mail: s.t.s.poon@qmul.ac.uk}
Richard P. Nelson,$^{1}$
Seth A. Jacobson,$^{3}$
Alessandro Morbidelli$^{4}$
\\
$^{1}$Astronomy Unit, Queen Mary University of London, London E1 4NS, UK\\
$^{2}$Royal Observatory Greenwich, London SE10 9NF, UK\\
$^{3}$Earth and Planetary Sciences, Northwestern University, Evanston, Illinois, USA \\
$^{4}$Observatoire de la C\^ote d'Azur, Laboratoire Lagrange, Bd. de l'Observatoire, CS 34229, F-06304 Nice Cedex 4, France
}

\date{Accepted 2019 November 19. Received 2019 October 8; in original form 2019 August 5}

\pubyear{2019}

\begin{document}
\label{firstpage}
\pagerange{\pageref{firstpage}--\pageref{lastpage}}
\maketitle

\begin{abstract}
NASA's Kepler mission discovered $\sim700$ planets in multi-planet systems containing 3 or more transiting bodies, many of which are super-Earths and mini-Neptunes in compact configurations. Using $N$-body simulations, we examine the in situ, final stage assembly of multi-planet systems via the collisional accretion of protoplanets. Our initial conditions are constructed using a subset of the Kepler 5-planet systems as templates. Two different prescriptions for treating planetary collisions are adopted. The simulations address numerous questions: do the results depend on the accretion prescription?; do the resulting systems resemble the Kepler systems, and do they reproduce the observed distribution of planetary multiplicities when synthetically observed?; do collisions lead to significant modification of protoplanet compositions, or to stripping of gaseous envelopes?; do the eccentricity distributions agree with those inferred for the Kepler planets? We find the accretion prescription is unimportant in determining the outcomes. The final planetary systems look broadly similar to the Kepler templates adopted, but the observed distributions of planetary multiplicities or eccentricities are not reproduced, because scattering does not excite the systems sufficiently. In addition, we find that $\sim 1$\% of our final systems contain a co-orbital planet pair in horseshoe or tadpole orbits. Post-processing the collision outcomes suggests they would not significantly change the ice fractions of initially ice-rich protoplanets, but significant stripping of gaseous envelopes appears likely. Hence, it may be difficult to reconcile the observation that many low mass Kepler planets have H/He envelopes with an in situ formation scenario that involves giant impacts after dispersal of the gas disc.
\end{abstract}




\section{Introduction}\label{sec:intro}
The Kepler mission discovered 4723 exoplanet candidates, of which 2302 have been confirmed as bona fide transiting planets  \citep{2010Sci...327..977B,2011ApJ...736...19B,2013ApJS..204...24B, 2014ApJS..210...19B,2015ApJS..217...16R,2015ApJS..217...31M, 2016ApJS..224...12C,2018ApJS..235...38T}. More than $70\%$ of Kepler planets have radii $1 \, 
R_{\oplus} \le R_{\rm p} \le 4 R_{\oplus}$, such that super-Earths and mini-Neptunes make up a large fraction of the known exoplanet population\footnote{All Kepler planetary data used in this paper are from \href{https://exoplanetarchive.ipac.caltech.edu}{NASA Exoplanet Archive} unless stated otherwise.}. 
A significant number of these planets are found in compact multiplanet systems, such as the 6-planet system Kepler-11 \citep{2011Natur.470...53L} and the 5-planet system Kepler-84 \citep{2014ApJ...784...45R}. The highest multiplicity system detected by Kepler where all planets are confirmed is Kepler-90, with eight planets transiting its host star \citep{2018AJ....155...94S}. Analyses of the Kepler data to determine occurrence rates of planets shows that systems of Earths and super-Earths with orbital periods $<100$ days are common around Solar-type stars \citep[e.g.][]{2013ApJ...766...81F,2013PNAS..11019273P}. A recent analysis suggests that the mean multiplicity of super-Earth systems with periods $<100$ days is $\sim 3$, approximately $1/3$ of Sun-like stars host compact planetary systems, and the mean number of planets per star is $\sim 1$ \citep{2018ApJ...860..101Z}.

The relative numbers of 1- to 8-planet systems discovered via transit detections are dependent on both the intrinsic multiplicities of the systems, and the mutual orbital inclinations of the planets that comprise the systems. In this work, we examine whether or not a simple model of the in situ, final stage assembly of planetary systems, involving dynamical instabilities and accretion through giant impacts among a large population of protoplanets after the gaseous protoplanetary disc has dispersed, is consistent with the Kepler observations. 

Previous $N$-body simulations have considered the formation of compact systems of planets from an earlier stage than we consider here, and include the influence of the protoplanetary disc and subsequent disc-planet interactions \citep[e.g.][]{2007ApJ...654.1110T, Hellary, 2014MNRAS.445..479C, 2014A&A...569A..56C,2016MNRAS.457.2480C}. One feature of these simulations is that chains of short period planets in mean motion resonances are a common outcome, contrary to what is observed in the Kepler planet population. More recent work, however, has indicated that these resonant chains can become dynamically unstable once the gas disc has dispersed, such that the final stages of planetary assembly involve mutual scattering and collisions between planets \citep{2012Icar..221..624M,2017MNRAS.470.1750I,2019arXiv190208772I,2019arXiv190302004C}. In addition to breaking the resonant chains, the gravitational scattering also raises the mutual inclinations and eccentricities, and allows under some circumstances for the simulations to produce planetary system multiplicities that are reported to be in agreement with the Kepler data. 

In contrast to the migration-driven formation scenario described above, there have also been $N$-body studies of in situ formation in a gas free environment \citep[e.g.][]{2012ApJ...751..158H, 2016ApJ...832...34M, 2017AJ....154...27M}. Here, the initial conditions consist of numerous protoplanets arranged in an annulus that undergo mutual scatterings and collisions on the way to assembling the final systems. 
While these calculations are in some ways similar to the final stages of the migration-driven scenarios when the break-up of the resonant chains occurs, they differ in some important respects. For example, the planets do not start in resonance, and the numbers of bodies involved in the collisional evolution is significantly larger. Hence, the number of collisions experienced by a typical planet is also larger during the evolution.

Our approach in this paper is similar to that used in the aforementioned in situ models, except we use a subset of the Kepler 5-planet systems as templates when constructing the initial conditions of the $N$-body simulations. We reconstruct the surface density distributions of the chosen planetary systems, and use this to define initial conditions consisting of numerous orbiting protoplanets.  The approach is therefore similar to the construction of a minimum mass exoplanet nebula model proposed by \citet{2013MNRAS.431.3444C}. The protoplanet systems are then evolved for $10^7$ yr in a gas-free environment. For each of the planetary systems we consider, we perform two sets of simulations. One uses a traditional hit-and-stick accretion prescription when collisions occur, and the other uses a more complex accretion prescription based on hydrodynamical simulations of colliding bodies \citep{2012ApJ...745...79L}. Hence, we are able to examine the influence of the accretion prescription on the outcomes of the simulations, similar to the recent study by \citet{2018MNRAS.478.2896M}.

Adopting a more complex collision model also allows us to track the impact energy during collisions, and we use this information to examine possible composition changes that the planets could potentially experience through the removal of volatile components. Using the relations between the collision energy and the final water content of the largest remnant after differentiated bodies composed of rock and water have collided \citep{2009ApJ...691L.133S, 2010ApJ...719L..45M}, we determine how much water could be removed from the planets during their collisional evolution. Although some individual collisions would likely lead to significant compositional changes, taken as a whole our results indicate that the compositions of water-rich super-Earths would not change significantly, if their final stages of evolution were similar to those occurring in the simulations. A similar analysis was also used to examine whether or not the impact energies during collisions could potentially remove putative H/He envelopes from the planets, by the conversion of impact energy into heat energy in the cores \citep{2019MNRAS.485.4454B}, and here we find that very significant erosion of gaseous envelopes should be expected.

The rest of this paper is structured as follow. In section \ref{sec:sim} we describe the simulation methods and the set-up of the initial conditions. In section \ref{sec:outcome} we present the main outcomes of our simulations, and in section~\ref{sec:coorbit} we examine the formation pathways of the co-orbital planets that arise in the simulations. In section~\ref{sec:comp} we post-process the simulation data and examine the changes to compositions that might arise during collisions, and in section~\ref{sec:env} we examine the stripping of gaseous envelopes that might arise. In section \ref{sec:synobser}, we discuss the results from synthetic observation of the final simulated systems, and examine in particular the distribution of system multiplicities and eccentricities that arises. Finally, we discuss our results and draw conclusions in section \ref{sec:discuss}.

\section{$N$-body simulation methods}
\label{sec:sim}
We use the $N$-body codes \texttt{MERCURY} \citep{1999MNRAS.304..793C} and \texttt{SyMBA} \citep{1998AJ....116.2067D} to undertake the simulations presented in this paper. Both codes use the Mixed Variable Symplectic (MVS) integration scheme \citep{1991AJ....102.1528W}, but whereas \texttt{MERCURY} handles close encounters by transitioning to a Bulirsch-Stoer method \citep{1992nrfa.book.....P}, \texttt{SyMBA} uses the Regularized MVS scheme \citep{1994Icar..108...18L}. More importantly for the work presented here, the versions of the two codes we employ handle collisions differently. \texttt{MERCURY} uses a simple hit-and-stick algorithm that conserves the total mass and linear momentum  when two bodies collide and accrete into a single object, whereas our version of \texttt{SyMBA} adopts the imperfect accretion algorithm from \citet{2012ApJ...745...79L}, which we describe below.

\subsection{Imperfect collision model}\label{subsec:symba}
For a detailed description of the \citet{2012ApJ...745...79L} collision model we refer the reader to that paper, and here we simply summarise the post-collision outcomes that are generated by it, along with a few salient details about the implementation. We note that the collision model was implemented in \texttt{Symba} by the authors\footnote{We warmly acknowledge the assistance of Zoe Leinhardt in this implemetation during a visit to the Observatoire de la C\^ote d'Azur}. We refer to the more massive body involved in the collision as the target, and the less massive object as the projectile. The range of outcomes includes the following: a perfect merger where a single body is formed with the total mass and momentum of the original two bodies; a single massive body remains whose (largest-remnant) mass is denoted $M_{\rm LR}$, along with collisional debris in the form of low mass `super-planetesimals' (gravitating particles that are not mutually interacting); two massive bodies remain with masses $M_{\rm LR}$ and $M_{\rm SLR}$ (mass of the second largest remnant), along with collision debris in the form of low mass `super-particles'; no massive bodies remain and all the mass is in the form of collision debris represented by low mass `super-planetesimals'. The following notation is used in the description below: $V_{\rm imp}$ is the impact velocity; $V_{\rm esc}$ is the escape velocity from the colliding bodies (or, more accurately, from the combined target mass and interacting mass of the projectile); $Q_{\rm R}$ is the impact energy; $Q_{\rm RD}^*$ is the catastrophic disruption energy, which by definition corresponds to the impact energy when the mass of the largest remnant contains half of the total mass of the colliding bodies; $b_{\rm crit}$ is the critical impact parameter that determines whether or not a collision is grazing ($b\ge b_{\rm crit}$) or non-grazing ($b < b_{\rm crit}$). The collision algorithm consists of a decision tree with the following possible outcomes:
\begin{itemize}
\item When $V_{\rm imp} < V_{\rm esc}$ we have a perfect merger
\item When $V_{\rm imp}$ exceeds the threshold for super-catastrophic disruption, both colliding bodies are destroyed and only collisional debris remains
\item When $V_{\rm imp}$ exceeds the threshold for catastrophic disruption or erosion, only one massive body remains and the rest of the mass is in the form of collisional debris. Here, $M_{\rm LR}$, the mass of the largest post-collision remnant, depends on $Q_{\rm R}$ and $Q_{\rm RD}^*$.
\item If $V_{\rm imp}$ is smaller than the threshold for erosion and $b < b_{\rm crit}$, then we have partial accretion where the target body gains mass from the projectile, which is itself completely disrupted into a number of low mass `super-planetesimals'. $M_{\rm LR}$ again depends on $Q_{\rm R}$ and $Q_{\rm RD}^*$
\item For $b \ge  b_{\rm crit}$,  in descending order of the impact velocities, we have the following outcomes that all preserve the mass of the target object and modify the mass of the projectile: hit-and-spray, where the projectile is completely disrupted into debris particles; hit-and-run, where the projectile mass is reduced and the remaining mass goes into debris particles; bouncing collision, where the projectile retains all of its mass and the collision is treated as an inelastic bounce; graze-and-merge collision, where a single body forms containing all the mass of the colliding objects.
\end{itemize}

The total mass before and after a collision, $M_{\rm Total}$, is conserved, which means the total mass of the post-collision bodies obey the relation $M_{\rm T,d}=M_{\rm Total}-(M_{\rm LR}+M_{\rm SLR})$, where $M_{\rm T,d}$ is the total mass in debris after the collision. The number of debris particles, $N_{d}$ is given by
\begin{equation}\label{eq:Nd}
    N_{d}= 
\begin{cases}
    \max\left(  \frac{M_{\rm T,d}}{10.0{\rm M}_{\rm Ceres}}, \min\left ( 38,\frac{M_{\rm T,d}}{0.1{\rm M}_{\rm Ceres}} \right )\right ),& \text{if } M_{\rm T,d} > 0\\
    0,              & \text{otherwise}
\end{cases}
\end{equation}
where ${\rm M}_{\rm Ceres}$ is the mass of Ceres. If the values of $M_{\rm T,d}/10.0{\rm M}_{\rm Ceres}$ and $M_{\rm T,d}/0.1{\rm M}_{\rm Ceres}$ are not even integers, they are rounded-up to the nearest even integers. 
With the known value of $M_{\rm T,d}$ and $N_{d}$, the mass is evenly distributed to each debris particle. 

If debris particles are formed after a collision,  they are evenly distributed in a circle on the plane of impact at a distance of one Hill radius ($R_{\rm Hill}$) from the collision centre of mass according to
\begin{equation}\label{eq:Nd}
\mathbf{r_{\rm d}}=\mathbf{r_{\rm com}}+R_{\rm Hill}\mathbf{\widehat{r_{\rm d}}},
\end{equation}
where $\mathbf{r_{\rm d}}$ is the initial position vector of the debris particles, $\mathbf{r_{\rm com}}$ is the position vector of the collision centre of mass, and $\mathbf{\widehat{r_{\rm d}}}$ is the position unit vector for the evenly distributed debris with respect to the collision centre of mass.
The velocities of the debris particles are simply assumed to be 5\% larger than $V_{\rm esc}$:
\begin{equation}\label{eq:Nd}
\mathbf{V_{\rm d}}=\mathbf{V_{\rm com}}+1.05\times V_{\rm esc}\mathbf{\widehat{r_{\rm d}}},
\end{equation}
where $\mathbf{V_{\rm d}}$ and $\mathbf{V_{\rm com}}$ are the initial velocity vector for the debris and the collision centre of mass velocity vector.

The accumulated effect of high energy collisions can lead to the creation of collisional debris in the form of thousands of `super-planetesimals' . These particles normally get reaccreted by the protoplanets during the simulations, but if, for example, a super-catastrophic collision occurs at the inner edge of our system, then a ring of planetesimals can form which have exceedingly long dynamical life times. This then causes the simulation run times to increase appreciably. To ameliorate this situation, we have introduced a scheme for removing such a ring of particles when it forms. This is motivated by the fact that the collision time in the ring is normally very short, and collisions between the planetesimals would be highly destructive, such that they would be ground down to dust which would then be removed by radiation pressure and/or Poynting-Robertson drag. The scheme calculates the collision time and reduces the masses of the planetesimals on that time scale, until the mass in the ring is negligible and the particles can be removed from the simulation. A more detailed description is given in appendix \ref{app:pfr}.

\subsection{Kepler multi-planet system templates}\label{subsec:select}
In this study we have selected a number of Kepler 5-planet multi-planet systems to provide templates for the initial conditions of the simulations, using the following criteria. We are interested in the compact systems, so we have chosen systems in which the known outermost planet has semi-major axis $\le 0.5$ AU. We have selected those Kepler systems where all five of the known planets are transiting. For example, the Kepler-122 system is not included due to one of its planets, Kepler-122f, being discovered by transit timing variations (TTVs) \citep{2014ApJ...787...80H}. When we began this project, Kepler-80 was listed as a 5-planet system \citep{2016AJ....152..105M}, but more recently it has been confirmed as a 6-planet system using deep learning by \citet{2018AJ....155...94S}. In spite of this recent announcement, we include this system using the five planets known before 2018. Kepler-296 is a binary system with two stars, Kepler-296A and -296B, that have a projected separation of $\sim70$~AU \citep{2015ApJ...809....7B}. All five planets are orbiting the same star (Kepler-296A). Given that the outermost planet, Kepler-296Af, orbits at $\sim0.255$AU, which is only $\sim0.36\%$ of the binary stars separation, the binary should have little influence on the dynamic stability and evolution of the planetary system \citep{1997AJ....113.1445W}, and hence we include this system.

As described below, we use the Kepler systems to construct individual mass surface density profiles, which are then used to produce initial conditions for the simulations consisting of 20 protoplanets. We impose selection criteria on these initial conditions that include a requirement that the inter-protoplanet separation is not too small or too large (i.e. $5 \le K \le 30$), where $K$ is the inter-protoplanet separation measured in units of the mutual Hill radius.  This avoids the evolution being dominated by collisions that occur at very early times before dynamical relaxation of the systems has had an opportunity to arise, or the converse where no collisions happen at all. Finally, we require the maximum value of the initial protoplanet mass to be $M_{\rm p} <6$~M$_{\oplus}$. After applying these criteria, eight systems were selected to be the templates. As listed in table \ref{tab:Kval}, these are Kepler-55, -80, -84, -102, -154, -169, -292 and -296.

\subsection{Surface density profiles from Kepler systems}\label{subsec:surfdensity}
In order to construct surface density profiles from the chosen Kepler systems, we need to know the semi-major axes and masses of the planets, which are not provided directly by the observations. The semi-major axis is obtained trivially from Kepler's 3$^{\rm rd}$ law
\begin{equation}\label{eq:Ttoa}
a=\sqrt[3]{\frac{GM_{\star}}{4\pi^2}P^2}.
\end{equation}
where $P$ is the measured planetary period and $M_{\star}$ is the mass of the host star. Numerous suggested relations between the observed planet radius, $R_{p}$, and the planet mass, $M_{p}$, have appeared in the literature.  In this study, we adopt the relation suggested by \citet{2011ApJS..197....8L}, 
based on fitting the Earth and Saturn:
\begin{equation}\label{eq:MRrelation}
M_p={\left(\frac{R_p}{R_{\oplus}}\right)}^{2.06}M_{\oplus}.
\end{equation}
We also considered the relation suggested by \citet{2016ApJ...825...19W} (which is the best-fit relation for the sample of RV-measured transiting sub-Neptunes with $1.5<R_{p}<4R_{\oplus}$). As discussed later in  Section~\ref{subsec:stable}, however, we find that obtaining $M_{p}$ from $R_{p}$ using this relation results in some of the selected Kepler systems being themselves dynamically unstable on relatively short time scales, hence we did not adopt this mass-radius relation in this study.

Once we have $M_p$ and $R_p$, the internal density, $\rho_p$, is given by
\begin{equation}\label{eq:density}
\rho_p=\frac{3M_p}{4\pi {R_{p}}^3}.
\end{equation}
In order to simplify the collision model for a given simulation, we adjusted the $\rho_p$ values within each individual system to be the same for different planets, obviating us from having to deal with collision outcomes involving planets with significantly different densities. We did this by constructing a mass weighted average of the planetary densities, as follows:
\begin{equation}\label{eq:avedensity}
\left \langle \rho_p \right \rangle=\frac{\sum M_p\rho_p}{\sum M_p}.
\end{equation}

\begin{figure}
\centering
\includegraphics[width=7.6cm]{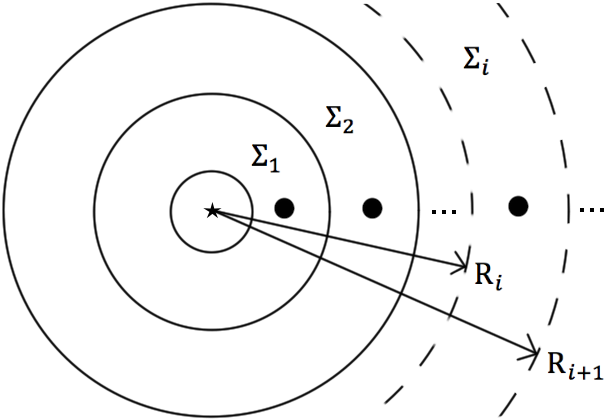}
\caption{Diagram illustrating method for calculating $\Sigma_{\rm fit}$, the surface density of each of the Kepler templates we have adopted, as described in the text.}\label{fig:plring}
\end{figure}

To find the surface density profile for the original Kepler system, we first define an annulus surrounding each planet. Figure \ref{fig:plring} shows an example planetary system and the annuli associated with each planet, where each annulus is defined by its inner and outer radius. For a general planet $i$, these are denoted as  $R_i$ and $R_{i+1}$. Here, $R_i$ is taken to be the midpoint between the semi-major axis of the planet, $a_i$, and its inner adjacent planet, $a_{i-1}$:
\begin{equation}\label{eq:Ri}
R_{i,i\neq1}=\frac{a_{i}+a_{i-1}}{2}.
\end{equation}
The innermost boundary is located at 
\begin{equation}
R_1=a_1-\left ( \frac{a_{2}-a_{1}}{2} \right ),
\end{equation} 
and when the planetary system has $n$ planets, the outermost boundary, $R_{n+1}$, is at
\begin{equation}
R_{n+1}=a_n+\left ( \frac{a_{n}-a_{n-1}}{2} \right ).
\end{equation} 
The area of the $i$-th annulus, $A_i$ is
\begin{equation}\label{eq:ringarea}
A_i=\pi\left (R_{i+1}^{2}-R_{i}^{2} \right ),
\end{equation} 
and the surface density of the annulus can be calculated using $M_{p,i}$ (the mass of the planet contained in the annulus obtained from equation \ref{eq:MRrelation}) and $A_i$ from equation~\ref{eq:ringarea}:
\begin{equation}\label{eq:Sigma}
\Sigma_i=\frac{M_{p,i}}{A_i}.
\end{equation} 

This gives the surface density at discrete radial locations around each star, and to obtain the surface density as a continuous function we simply fit the five $\Sigma_i$ values with a smooth function.
A $4$-th order polynomial can always be found that passes through 5 real data points, but this approach has not been used here because it often gives negative values of $\Sigma$ at some locations. Instead, we have fitted the $\Sigma_i$ using four different model functions, namely:
\begin{subnumcases}{\Sigma_{\rm fit}(a)=\label{eq:Sigmafit}}
    c_1a^{c_2}+c_3 \label{subeq:power}\\
    c_1a^3+c_2a^2+c_3a+c_4 \label{subeq:poly}\\
    c_1\exp(c_2a)+c_3 \label{subeq:exp}\\
    c_1+c_2\cos(c_4a)+c_2\sin(c_4a),\label{subeq:FS}   
\end{subnumcases}
where $\Sigma_{\rm fit}(a)$ is the fitted surface density profile as a function of semi-major axis $a$, equation (\ref{subeq:power}) is the power fitting model; (\ref{subeq:poly}) is the polynomial fitting model; (\ref{subeq:exp}) is the exponential fitting model; (\ref{subeq:FS}) is the Fourier series fitting model. The $c_i$'s are the fitting coefficients. The selection criteria for which model to choose are: 1) the model that provides the best least-squares fit among all models;  2) no negative values between $\Sigma_{\rm fit}(a=R_1)$ and $\Sigma_{\rm fit}(a=R_n)$. Appendix \ref{app:Sigmafit} provides additional details about the coefficients used to obtain $\Sigma_{\rm fit}$ in this study.

\subsection{Constructing initial conditions for the simulations}\label{subsec:IC}
The motivation behind this study is to determine whether or not dynamical instabilities and giant impacts among a large population of protoplanets can lead to final planetary systems that are similar to the compact Kepler multi-planet systems. The initial conditions of the $N$-body simulations consist of 20 protoplanets that are constructed from the $\Sigma_{\rm fit}$ obtained for each of the Kepler systems. 

\begin{figure}
\centering
\includegraphics[width=8.3cm]{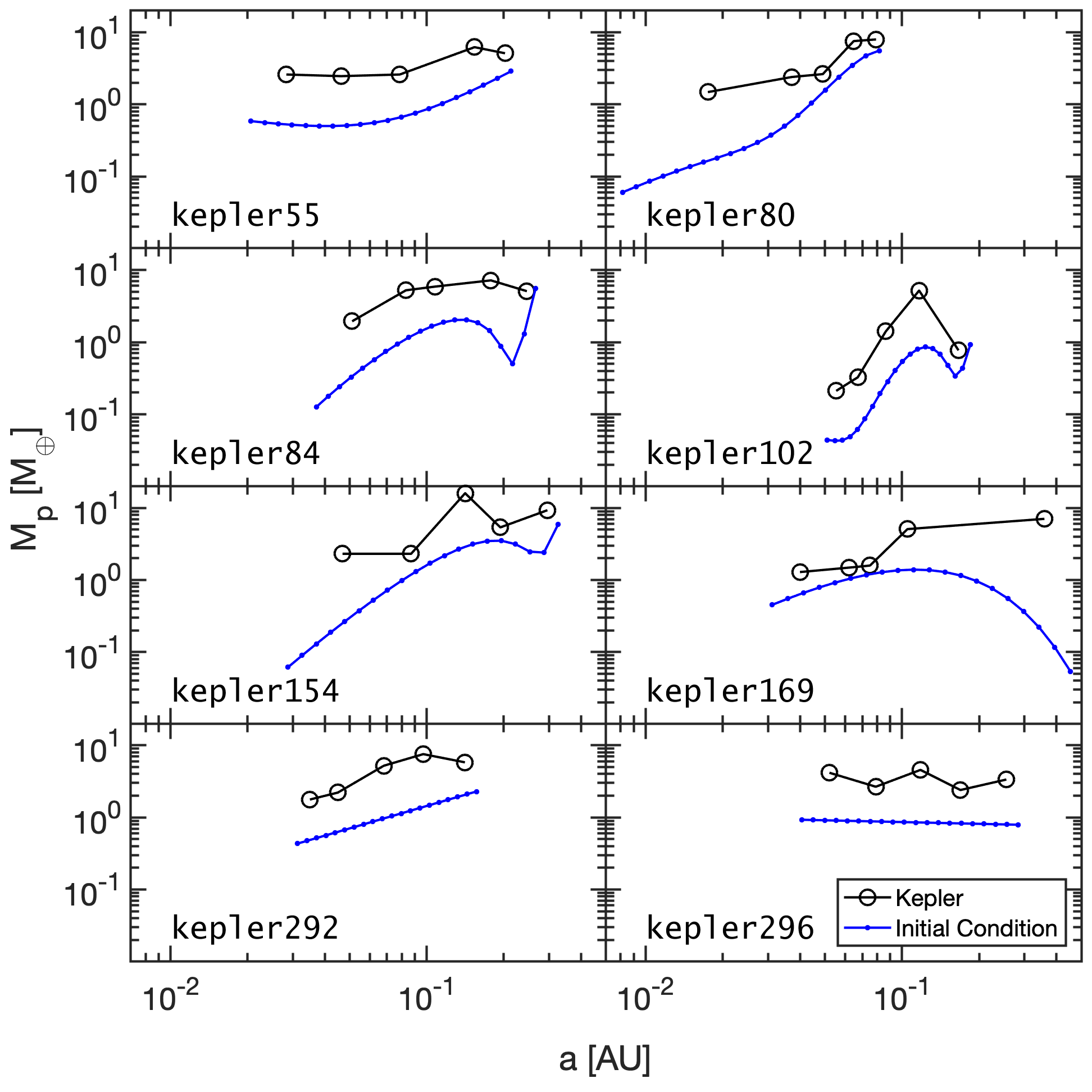}
\caption[$M_p$-$a$ relation for the initial 20 protoplanets templates]{$M_p$ versus $a$ for all 8 Kepler templates. Initial masses of the 20 protoplanets are marked in blue dots, and the original Kepler planet masses are marked in black circles.}\label{fig:initmass}
\end{figure}

First we assume the semi-major axes of the new protoplanets are distributed following a power law, which is achieved by defining 20 new annuli with appropriate boundaries. The radius of the innermost boundary $R_1$ and outermost boundary $R_{n+1}$  always remain at the same position as the original Kepler system. The radius of the $i$-th boundary in between $R_1$ and $R_{n+1}$ can be calculated by
\begin{equation}\label{eq:ri}
R_{i+1}=f \times R_i,
\end{equation}
where $f$ is a constant distance ratio and is given by
\begin{equation}
f=\sqrt[n]{\frac{R_{n+1}}{R_1}}.
\end{equation}
Once the $R_i$ are obtained from equation \ref{eq:ri}, the semi-major axis of the $i$-th protoplanet, $a_i$, is set at the mid-point between $R_i$ and $R_{i+1}$. The mass, $M_{p,i}$, of the protoplanet in position $a_i$ is calculated according to
\begin{equation}
M_{p,i}=2\pi R_i\left( R_{i+1}-R_i \right)\Sigma_{\rm fit}(a_i).
\end{equation}
Using this process to find the $M_{p,i}$ values may result in a system with total mass that differs from the  original Kepler system, in which case the mass of each protoplanet is scaled appropriately. Figure \ref{fig:initmass} shows the masses of the initial 20 protoplanets with respect to their semi-major axis for each system template. We also assumed the density of the protoplanets are $\rho_p=\left \langle \rho_p \right \rangle$ throughout the whole system. With the new value of $\rho_p$ and $M_p$, the planetary radii, $R_p$, adopted in the simulations can be obtained from equation~\ref{eq:density}.

The initial eccentricities, $e$, inclinations, $I$, arguments of pericenter, $\omega$, longitudes of ascending node, $\Omega$, and mean anomalies, $M$, also need to be defined when setting the initial conditions of the simulations. The values of $e$ and $I$ are uniformly distributed within a range  $0\leq e\leq e_{\rm max}$ and $0\leq I\leq I_{\rm max}$, where $e_{\rm max}$ and $I_{\rm max}$ are defined below. The values of $\omega$, $\Omega$, and $M$ are distributed uniformly in the range $0\leq (\omega,\Omega,M)\leq 2\pi$.

Two sets of $e_{\rm max}$ and $I_{\rm max}$ values are used here to investigate the effect of the initial eccentricities and inclinations. The first (higher initial value) set has $e_{\rm max}=0.02$ and $I_{\rm max}=0.01$ rad, while the second (lower initial value) set has $e_{\rm max}=0.002$ and $I_{\rm max}=0.001$ rad. In each $e_{\rm max}$-$I_{\rm max}$ set, 10 simulations were run, with different random number seeds being used to generate the values of $\omega$, $\Omega$, and $M$. Hereafter, the `higher set' refers to the runs with ($e_{\rm max}$, $I_{\rm max}$)=(0.02, 0.01), and `lower set' refers to the runs with ($e_{\rm max}$, $I_{\rm max}$)=(0.002, 0.001). Each higher and lower set was run using both perfect and imperfect collision models using the \texttt{MERCURY} and \texttt{SyMBA} $N$-body codes, respectively. Hence, there are 40 simulations for each Kepler system template.

The central bodies of each system have their masses and radii taken from the Kepler data. Each simulation runs for $10^7$ yr. The time steps used in the simulations are set to be $1/20^{\rm th}$ of the shortest orbital period \citep{1998AJ....116.2067D}. 

As discussed above in Section~\ref{subsec:select}, one of the criteria used to constrain our initial conditions is that the mutual separation between neighbouring protoplanets must satisfy $5 \le K \le 30$,
where $K$ is the inter-protoplanet separation measured in units of the mutual Hill radius. The mutual Hill radius for a pair of adjacent planets is defined by
\begin{equation}\label{eq:R_Hi}
R_{H,i}=\frac{a_{i}+a_{i+1}}{2}\left ( \frac{M_{p,i}+M_{p,i+1}}{3M_{\star}} \right )^{\frac{1}{3}}.
\end{equation} 
The dimensionless number $K$ can then be expressed as
\begin{equation}\label{eq:Ki}
K_i=\frac{a_{i+1}-a_{i}}{R_{Hi}},
\end{equation}
where $K_i$ is the $K$-value for the $i$-th pair of adjacent planets in the system.

For a Kepler planetary system, applying the value of $a$ obtained from equation \ref{eq:Ttoa}, $M_p$ from equation \ref{eq:MRrelation}, and $M_{\star}$ from the Kepler data, $K$ can be directly calculated by equation \ref{eq:Ki}. For the selected Kepler planetary systems in this study (see section \ref{subsec:select}), the $K_{i}$ values for each planet pair and mean $K$ value of each system, $ \bar K$, are listed in table \ref{tab:Kval}. The overall mean $K$ value across all selected systems has the value $\langle \bar K \rangle \approx 19.4$. This value is about the same as the typical average $K$ value for Kepler multiplanetary systems (see Section \ref{sec:intro}) 

\begin{table}
	\centering
	\caption{$K$-values of the selected Kepler 5-planet systems. $K_1$ to $K_4$ denote the $K$-value from the 1st to 4th pair of adjacent planets, respectively, where the 1st pair is the innermost pair  and 4th pair is the outermost pair (penultimate and outermost planet). $\bar K$ denotes the arithmetic mean of $K$ for the system. Underlined values are the minimum $K$ values in the system, $K_{\rm min}$. The minimum $K_{\rm min}$ in the table is 7.2, which is greater than the $K_{\rm min}= 7.1$ required to give 100\% stable rates in a 5-planet system for up to $10^6$ years from \citet{2018arXiv180908499W}.}
	\label{tab:Kval}
	\begin{tabular}{lrrrrr} 
		\hline
		\hline \\
		System & $K_{1}$ & $K_{2}$ & $K_{3}$ & $K_{4}$ & $\bar K$\\ \\
		\hline
		Kepler-55  & 23.9 & 25.6 & 26.7 & \underline{10.3} & 21.6 \\
		Kepler-80  & 41.2 & 14.5 & 11.4 & \underline{7.2}  & 18.6 \\
		Kepler-84  & 24.6 & \underline{11.6} & 21.0 & 13.9 & 17.8 \\
		Kepler-102 & 22.1 & 19.4 & \underline{14.8} & 18.0 & 18.6 \\
		Kepler-154 & 34.5 & 17.4 & \underline{10.9} & 16.4 & 19.8 \\
		Kepler-169 & 29.2 & \underline{12.4} & 16.8 & 45.3 & 26.0 \\
		Kepler-292 & 15.1 & 20.0 & \underline{14.4} & 14.9 & 16.1 \\
		Kepler-296 & 17.2 & 16.2 & \underline{14.8} & 18.0 & 16.5 \\
		\hline
		\hline
	\end{tabular}
\end{table}

Finally, we comment that the initial conditions of the simulations presented in this paper represent the state of the system after substantial evolution has already taken place, and once the gaseous protoplanetary disc has been dispersed. For discussion of possible scenarios leading to these initial conditions, involving the accretion of planetesimals, boulders and/or pebbles onto planetary embryos embedded within the gas disc, we refer the reader to the following papers that present the results of $N$-body simulations of these earlier epochs of planet formation \citep{2014MNRAS.445..479C, 2016MNRAS.457.2480C, 2019A&A...627A..83L,2019arXiv190208772I}.

\begin{figure*}
\centering
\includegraphics[width=16cm]{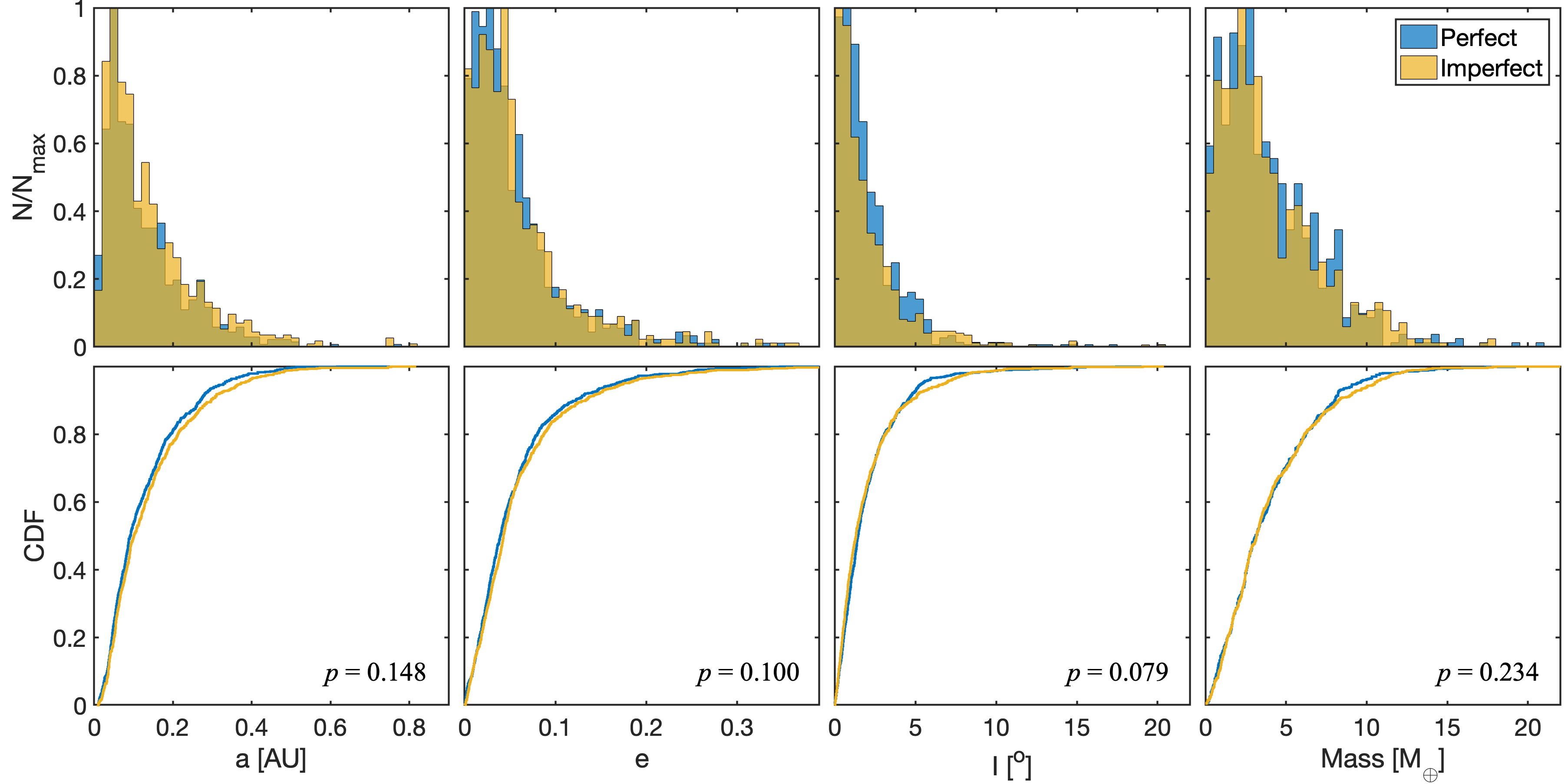}
\caption[All simulation results of $a$, $e$, $I$,and $M_p$]{The normalised distributions (top panel) and the cumulative distribution functions (CDFs, bottom panel) of the  semi-major axes, eccentricities, inclinations, and planet masses of the planets obtained in all simulations. Perfect collision simulations are shown in blue, and imperfect collision simulations are shown in yellow.}\label{fig:allaeimass}
\end{figure*}

\section{Results}\label{sec:outcome}
\subsection{Stability of the original Kepler multi-planet systems}\label{subsec:stable}
Before embarking on a study of the formation of the Kepler systems considered in this paper, we begin by considering the dynamical stability of the observed systems themselves. This acts as a consistency check on the mass-radius relation used to construct the initial conditions for the formation simulations, given by equation~(\ref{eq:MRrelation}) \citep{2011ApJS..197....8L}. Since the first multiplanet systems discovered by Kepler were confirmed, a number of mass-radius relations have been suggested. For example, \citet{2013ApJ...772...74W} suggested $M_{p}=3M_{\oplus}(R_{p}/R_{\oplus})$, \citet{2014ApJ...783L...6W}  suggested $M_{p}=2.69M_{\oplus}(R_{p}/R_{\oplus})^{0.93}$, and \citet{2016ApJ...825...19W} suggested $M_{p}=2.7M_{\oplus}(R_{p}/R_{\oplus})^{1.3}$. Clearly, given that the masses obtained from each of these relations vary for specific values of the planetary radii, stability of the confirmed system is not guaranteed to hold under all these relations. Checking the stability hence provides some constraint on the mass-radius relation that applies. 

We carried out a stability check for all Kepler 5-planet systems, including those which do not obey the selection criteria mentioned in Section~\ref{subsec:select}, comparing the relation suggested by \citet{2016ApJ...825...19W} and that provided by \citet{2011ApJS..197....8L}. We performed $N$-body simulations using \texttt{Mercury}, and adopted initial conditions that assumed the system planets are initially on circular and coplanar orbits ($e=0$ and $I=0$ for all planets). The initial values of the mean anomalies of the planets were assigned randomly, and 10 different realisations were run for each system. We found that the relation from \citet{2011ApJS..197....8L} provides stable systems for all 5-planet systems over a 10 Myr run time, while the mass-radius relation provided by \citet{2016ApJ...825...19W} fails to produce stable systems for some Kepler systems over the same time scale (e.g. the systems Kepler-32 and -33, which were validated by \citet{2011ApJS..197....8L}). For this reason, the relation provided by \citet{2011ApJS..197....8L} is the one we used to construct the initial conditions for the 20-protoplanet simulations described in the following sections.

\subsection{Results of the formation simulations}\label{subsec:SP}
To recap, two sets of $N$-body simulations were performed for each Kepler template, one assuming perfect accretion using a simple hit-and-stick model, and the other adopting the imperfect accretion algorithm of \citet{2012ApJ...745...79L}. For each Kepler template, we considered two initial distributions of the eccentricites and inclinations, a `High set' and a `Low set', and for each of these we computed 10 different realisations of the initial conditions by varying the random number seeds used to create the initial conditions. Hence, we ran 40 $N$-body simulations for 10 Myr for each Kepler template.

We begin our discussion of the results by first considering how the simulation outcomes considered as a whole vary when considering the perfect and imperfect accretion prescriptions. We then look at the simulations in more detail by considering how the outcomes vary between the different Kepler templates, focussing on the resulting planet masses, orbital elements, period ratios, $K$-values and system architectures that emerge from the simulations.

\subsubsection{Comparison between perfect and imperfect accretion across all runs}\label{subsec:PandImp}
The distributions of the semi-major axes, eccentricities, inclinations and masses are shown in the histograms and cumulative distribution functions (CDFs) in figure~\ref{fig:allaeimass}. Later in the paper we discuss the mutual separations between pairs of planets, and the $K$-values are shown in figure~\ref{fig:allK}. By-eye inspection suggests that the distributions are in good agreement when comparing the perfect and imperfect accretion models, and applying the Kolmogorov-Smirnov test (K-S test) yields $p$-values of 0.148 for the semi-major axis distributions, 0.100 for the eccentricities, 0.079 for the inclinations, and 0.234 for the masses. Hence, the null hypothesis that the data plotted in figure~\ref{fig:allaeimass} for the perfect and imperfect collision simulations are drawn from the same underlying distribution cannot be rejected with a $>95\%$ confidence level.

\begin{figure*}
\centering
\includegraphics[width=16cm]{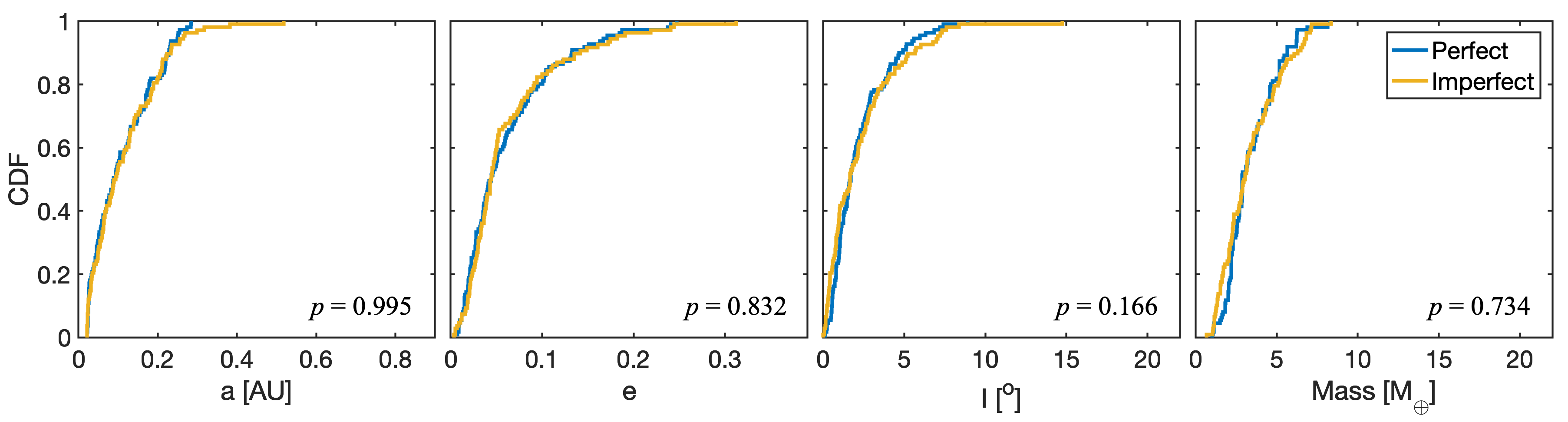}
\caption[All simulation results of $a$, $e$, $I$, and $M_p$ for \texttt{Kepler55} template]{Cumulative distributions, from all simulations of the \texttt{Kepler55} template, of the same 4 parameters listed in figure \ref{fig:allaeimass}. They are (in order from left to right) $a$, $e$, $I$, and $M_p$. The blue and yellow lines are for the perfect collision and imperfect collision simulations, respectively.}\label{fig:all_accumulate_p_55}
\end{figure*}

Agreement between the perfect and imperfect accretion runs can also be seen when looking at individual system templates. For example, figure~\ref{fig:all_accumulate_p_55} shows the CDFs for the same parameters shown in figures~\ref{fig:allaeimass}, but only for the \texttt{Kepler55} system template, and again it can be seen that the distributions are very similar. Here the K-S test yields $p$-values of 0.995 for the semi-major axis distributions, 0.832 for the eccentricities, 0.166 for the inclinations, and 0.734 for the masses.

In summary, based on the global properties of the final planetary systems that are formed, we can conclude that the differences produced by the perfect and imperfect accretion prescriptions are small, and do not have a statistically significant influence on the outcomes of the simulations.

\subsubsection{Instabilities and multiplicities}\label{subsubsec:MandI}
\begin{figure}
\centering
\includegraphics[width=8.4cm]{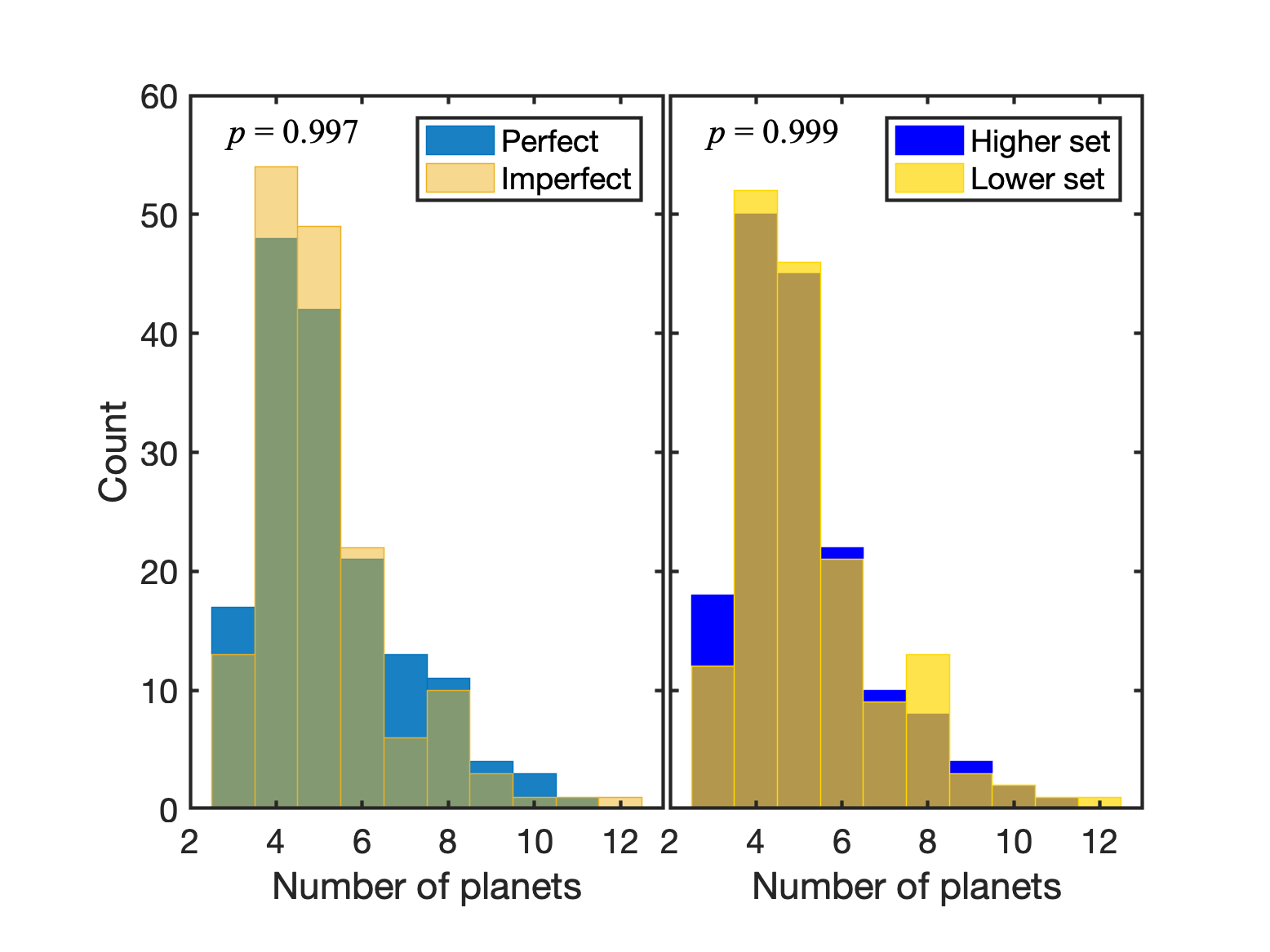}
\caption[System multiplicities]{Multiplicity distributions from all simulations. The left panel compares the distributions obtained in the perfect and imperfect collision simulations. The right panel compares the distributions obtained in the high and low initial eccentricity/inclination simulations (see section \ref{subsec:IC} for the definitions of these simulation sets).}\label{fig:planetnumber}
\end{figure}

All simulations resulted in dynamical instabilities that led to mutual scattering and giant impacts. Figure \ref{fig:planetnumber} shows the distributions of the multiplicities of all final planetary systems. The maximum number of planets remaining after 10 Myr was 12 and the minimum was 3.  No single or double planet systems were formed. Having 4 or 5 planets remain in the system is the most common outcome. Our multiplicity distribution appears to agree with the distribution obtained by \citet{2013ApJ...775...53H}, where they also obtain a minimum multiplicity of 3, and a peak in the multiplicity distribution at 4 or 5 planets. The mean value of the number of planets obtained in the imperfect collision simulation is $\left \langle N_{I} \right \rangle=5.21$ and for perfect collisions it is $\left \langle N_{P} \right \rangle=5.06$. For the higher initial value set $\left \langle N_{\rm H} \right \rangle=5.05$, and for the lower initial value set $\left \langle N_{\rm L} \right \rangle=5.22$. It is noteworthy how close these values are to 5, given that our template systems all contain 5 planets, indicating that the initial conditions constructed from the templates are able to reproduce the desired multiplicity on average. In a recent analysis of the Kepler data, \citet{2018ApJ...860..101Z} concluded that the mean number of super-Earths in compact systems around solar-type stars is approximately 3, with the fraction of stars hosting planetary systems being approximately 0.3. This suggests that the 5-planet systems we have chosen for this study may not be representative of the Kepler planets as a whole, even if we allow for the fact that the Kepler systems contain unseen planets by virtue of them being on orbits that are inclined to the line of sight. 

The K-S test applied to the CDFs derived from the data in figure~\ref{fig:planetnumber} gives a $p$-value of 0.997 for the perfect and imperfect collision models, and 0.999 for the runs with the higher and lower initial eccentricity/inclination values.
The small difference between $\left \langle N_{I} \right \rangle$ and $\left \langle N_{P} \right \rangle$, together with the large $p$-value, shows that our runs are in accord with the conclusions reached by \citet{2018MNRAS.478.2896M}, namely that assuming either perfect or imperfect collisions has little impact on the final multiplicities. The similar values for $N_{\rm H}$ and $N_{\rm L}$ shows that the initial value of $e$ and $I$ also has a limited impact on the final multiplicities, at least for the range of values adopted here.

\begin{figure}
\centering
\includegraphics[width=8.4cm]{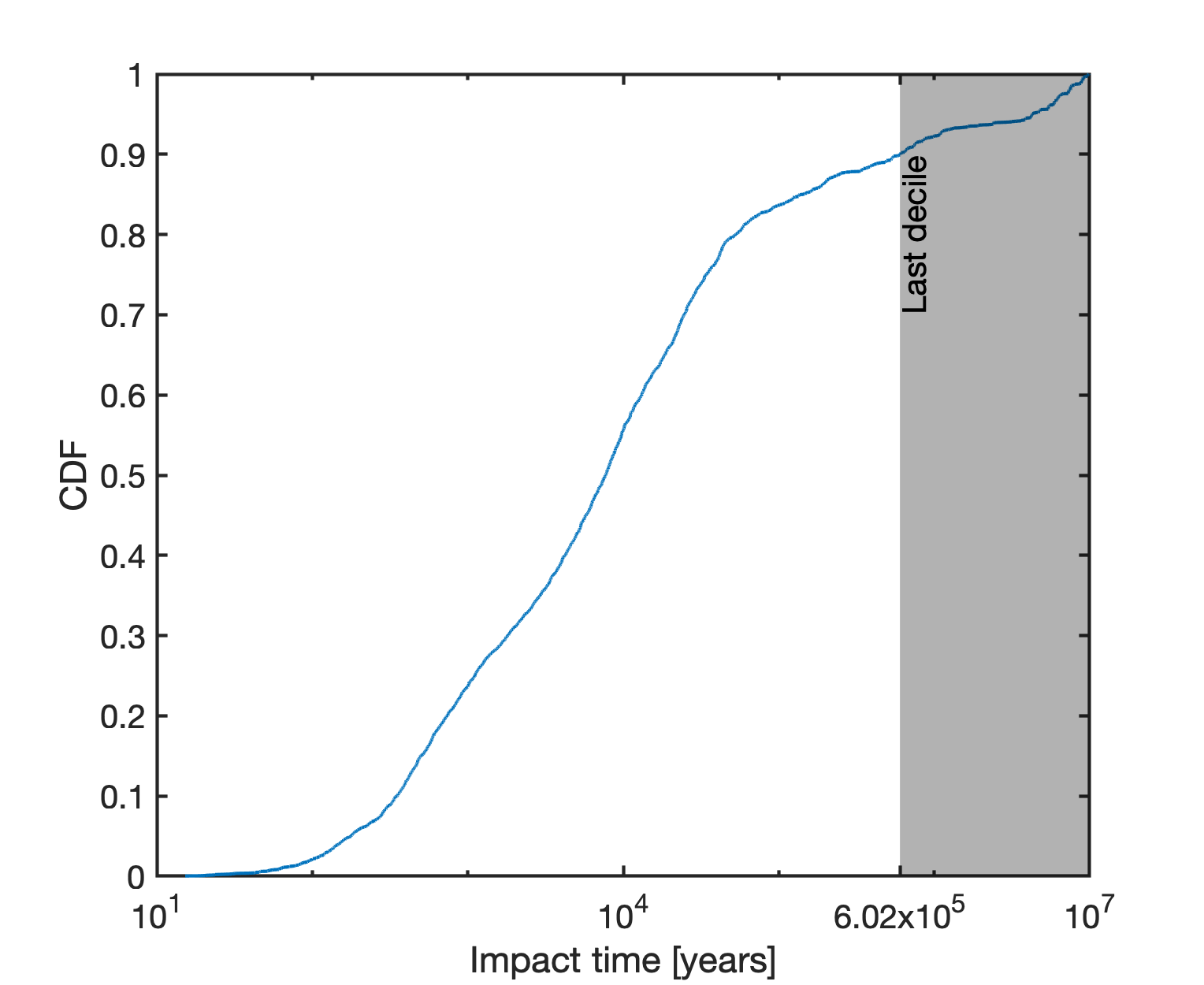}
\caption[Giant impact time]{Cumulative distribution of all the giant impact events with respect to time during the imperfect collision simulations. Grey area denotes the latest 10\% of the collisions.}\label{fig:impacttime}
\end{figure}

Figure \ref{fig:impacttime} shows the CDF for the occurrence times of all giant impact events detected during the imperfect collision simulations. More than $90\%$ of the giant impacts happened before 1 Myr (the white area in the figure), and 50\% of the impacts occurred within $10^4$ yr. Given that the planetary systems are centred around $a\sim 0.1$ AU, this latter figure corresponds to $\sim 3 \times 10^5$ dynamical times, indicating that the initial conditions do not result in excessively short accretion times. Instead, the systems have time to undergo substantial dynamical relaxation during the epoch of accretion. Furthermore, the fact that only 10\% of the collisions occur after 1 Myr indicates that our run times of 10 Myr are long enough to have formed long-term stable systems in most cases. However, the fact that some collisions are occurring at late times also indicates that some of our final planetary systems would have evolved further if the integrations had been extended. Finally, we note that with 90\% of the collisions occurring in the simulations within 1 Myr, this implies that if the protoplanets we consider in the initial conditions were formed within the life time of the gaseous protoplanetary discs, then substantial collisional evolution would have likely occurred while the gas was still present as disc life times are typically 3 Myr \citep{Haisch2001}. \citet{2015A&A...578A..36O} have shown that under such conditions the effects of migration cannot be ignored and strongly influence the architectures of the resulting systems.

\subsubsection{Eccentricities, inclinations and masses}\label{subsubsec:EIM}
Table \ref{tab:all_element} lists the mean values of $e$, $I$ and the $K$-values for each subset of runs associated with each of the Kepler templates. Hence, the averaging has been performed over the final outcomes of the 10 simulations associated with perfect/imperfect accretion, and high and low initial eccentricities/inclinations. While there is some variation of the mean eccentricities and inclinations when comparing the different initial eccentricities/inclinations and the accretion prescription (particularly for \texttt{Kepler80}), the largest variation is observed when comparing the different Kepler templates. For example, the \texttt{Kepler55} runs all give $\left \langle e \right \rangle \sim 0.06$, where the \texttt{Kepler296} runs give higher values distributed around $\left \langle e \right \rangle \sim 0.09$. 
\begin{table*}
\centering
\caption{Mean values of $e$, $I$ and $K$-values from the different simulation subsets from all 8 Kepler templates. The numbers in parentheses are the standard deviations about the respective means.}
\label{tab:all_element}
\begin{tabular}{cc|cccc|cc|cc|c}
\hline
\hline
System              & Kepler  & Imperfect      & Imperfect      & Perfect      & Perfect      & All     & All     & All     & All     & All       \\
Elements              & System & <Higher>      & <Lower>      & <Higher>       & <Lower>      & <Higher>      & <Lower>     & Imperfect     & Perfect     & Sets      \\
\hline                    
$\left\langle e\right\rangle$ & 55  & 6.32$_{(5.71)}$  & 6.27$_{(5.50)}$  & 6.03$_{(4.46)}$ & 6.34$_{(5.97)}$ & 6.18 & 6.31  & 6.30  & 6.19 & 6.24 \\
		  ($\times 10^{-2}$)  & 80  & 3.21$_{(4.47)}$  & 5.09$_{(5.50)}$  & 1.85$_{(2.25)}$ & 1.88$_{(1.90)}$ & 2.53 & 3.49 & 4.15 & 1.87 & 3.01 \\
  							  & 84  & 7.02$_{(4.86)}$  & 5.20$_{(4.56)}$  & 6.32$_{(6.14)}$ & 5.91$_{(4.39)}$ & 6.67 & 5.56  & 6.11  & 6.12 & 6.11 \\
   							  & 102 & 5.55$_{(6.64)}$  & 3.74$_{(2.81)}$  & 5.64$_{(4.46)}$ & 6.27$_{(7.30)}$ & 5.60 & 5.01  & 4.65  & 5.96 & 5.30 \\
    						  & 154 & 8.46$_{(6.65)}$  & 5.25$_{(5.68)}$  & 7.23$_{(7.02)}$ & 7.12$_{(7.17)}$ & 7.85 & 6.19  & 6.86  & 7.18 & 7.02 \\
 							  & 169 & 4.42$_{(2.72)}$  & 5.49$_{(6.22)}$  & 4.68$_{(4.06)}$ & 4.70$_{(6.92)}$ & 4.55 & 5.10  & 4.96  & 4.69 & 4.82 \\
 							  & 292 & 5.43$_{(3.08)}$  & 5.04$_{(4.00)}$  & 6.65$_{(5.20)}$ & 5.72$_{(3.60)}$ & 6.04 & 5.38  & 5.24  & 6.19 & 5.71 \\
 							  & 296 & 10.28$_{(9.86)}$ & 11.40$_{(11.05)}$ & 7.60 $_{(5.72)}$ & 9.90$_{(6.64)}$ & 8.94 & 10.65 & 10.84 & 8.75 & 9.80 \\
\hline
$\left\langle I\right\rangle$ 		& 55  & 4.34$_{(4.85)}$  & 3.74$_{(2.90)}$  & 4.31$_{(3.55)}$ & 3.47$_{(2.61)}$ & 4.33 & 3.61  & 4.04  & 3.89 & 3.97 \\
				($\times 10^{-2}$)  & 80  & 1.18$_{(1.20)}$  & 1.67$_{(2.45)}$  & 2.72$_{(3.89)}$ & 1.41$_{(1.96)}$ & 1.95 & 1.54  & 1.43  & 2.06 & 1.74 \\
				  $[\textrm{rad}]$  & 84  & 3.38$_{(2.86)}$  & 3.63$_{(3.51)}$  & 3.18$_{(3.61)}$ & 3.97$_{(4.39)}$ & 3.28 & 3.80  & 3.51  & 3.58 & 3.54 \\
								    & 102 & 2.52$_{(3.03)}$  & 2.38$_{(2.45)}$  & 3.65$_{(4.30)}$ & 3.93$_{(3.64)}$ & 3.08 & 3.15  & 2.45  & 3.79 & 3.12 \\
								    & 154 & 5.00$_{(5.68)}$  & 5.62$_{(3.82)}$  & 4.94$_{(6.94)}$ & 5.38$_{(6.33)}$ & 4.97 & 5.50  & 5.31  & 5.16 & 5.23 \\
								    & 169 & 2.38$_{(1.92)}$  & 3.48$_{(3.41)}$  & 3.42$_{(2.32)}$ & 3.00$_{(2.30)}$ & 2.90 & 3.24  & 2.93  & 3.21 & 3.07 \\
								    & 292 & 4.19$_{(2.95)}$  & 2.94$_{(2.66)}$  & 2.33$_{(1.87)}$ & 2.87$_{(2.42)}$ & 3.26 & 2.90  & 3.56  & 2.60 & 3.08 \\
								    & 296 & 5.90$_{(3.56)}$  & 7.89$_{(7.91)}$  & 5.32$_{(2.76)}$ & 5.90$_{(4.67)}$ & 5.61 & 6.90  & 6.89  & 5.61 & 6.25 \\
\hline
$\left\langle K\right\rangle$ & 55  & 21.1$_{(6.43)}$  & 21.2$_{(6.91)}$  & 22.6$_{(6.25)}$ & 20.9$_{(7.17)}$ & 21.8 & 21.1  & 21.2  & 21.7 & 21.4 \\
 							  & 80  & 21.5$_{(11.37)}$  & 22.8$_{(11.09)}$  & 20.6$_{(10.40)}$ & 19.0$_{(10.40)}$ & 21.0 & 20.9  & 22.2  & 19.8 & 21.0 \\
 							  & 84  & 21.2$_{(4.80)}$  & 19.2$_{(5.42)}$  & 20.8$_{(4.79)}$ & 19.8$_{(4.81)}$ & 21.0 & 19.5  & 20.2  & 20.3 & 20.3 \\
 							  & 102 & 21.4$_{(6.54)}$  & 19.1$_{(6.01)}$  & 20.6$_{(5.97)}$ & 21.9$_{(4.59)}$ & 21.0 & 20.4  & 20.3  & 21.1 & 20.7 \\
 							  & 154 & 20.7$_{(6.82)}$  & 19.0$_{(6.46)}$  & 20.4$_{(7.43)}$ & 18.7$_{(8.16)}$ & 20.6 & 18.9  & 19.9  & 19.6 & 19.7 \\
 							  & 169 & 21.1$_{(6.23)}$  & 22.5$_{(7.61)}$  & 22.2$_{(6.74)}$ & 21.9$_{(9.70)}$ & 21.6 & 22.2  & 21.8  & 22.0 & 21.9 \\
 							  & 292 & 19.5$_{(3.87)}$  & 18.5$_{(4.23)}$  & 18.7$_{(4.40)}$ & 17.3$_{(5.28)}$ & 19.1 & 17.9  & 19.0  & 18.0 & 18.5 \\
 							  & 296 & 24.3$_{(7.47)}$  & 22.9$_{(9.15)}$  & 22.4$_{(5.69)}$ & 21.4$_{(4.37)}$ & 23.4 & 22.2  & 23.6  & 21.9 & 22.8 \\
\hline
\hline                  
\end{tabular}
\end{table*}

\begin{figure*}
\centering
\includegraphics[width=16cm]{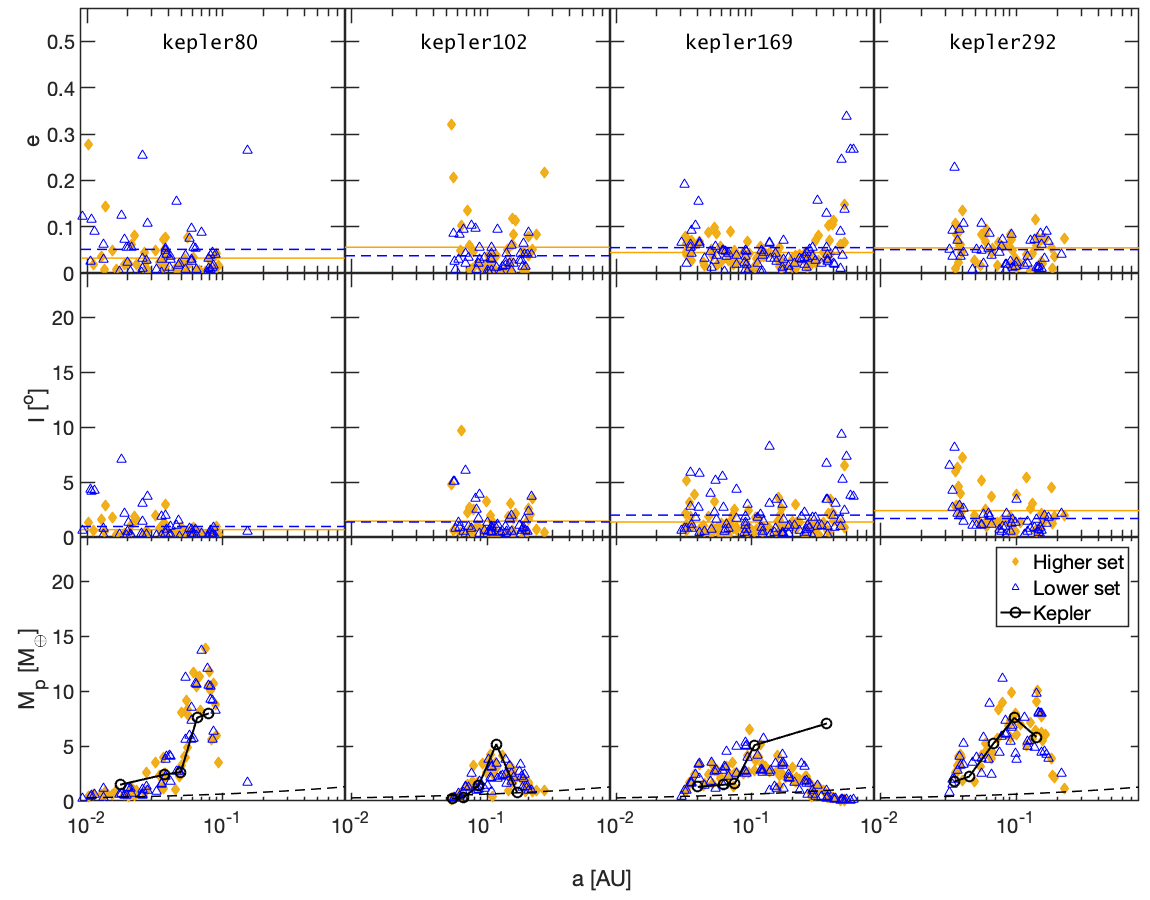}
\caption[All imperfect collision results for \texttt{Kepler80}, \texttt{102}, \texttt{169}, and \texttt{292}]{All imperfect collision simulation results from (from left to right) \texttt{Kepler80}, \texttt{102}, \texttt{169}, and \texttt{292} templates. The scatter plots show the eccentricities (top panel), inclinations (middle panel), and planet masses (bottom panel) with respect to their semi-major axis. The orange-diamonds data are from the higher initial eccentricity set, and the blue-triangle data are from the lower initial eccentricity set. The horizontal lines in each subplot show the mean values of the data in their respective colour (also plotted as solid lines for the higher initial eccentricity set and dashed lines for the lower initial eccentricity set). The black circles in the bottom panel denote the masses and semi-major axes of the observed Kepler planets. The black dashed lines indicate the detection limit applied when undertaking the synthetic transit observations described in section~\ref{subsec:obsmulti}.}\label{fig:allseim}
\end{figure*}

The top and middle panels of figure~\ref{fig:allseim} show the values of $e$ and $I$ with respect to semi-major axis $a$ from four of the simulation templates performed with the imperfect collision model (see figure \ref{fig:appseim} in appendix \ref{app:remain} for the results of the other 4 templates that adopt imperfect collisions, and figures \ref{fig:allmeim} and \ref{fig:appmeim} for all 8 templates that use the perfect collision model). The plots show that the final distributions arising from the higher set (blue triangles) and the lower set (yellow diamonds) are very similar. This is not surprising as the initial eccentricities and inclinations in all of these runs are considerably smaller than the mean values at the ends of the simulations. Hence, the final values are determined by planet-planet scattering and collisional damping, and little memory is retained of the original eccentricity and inclination values. This would not be the case if the initial eccentricities and inclinations had been comparable to or larger than the values obtained from dynamical relaxation \citep{2017AJ....154...27M}. It is common to see that $e$ and $I$ have relatively high values near the inner and outer edges of the systems, perhaps best illustrated by the \texttt{Kepler102} and \texttt{Kepler169} templates. This feature was already noted by \citet{2013ApJ...775...53H} in their study of in situ formation of super-Earths, and arises because bodies at the edge of the initial annulus of protoplanets are scattered outwards and do not experience collisions that tend to damp the eccentricities and inclinations. 

The bottom panels of Figure \ref{fig:allseim} show the final planet masses versus their semi-major axes. A striking feature of these plots is how the simulated systems (denoted by blue triangles for higher set runs, and yellow diamonds for lower set runs) generally match the observed Kepler systems (denoted by black circles joined by solid lines). Hence, based on this comparison alone, it is reasonable to conclude that the initial conditions and formation histories that we simulate here might be reasonable approximations to those that applied to the actual Kepler systems we have used as templates. The exception is \texttt{Kepler169}, where the outer regions of these simulations failed to generate significant collisional growth because the initial masses of the protoplanets there, generated by the method mentioned in section \ref{subsec:IC}, were too small (giving initial $K$-values $>$15), leading to instability times longer than the 10 Myr simulation run times.

In view of this, we extended the run times of the \texttt{Kepler169} template simulations to 100 Myr, using the perfect accretion routine (see figure \ref{fig:169_10100_eim} in appendix \ref{app:remain} for the comparison). As expected, the outer regions of the systems experienced increased growth and provided better agreement with the original Kepler masses. This suggests that a better strategy for future work would be to run simulations for a set number of orbits measured at the outer edges of the systems, instead of a fixed number of years as was done in this work.
 
 Although the final planet masses in the simulations match their Kepler templates on average, it is worth noting that when we consider the CDF of planet masses later in this paper, and compare it with that obtained from the original Kepler template systems (see figure~\ref{fig:mass_cdf}), the agreement is not good because the simulations produce a range of planetary systems, some of which have higher multiplicity than 5 and hence contain planets with relatively low masses. 

As with the eccentricities and inclination distributions discussed above, there are no systematic differences in the final planetary masses when comparing the high and low initial eccentricity/inclination subsets of runs. 

\begin{figure}
\centering
\includegraphics[width=8.4cm]{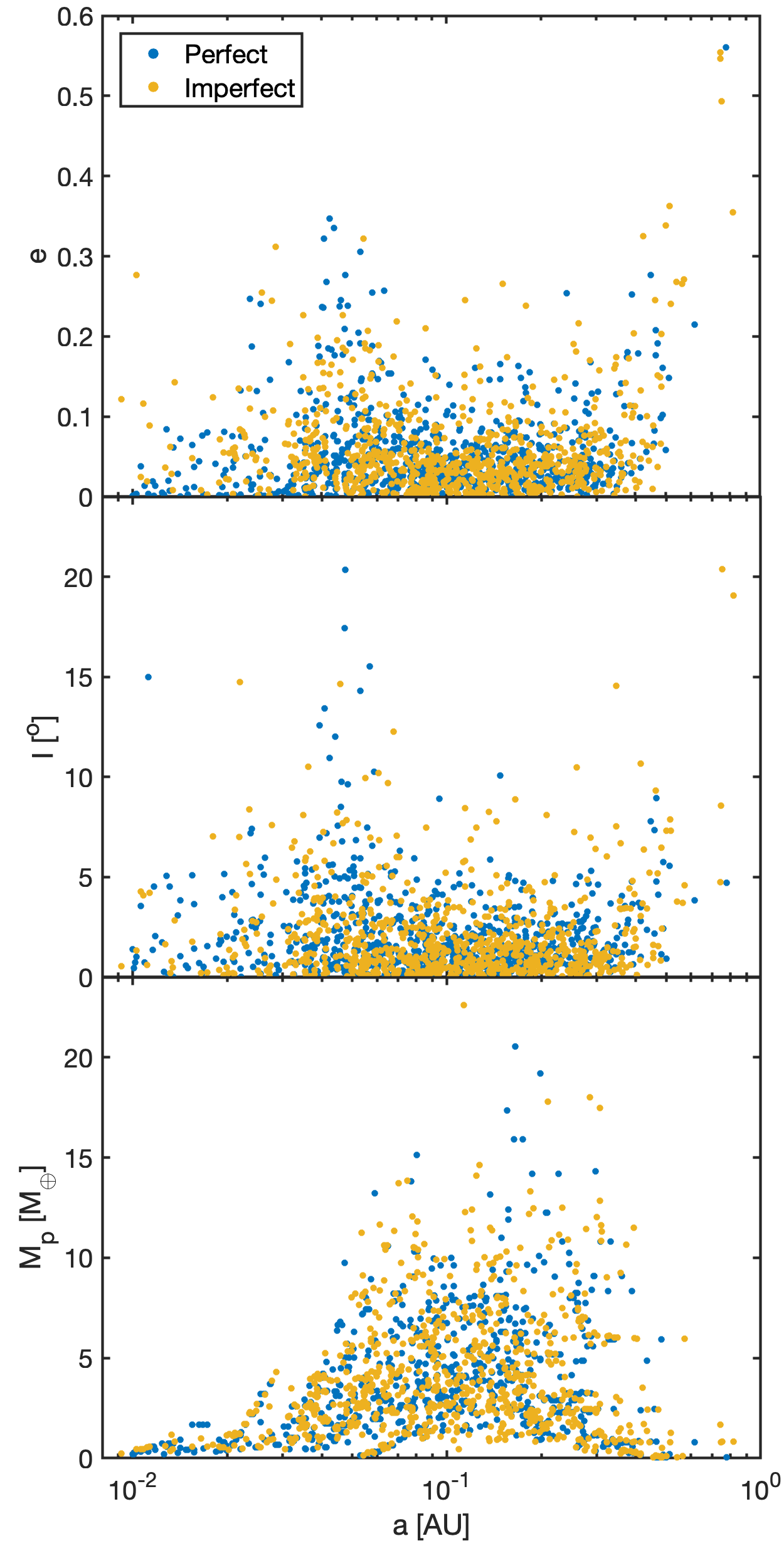}
\caption[Comparison between $e$, $I$, and $M_{\rm p}$ arising from perfect and imperfect collision simulations]{Scatter plots comparing $e$, $I$ and $M_{\rm p}$ as a function of $a$ arising from perfect (blue points) and imperfect (yellow points) collision simulations.}\label{fig:compareMandS_eim}
\end{figure}

\begin{figure}
\centering
\includegraphics[width=8.4cm]{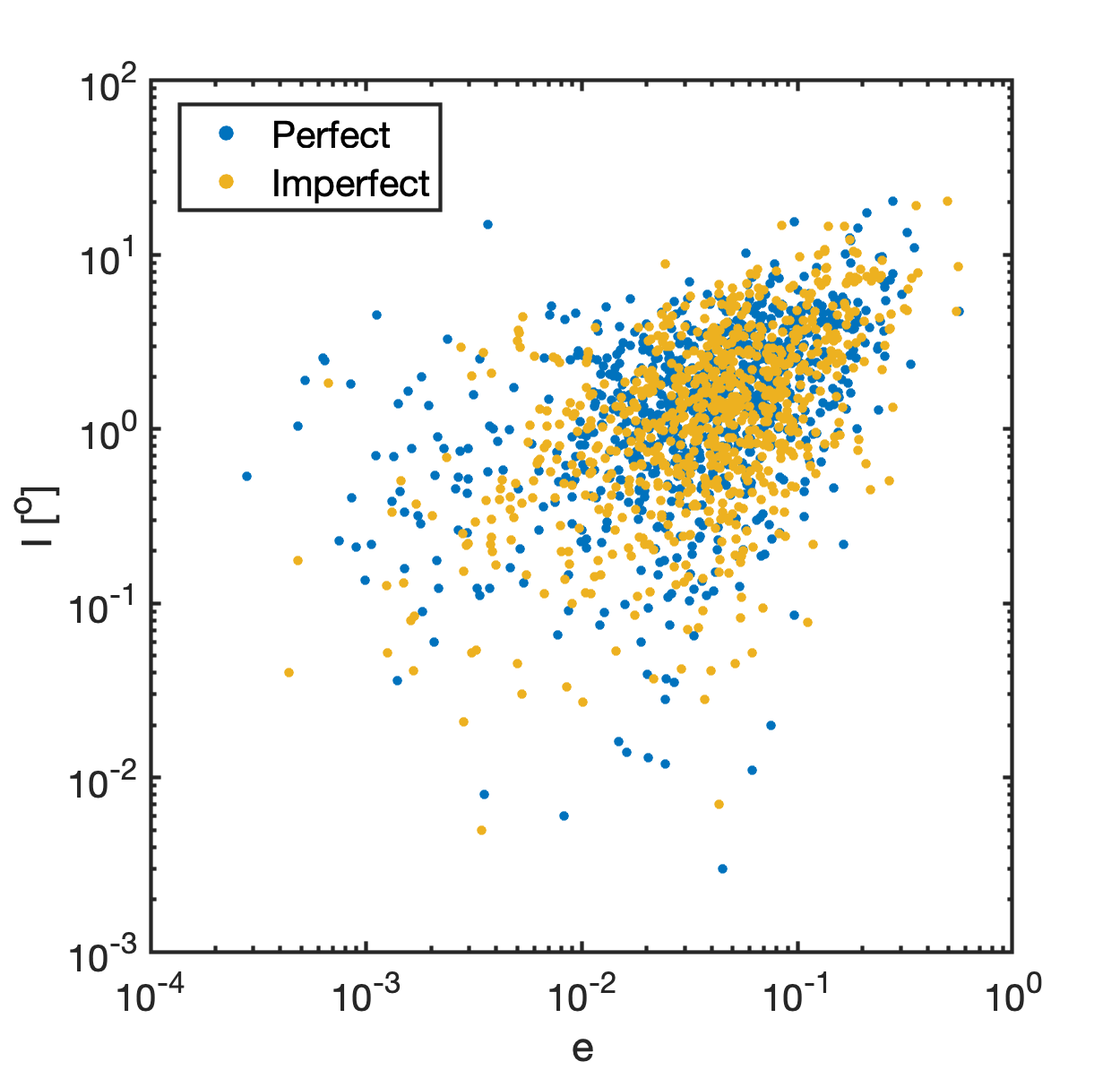}
\caption[Comparison of $e$ and $I$ by perfect and imperfect collision simulations]{Scatter plot showing correlation between $e$ and $I$ from all runs.  Perfect collision results are shown by blue points, and imperfect collision results are shown by yellow points.}\label{fig:compareMandS_ei}
\end{figure}

The high values of $e$ and $I$ at the edges of the system due to planets being scattered but experiencing fewer collisions there, discussed above and noted by \citet{2013ApJ...775...53H}, can also be seen in Figure~\ref{fig:compareMandS_eim}, which shows the final planets from all runs in the $a$-$e$ plane (top panel), $a$-$I$ plane (middle panel), and $a$-$M_{\rm p}$ plane (bottom panel).   We also see from the lower panel that higher mass planets occupy the centre of the $a$-$M_{\rm p}$ plane, where collisional growth occurs more frequently, with lower mass planets being present at the edges of the annuli  where collisions occur less frequently. Figure~\ref{fig:compareMandS_ei} shows that $e$ and $I$ are strongly correlated, as expected for systems that have undergone dynamical relaxation. 

\subsubsection{Period ratios and $K$-values}\label{subsubsec:PeriodandK}
\begin{figure}
\centering
\includegraphics[width=8.4cm]{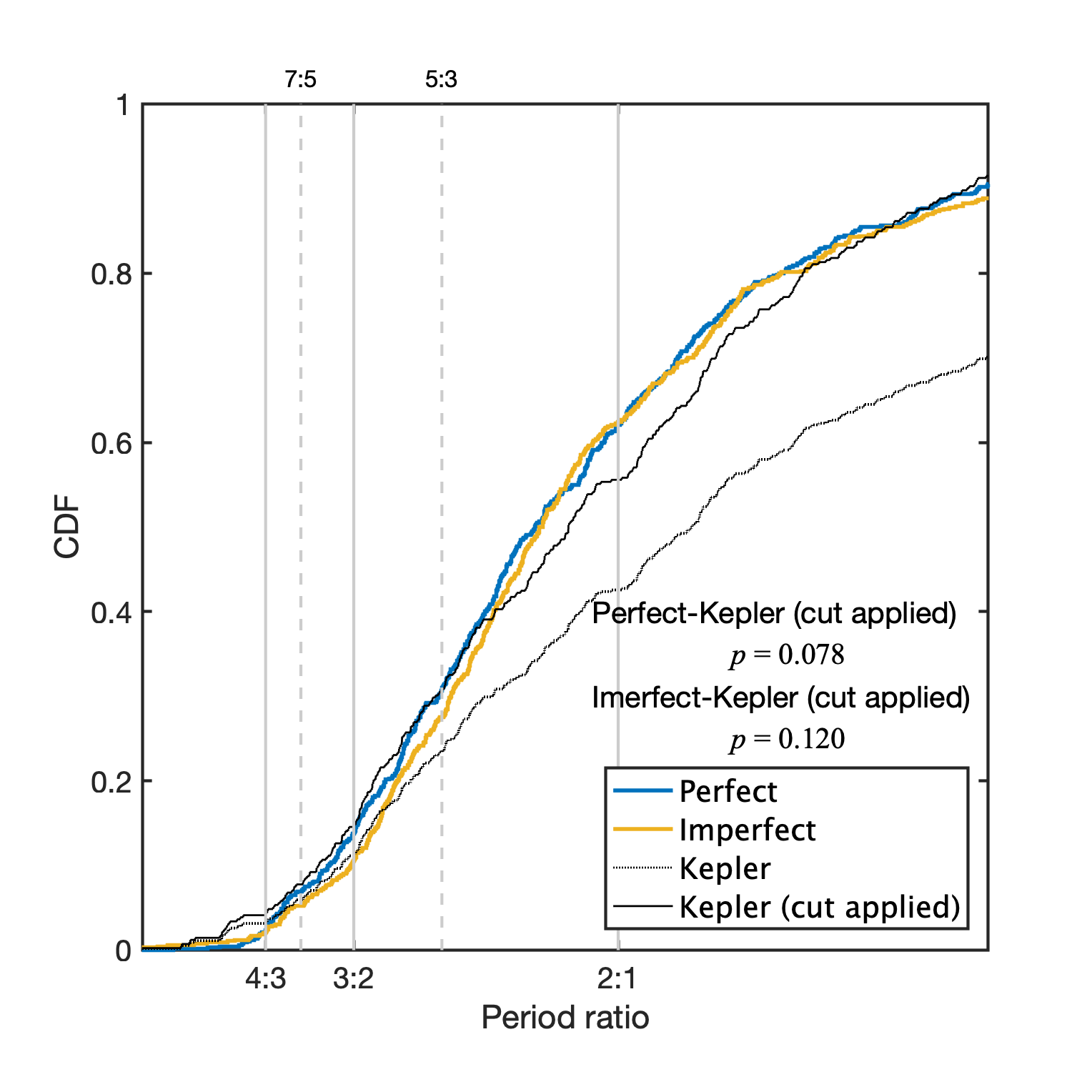}
\caption[Cumulative period ratio]{CDFs of the period ratios of all adjacent planet pairs. The blue and yellow lines correspond to the perfect and imperfect collision models, respectively. The black dotted line includes all the original Kepler planet pairs. The solid black line shows the original Kepler planet pairs with a cut off for period ratio > 3.}\label{fig:periodratio_cumulative}
\end{figure}

Figure \ref{fig:periodratio_cumulative} shows that the perfect and imperfect accretion simulations provide similar cumulative distributions of the period ratios between neighbouring planet pairs. We can compare these with the distribution of period ratios for the actual Kepler multi-planet systems. In section~\ref{sec:PeriodRatios} below, we also compare the period ratios obtained from the simulations when they are synthetically observed with the Kepler data, but here we focus on the intrinsic period ratios. For period ratios smaller than 4:3, we see that the Kepler data shows an excess compared to the simulations. A K-S test performed on a subset of period ratios between 5:4 and 4:3 gives $p$-values of  $1.69\times10^{-4}$ and 0.029 when comparing the Kepler data with the perfect and imperfect collision simulations, respectively, demonstrating that the distributions are different. Hence, some process occurred during the formation of at least some Kepler systems that allowed the survival of more compact architectures, which are nonetheless non-resonant. Dynamical relaxation and collisional evolution in the absence of any dissipative process clearly results in such closely separated planet pairs being destabilised, suggesting those Kepler systems with particularly compact configurations formed in a dissipative environment and did not undergo dynamical instability in spite of the close proximities of the planets. One such system not considered here that displays this property is Kepler-11 \citep{2014ApJ...795...32M}.

The full Kepler data set, without any limits in period ratio being applied, clearly contains too many planet pairs with large period ratios compared to the simulation intrinsic outcomes, and a K-S test comparing the data and simulations results in $p$-values $<0.05$.
Again, such system architectures do not naturally arise from a formation scenario in which even a wide annulus of protoplanets undergoes dynamical instability and collisional growth, since this mode of evolution results in neighbouring planets being separated by $\sim 20$ mutual Hill radii (see discussion below).
Instead, additional processes would need to be invoked which either cause the initial distribution of protoplanets to have a more complex structure involving concentrations around particular orbital radii, or which involve orbital migration because formation occurred in the presence of either a gas or planetesimal disc.  
Considering planet pairs in the Kepler data with a maximum period ratio of 3:1 results in much better agreement between the observational data and simulations, with $p$-values of 0.078 and 0.120 for perfect and imperfect collisions respectively. However, we also note here that the synthetically observed systems, described in section~\ref{sec:PeriodRatios}, provide a distribution of period ratios that is quite different to that obtained from the simulations directly, due to the fact that mutual inclinations between the planets lead to some planets not being detected during the observations. This has the effect of increasing the numbers of systems with large period ratios.

One feature within the Kepler multiplanet systems that the simulations do not reproduce particularly well is the known peaks in occurrence rates of planet pairs just outside of the 3:2 and 2:1 resonances (seen in particular as a flattening and then rise in the CDF at around period ratio 2:1 in figure \ref{fig:periodratio_cumulative}). The cumulative distribution for the simulation data shows a very modest inflection around the 2:1 resonance, but it is not as pronounced as in the Kepler data, and is not statistically significant. It is noteworthy that \citet{2013ApJ...770...24P} were able to reproduce the resonance features using 3-body integrations that resulted in final systems of two planets for planet masses $20 \le M_{\rm p} \le 100$ M$_{\oplus}$. The end states of our simulations always have more than two planets, and the final planet masses are typically $<10$ M$_{\oplus}$. These properties likely serve to reduce the prominence of features in the period ratio distribution near first order mean motion resonances, and leave open the question of what dynamical processes have given rise to the near-resonance features in the period ratio distributions of the Kepler planets.

\begin{figure}
\centering
\includegraphics[width=8.4cm]{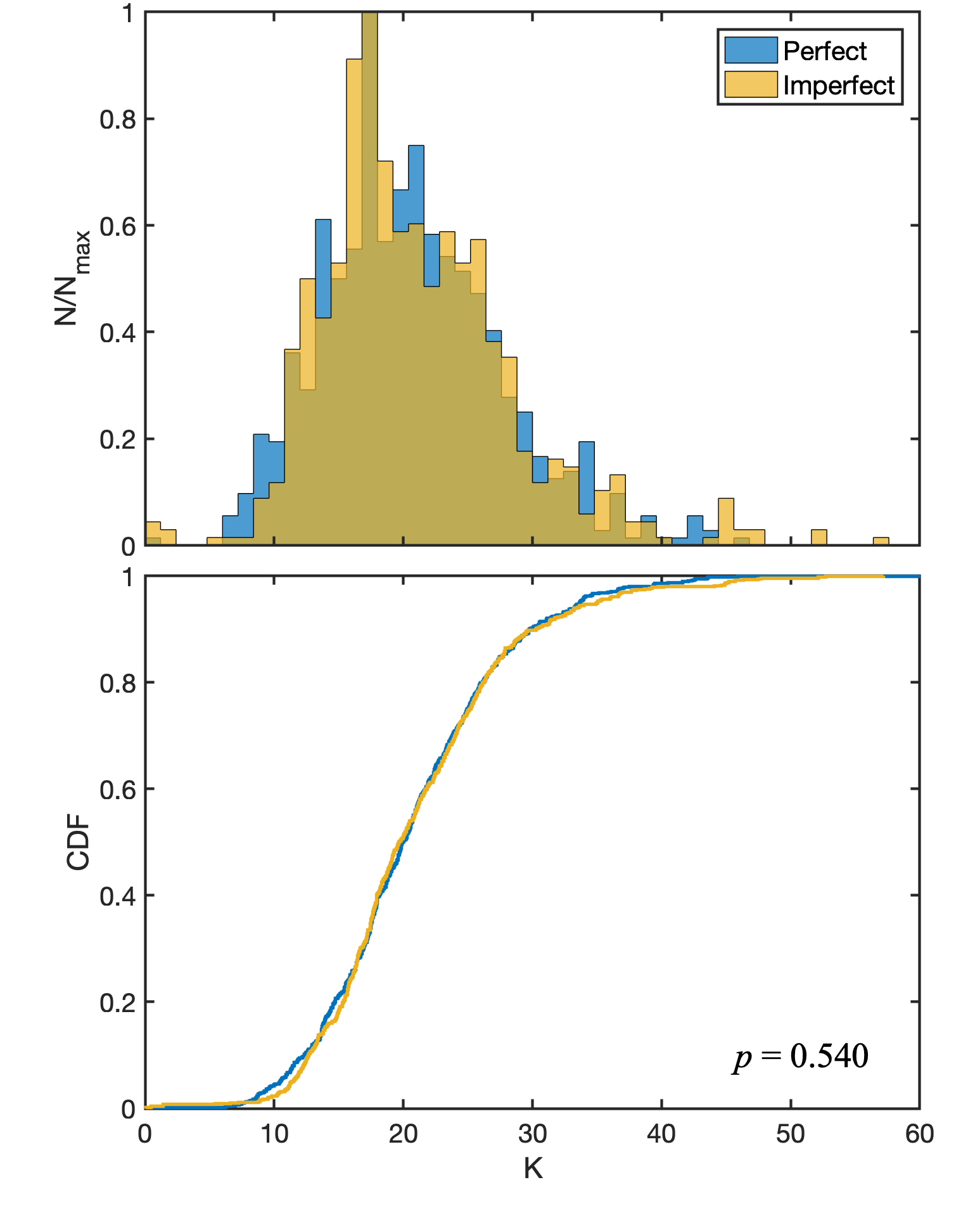}
\caption[Cumulative K-value]{The normalised distributions (top panel) and the cumulative distributions (bottom panel) of the $K$-value of the planet pairs obtained in all our simulations. Perfect collision simulations are shown in blue, and imperfect collision simulations are shown in yellow.}\label{fig:allK}
\end{figure}

\begin{figure*}
\centering
\includegraphics[width=16cm]{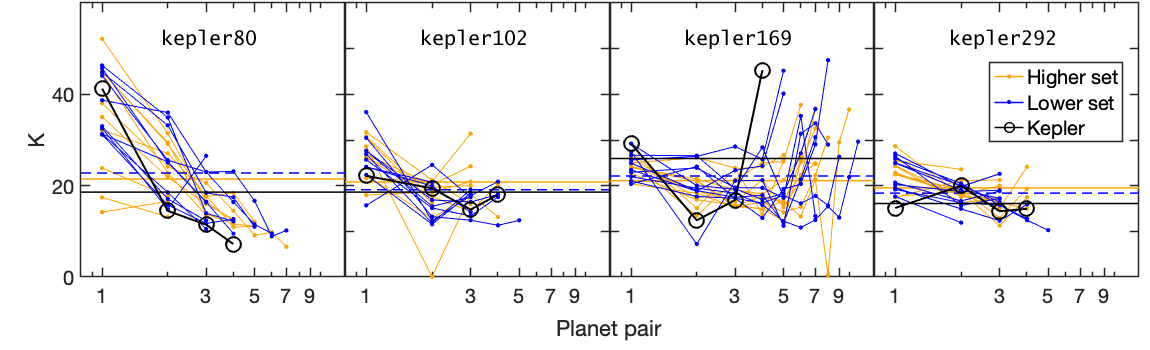}
\caption[All $K$-values from imperfect collision simulations of \texttt{Kepler80}, \texttt{102}, \texttt{169}, and \texttt{292}]{All $K$-values for neighbouring planet pairs from imperfect collision simulations of (from left to right) \texttt{Kepler80}, \texttt{102}, \texttt{169}, and \texttt{292} templates. The orange diamonds are from the higher initial eccentricity set and the blue triangles are from the lower initial eccentricity set. The horizontal lines in each subplot are the mean values of the data in their respective colour (also plotted as solid lines for the higher initial eccentricity set and dashed lines for the lower initial eccentricity set). The black circles denote the $K$-values of the original planets pairs, as listed in table \ref{tab:Kval}.}\label{fig:allsK}
\end{figure*}

The distributions of the $K$-values from all our simulations can be seen in figure \ref{fig:allK}. Both collision models result in similar distributions, with a $p$-value of 0.54, with 50\% of systems having $10 \le K \le 20$, and the maximum value of $K$ being $\sim 50$. Although not obvious in the plot, the number of planets with $K<10$ decreases to zero and then rises again close to $K=0$, with these latter planets surviving because they are protected by a 1:1 resonance. \citet{2018arXiv180908499W} have presented similar simulation results on this feature in the $K$-value distribution and its relation to the 1:1 resonance (period ratio < 1.05), and we discuss these co-orbital systems further in the next section. Figure~\ref{fig:allsK} shows the value of $K$ for each planet pair using the same four Kepler templates shown in figure \ref{fig:allseim}. The distributions of the $K$-values are similar, independent of whether we consider the high or low eccentricity/inclination set.  And they also show similar $K$-values compared to the original Kepler systems (shown by the black circles joined by lines), based on the adopted mass-radius relation, although again the \texttt{Kepler169} system is an exception (see figure \ref{fig:169_10100_K} in appendix \ref{app:remain} for a comparison to systems that were evolved for 100 Myr instead of 10 Myr). 

\section{Co-orbital planet pairs}\label{sec:coorbit}
In section \ref{subsubsec:PeriodandK}, we stated that a small number of planets have very small $K$-values, and these are shown in the figure~\ref{fig:allK}. Further investigation has shown that these planets have been captured into 1:1 co-orbital resonances, and these co-orbital planets make up about 1\% (4 out of 320 simulations) of the total number of planet pairs.

\subsection{Stability}
\begin{figure}
\centering
\includegraphics[width=8.4cm]{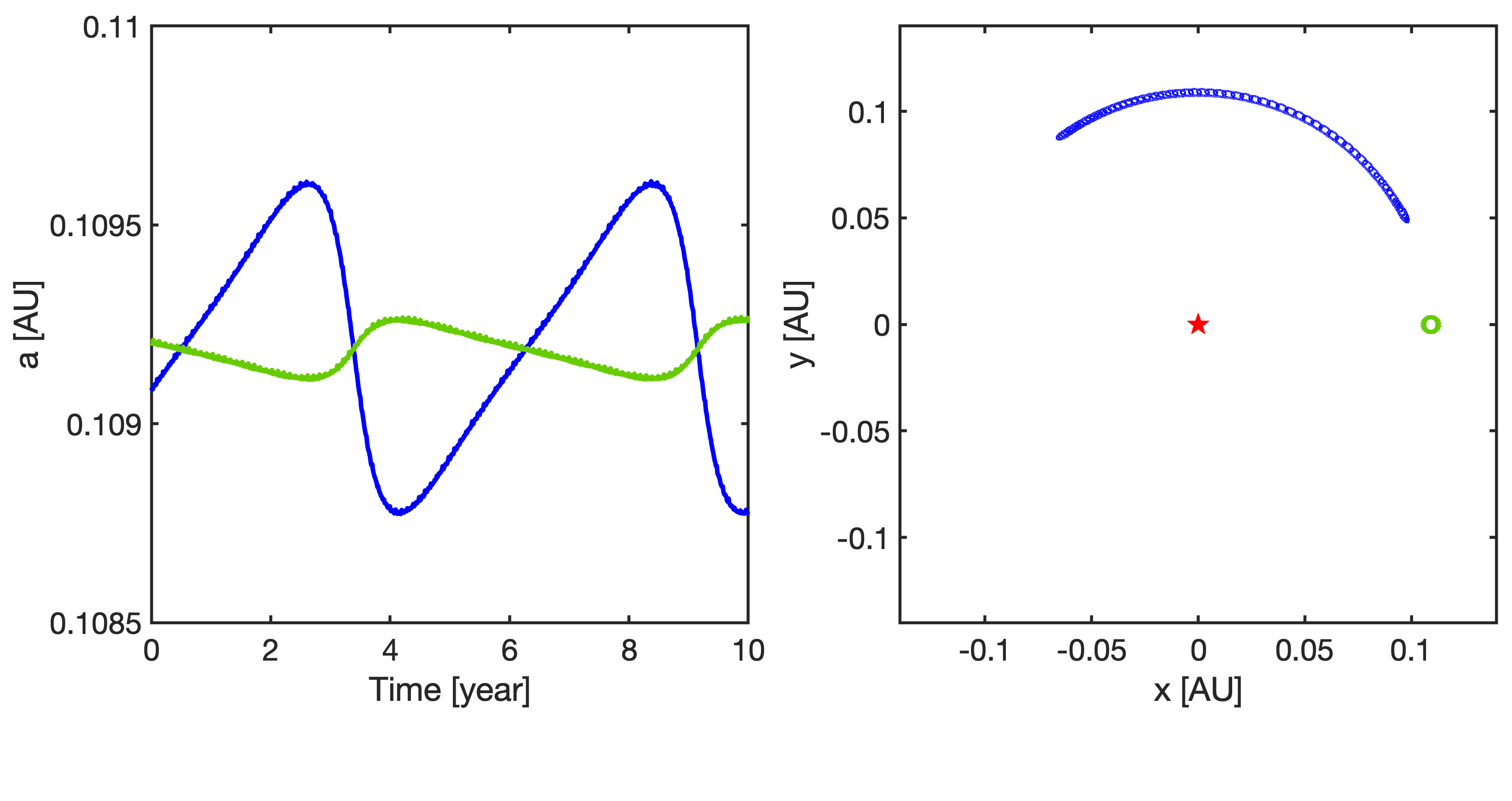}
\caption[Semi-major axis and orbit in co-rotating frame]{Left panel shows the semi-major axis evolution of the co-orbital planet pair. The right panel shows the orbit trajectory of the same planet pair in a frame that corotates with the planet denoted by the green open circle.}\label{fig:corotate}
\end{figure}

In spite of the very small $K$-values, the 1:1 resonance protects co-orbital planet pairs from instability. In general both stable tadpole orbits, which involve libration around the L4/L5 points, and horseshoe orbits are permitted \citep{1981Icar...48....1D,1981Icar...48...12D}, and we see examples of both of these orbit types in the simulations. As the simulation run times are 10 Myr, and the co-orbital pairs are found to form early in some simulations, we find tadpole and horseshoe orbits which are stable for 9.5$\times10^6$ years. This is in spite of the co-orbital pairs being in systems of high multiplicity, where the resonance configuration is subject to external perturbations. Previous studies have shown that co-orbital planet pairs can be stable for up to $10^9$ yr \citep{2000MNRAS.319...63T}. Figure \ref{fig:corotate} shows an example of the semi-major axis versus time during the last 10 years of one simulation (the left panel) and the corresponding orbital trajectory of the 1:1 resonance planet pair from the same model in a frame that co-rotates with one of the planets (the right panel). From the semi-major axis plot, we can see the 2 planets undergo a periodic exchange of radial location, and the co-rotating plot shows that the orbit is a tadpole orbit.

In other simulations that produce 1:1 resonant planet pairs, we see similar characteristics in the semi-major axis evolution. Although it is generally expected that horseshoe orbits are not as stable as tadpole orbits \citep{1981Icar...48...12D}, the planet pairs in horseshoe orbits produced in the simulations are found to be stable over the runs times we consider.  The maximum value of the period ratio among all the co-orbital pairs is $\sim$1.05:1, in agreement with the simulations by \citet{2018arXiv180908499W}, which show that period ratios in the range 1.05 to 1.1 are unstable (independent of whether or not the system has two planets or a higher multiplicity).

\subsection{Formation}
\begin{figure}
\centering
\includegraphics[width=8.4cm]{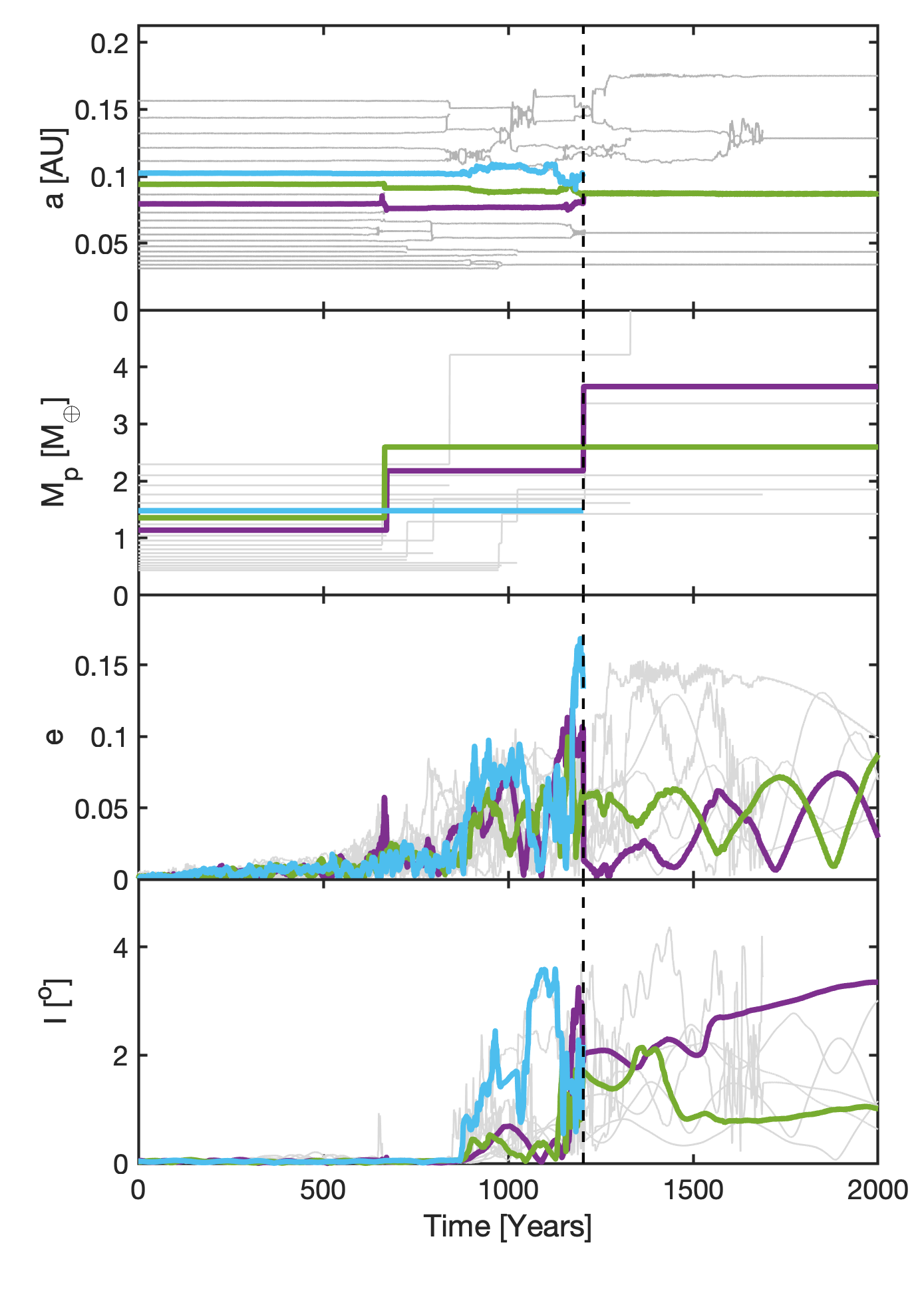}
\caption[An example of co-orbit planet pair formation]{An example of the formation of a co-orbital planet pair by an inelastic collision. Shown are the evolution of the semi-major axes (top panel), planet masses, (second panel) eccentricities (third panel), and inclinations (fourth panel). The time interval shown is the first 2,000 years of the simulation, and the collision occurred after $\sim$1,200 years (dashed line). The  representing the three bodies involved in the encounter are marked in blue, green, and purple, where we name them planet-b, planet-g, and planet-p, respectively.}\label{fig:coorbitaeim}
\end{figure}

All the co-orbital planet pairs form fairly early in our simulations (within a few thousand years). Figure \ref{fig:coorbitaeim} demonstrates the formation of a co-orbital pair by a 2-body collision event in the perfect collision model. The top panel shows the evolution of the semi-major axis during the first 2,000 yr of the simulation. We can see the collision happened around 1,200 yr (black dashed line). The 3 bodies involved in the encounter are marked by blue, green and purple lines (labelled as planet-b, planet-g and planet-p, respectively, from now on). In this encounter, planet-g and planet-p are the surviving planets and they form a co-orbital tadpole orbit. The second panel of the figure shows the change of mass of the bodies during the evolution. We can see planet-b collided with planet-p during the encounter, and the resulting body ended up with the appropriate energy and angular momentum so that it could settle into a co-orbital configuration with planet-g. The figure also shows the eccentricity and inclination evolution in the bottom two panels. The eccentricity evolution shows the angular momentum and energy exchange  \citep{2011MNRAS.410..455F,2013CeMDA.117...41F} between planet-g and planet-p. The fluctuations of these two orbital elements are seen to be reduced after the encounter compared to before it, because of the collision and formation of the co-orbital pair. The two co-orbital planets are mutually inclined by approximately 2$^{\circ}$, and hence it cannot be guaranteed that both planets would be detected in a photometric survey searching for transiting planets.

It is clear that formation of a co-orbital planet pair, involving two planets that were initially well-separated in orbital radius, requires energy and angular momentum loss from one of the planets. 
Within our simulations, there are three possible ways to achieve this energy loss: 1) an inelastic collision between two bodies resulting in the composite body having the appropriate energy and angular momentum to form a co-orbital pair with a third planet; 2) interaction with collision debris (in the form of multiple planetesimals) formed from an earlier collision, leading to the requisite change in energy and angular momentum by the members of the co-orbital pair; 3) a 3-body encounter in which energy and angular momentum from at least one planet in the co-orbital pair is given to a third body. The case illustrated in figure \ref{fig:coorbitaeim} corresponds to the first of these formation scenarios. Neither the interaction with debris nor the 3-body encounter formation mechansims were observed in the simulations, although simulations using the perfect and imperfect accretion routines both resulted in the formation of co-orbital pairs. All systems that formed co-orbital pairs did so early in the simulations, when the space density of planets and the probability of capture due to kinetic energy lost of the colliding planets were at their highest. 

\subsection{Resonance-induced TTV}
The co-orbital planet pair (planet-p and planet-g) shown in figure \ref{fig:coorbitaeim} survived to the end of the 10 Myr evolution. The final mutual inclination of this planet pair is $\sim2^{o}$, which makes it unlikely that both planets would be detected directly during a transit survey. On the other hand, this type of 1:1 resonance pair would induce transit timing variations (TTV) on each other, which might provide a signal indicating the presence of the other non transiting co-orbital planet. 

\begin{figure}
\centering
\includegraphics[width=8.4cm]{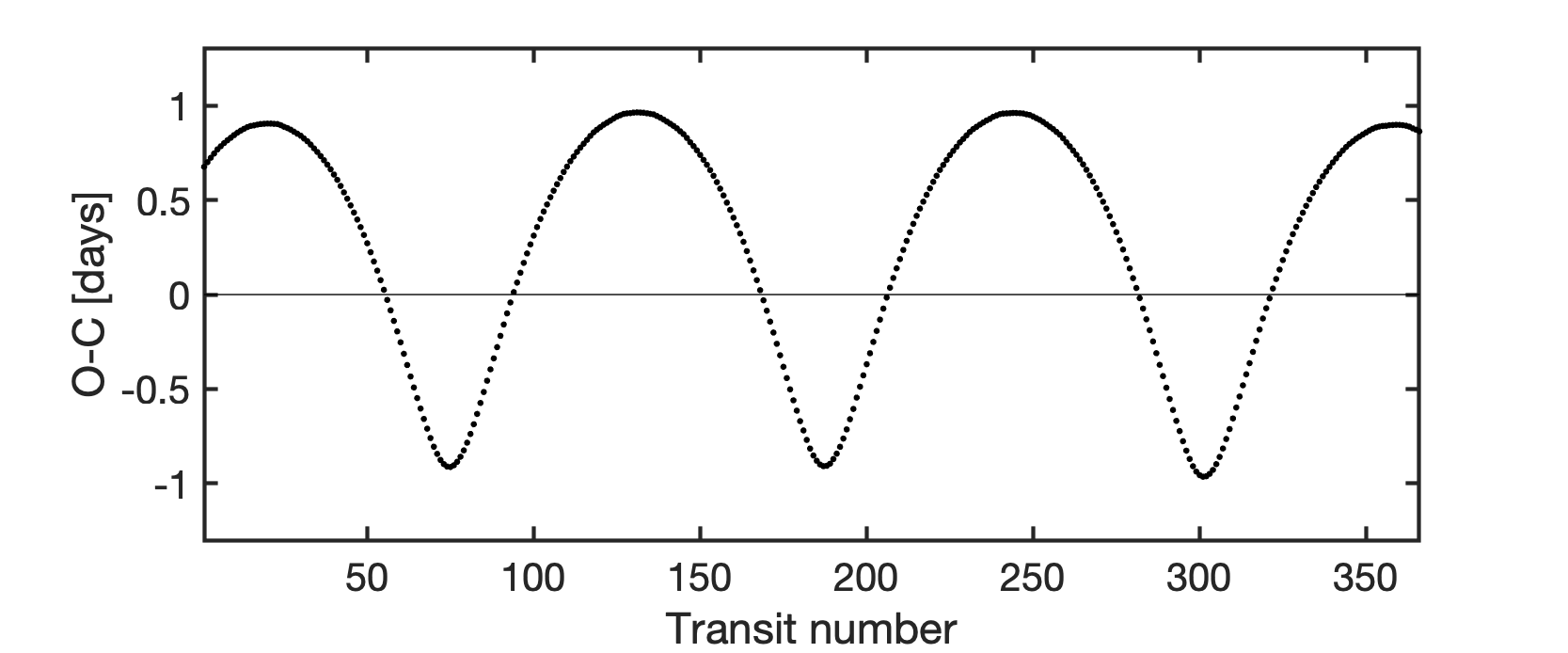}
\caption[TTV Example]{An example synthetic TTV signal for the 1:1 co-orbital planet pair discussed in the text. It shows the difference between the observed and calculated (predicted) times of planet-g's transit events for 10 years after the end of the formation simulation. A total of 366 transit events are observed within this 10 year period.}\label{fig:TTV}
\end{figure}

Figure \ref{fig:TTV} demonstrates the TTV signal expected for planet-g during a 10 year period after the end of the simulation. Here, we have calculated the mean orbital period of planet-g over this 10 year period ($P_{g}\approx 9.9872$ days), and have then computed the Observed$-$Calculated ($O-C$) times for the transits of planet-g. The amplitude of the TTV signal reaches $\pm0.9639$ day, and the maximum difference between adjacent periods is $\sim15$ minutes. If it was possible to pick up the transit signal of planet-g and confirm it as a planet within a transit survey, then the TTV signal would provide strong evidence of the presence of the other planet (e.g. planet-p in this case). However, we note that such a strong TTV signal might also provide a barrier to detecting co-orbital planets in the automated pipelines of transit surveys which adopted schemes such as box-least squares with fixed orbital periods, particularly for systems with low signal to noise.

All 4 of the co-orbital planet pairs formed in our simulations are mutually inclined and lead to a similar situation to that discussed above. So far, no confirmed co-orbital planets have been found. We note that simulations involving dynamical relaxation within a protoplanetary disc also suggest that 1:1 co-orbital planets are a natural outcome \citep{2006A&A...450..833C,2008A&A...482..677C}, but in that case the co-orbital pairs are expected to be co-planar and hence would both be detected directly in transit surveys. The strong TTV signal we demonstrated here might provide an explanation of why co-orbital pairs have not been found, and also provide a means of detecting non-coplanar co-orbital systems.

\section{Collision-induced composition changes}\label{sec:comp}
Recent analyses of the distribution of planetary radii for planets discovered by Kepler indicate the presence of a valley in the distribution for radii $1.6 \lesssim R_{\rm p} \lesssim 2$ R$_{\oplus}$ \citep{2017AJ....154..109F, 2018AJ....156..264F, 2018MNRAS.479.4786V}. The location of this valley has been widely interpreted as providing evidence that complete photoevaporation of hydrogen-helium envelopes from core dominated super-Earths has unveiled a population of bodies whose densities are consistent with them having Earth-like compositions \citep{2017ApJ...847...29O,2018ApJ...853..163J}. The lack of clear evidence for the solid cores having densities consistent with having significant ice fractions suggests that the observed cores did not migrate to their current locations after formation beyond the ice line.  An alternative possibility that we explore here is that high energy collisions during giant impacts may have changed the compositions of previously water-rich cores by stripping off the volatile outer layers.

For head-on collisions, the specific collision energy, $Q_{\rm R}$, can be calculated according to
\begin{equation}\label{eq:QR}
Q_{\rm R}=\frac{1}{2}\frac{\mu V_{\textrm{imp}}^{2}}{M_{\textrm{Total}}},
\end{equation}
where the reduced mass $\mu=(M_{1}M_{2})/(M_{1}+M_{2})$ (with $M_{1}$ and $M_{2}$ being the masses of the target and projectile, respectively), $M_{\textrm{Total}}$ is the total mass of the two colliding bodies, and $V_{\rm imp}$ is the relative impact velocity. \citet{2009ApJ...691L.133S} give a catastrophic disruption threshold energy, $Q_{\rm RD}^{*}$, which depends on the sizes of the colliding objects and $V_{\rm imp}$. $Q_{\rm RD}^{*}$ is defined as the energy needed to leave the largest remnant with less than 50\% of $M_{\rm Total}$ and is given by the relation 
\begin{equation}\label{eq:QRD_star}
Q_{RD}^{*}=\frac{1}{10^4}R_{C1}^{1.2}V_{\textrm{imp}}^{0.8},
\end{equation}
where $R_{C1}$ is the radius of a spherical body containing all of the colliding mass with density $\rho_{1}=1$~g~cm$^{-3}$, given by 
\begin{equation}\label{eq:RC1}
R_{C1}=\sqrt[3]{\frac{3M_{\textrm{Total}}}{4\pi \rho_{1}}}.
\end{equation}
With the masses of the two colliding bodies known, and the impact velocities measured from the $N$-body simulations, we can obtain $Q_{\rm R}$ and $Q_{\rm RD}^{*}$ directly. \citet{2010ApJ...719L..45M}, considered collisions involving differentiated bodies, with half of the mass of the colliding planets being water ice and the other half being rock, and found that the mass fraction of the core of the largest remnant, $M_{\textrm{core}}/M_{\textrm{lr}}$, can be fit by the expression
\begin{equation}\label{eq:McoreMlr}
\frac{M_{\textrm{core}}}{M_{\textrm{lr}}}=0.5+0.25\left( \frac{Q_{\rm R}}{Q_{\rm RD}^{*}}\right)^{1.2}.
\end{equation}
This power law of $Q_{\rm R}/Q_{\rm RD}^{*}$ is the best fit from their smoothed particle hydrodynamics simulations.

\begin{figure}
\centering
\includegraphics[width=8.4cm]{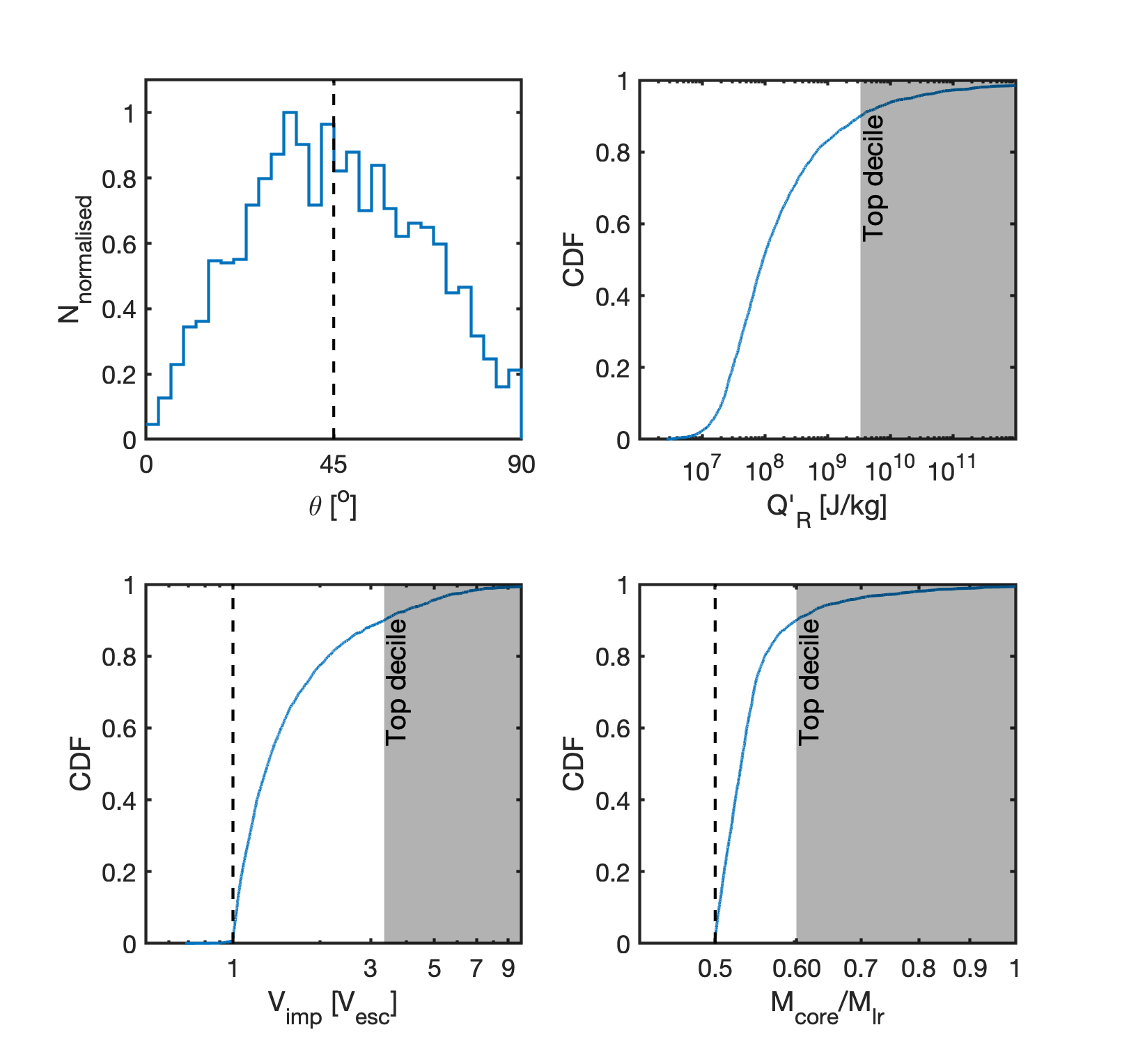}
\caption[lnformation of all the giant impact event in imperfect collision simulation]{Information concerning collisions occurring across all imperfect accretion simulations. The top left panel shows the distribution of impact angles. The top right panel shows the cumulative distribution of the impact energies, calculated using equation \ref{eq:Q'R}. The bottom left panel shows the cumulative distribution of the impact velocities in units of the escape velocity, and the bottom right panel shows the cumulative distribution of the ratio of $M_{\textrm{core}}/M_{\textrm{lr}}$, calculated using equation~\ref{eq:McoreMlr_b}. The grey area in each plot denotes the region where the cumulative number is within 10\% of the maximum.}\label{fig:impactinfo}
\end{figure}

The upper left panel of figure \ref{fig:impactinfo} shows the distribution of the recorded impact angles, $\theta$, in all the giant impact events from our imperfect accretion routine. It is clear that the majority of the collisions are not head-on. For a more realistic analysis for the composition changes, we also consider the effect of off-centre collisions. \citet{2012ApJ...745...79L} provide a correction to $Q_{\rm R}$ and $Q_{\rm RD}^{*}$ that allows them to be applied to off-centre collisions by considering the fraction of the projectile mass that directly intersects the target during a collision. The mass fraction of $M_2$ involved in the collision is defined as $\alpha$ (i.e. $\alpha=M_{2,\rm involved}/M_{2}$). After defining $\alpha$, the interacting reduced mass, $\mu_{\alpha}$, can be calculated by 
\begin{equation}
\mu_{\alpha}=\frac{\alpha M_{1}M_{2}}{\alpha M_{2}+M_{1}},
\end{equation}
where $\alpha$ can be calculated directly from the information recorded during the simulations by
\begin{equation}
\alpha=\frac{3R_{2}\left [ R_{\textrm{Total}}-R_{\textrm{Total}}\sin{\theta} \right ]^{2}-\left [  R_{\textrm{Total}}-R_{\textrm{Total}}\sin{\theta} \right ]^{3}}{4R^{3}_{2}},
\end{equation}
where $R_{\textrm{Total}}=R_1 + R_2$, and $\alpha=1$ when $R_{1}-R_{2}>R_{\rm Total}\sin{\theta}$. The specific impact energy with off-centre collision correction, $Q^{'}_{\textrm{R}}$, can then be calculated according to
\begin{equation}\label{eq:Q'R}
Q^{'}_{\textrm{R}}=\frac{\mu}{\mu_{\alpha}}Q_{\textrm{R}}.
\end{equation}
And similarly, the catastrophic disruption threshold energy with off-centre collision correction, $Q^{'*}_{\textrm{RD}}$, can be calculated by
\begin{equation}\label{eq:Q'*RD}
Q^{'*}_{\textrm{RD}}=\left ( \frac{\mu}{\mu_{\alpha}}\right )^{2-\frac{3\bar{\mu}}{2}} Q^{*}_{\textrm{RD}},
\end{equation}
where $\bar{\mu}$ is the velocity exponent in the coupling parameter \citep{1987JGR....92.6350H,1990Icar...84..226H}. \citet{2012ApJ...745...79L} and its follow-up study by 
\citet{2012ApJ...751...32S}, suggested the range of values of $\bar{\mu}$ is between 0.33 to 0.37. The middle value ($\bar{\mu}=0.35$) is adopted in equation \ref{eq:Q'*RD} for our calculations. With the new value of $Q^{'}_{\textrm{R}}$ and $Q^{'*}_{\textrm{RD}}$ calculated by equation \ref{eq:Q'R} and \ref{eq:Q'*RD}, respectively, the mass ratio between the core and and the largest remnant from equation \ref{eq:McoreMlr} is modified to
\begin{equation}\label{eq:McoreMlr_b}
\frac{M_{\textrm{core}}}{M_{\textrm{lr}}}=0.5+0.25\left( \frac{Q^{'}_{\rm R}}{Q^{'*}_{\rm RD}}\right)^{1.2}.
\end{equation}

Our imperfect accretion simulations record all of the data needed to calculate $M_{\textrm{core}}/ M_{\textrm{lr}}$, and hence determine whether or not the giant impacts occurring in the simulations would have been likely to lead to significant compositional changes if our protoplanets were differentiated bodies consisting of $\sim 50\%$ rock and $\sim 50\%$ water ice, as considered by \citet{2010ApJ...719L..45M}. The results of our analysis are shown in figure \ref{fig:impactinfo}. In the upper left panel, the histogram shows the impact angles, which are peaked at the value around $45^{\circ}$, as expected \citep{1962pam..book.....K}. The lower left panel records the cumulative distribution of $V_{\textrm{imp}}$ in terms of the mutual surface escape velocity, $V_{\textrm{esc}}$, of all collisions experienced across all of the imperfect accretion simulations. The upper right panel shows the cumulative value of $Q^{'}_{\rm R}$, and the lower right panel shows the resulting estimates of $M_{\textrm{core}}/M_{\textrm{lr}}$ arising from each collision, calculated from equations \ref{eq:Q'R} and \ref{eq:McoreMlr_b}. The data suggest that only 10\% of the giant impacts in our simulations would lead to a greater than 10\% mass loss from the protoplanet, where this mass loss would correspond to partial stripping of the putative water-rich mantle. It is very uncommon to have a collision which can cause the protoplanet to have a mass loss of up to 45\% (i.e. 90\% of the water/ice content), so we conclude that while moderate compositional changes would be likely to have occurred if the Kepler multi-planet systems underwent a final stage of assembly involving giant impacts, the changes would have been insufficient to explain the location of the valley in the distribution of planetary radii discussed above. 

\section{Collision-induced atmospheric loss}\label{sec:env}
Observations and structure models of exoplanets suggest that many of the low and intermediate mass planets observed by Kepler have hydrogen-helium (H/He) envelopes \citep[e.g.][]{2014ApJ...792....1L}. These H/He envelopes must have been accreted while the planets were embedded in the gaseous protoplanetary disc \citep[e.g.][]{2014ApJ...791..103B, 2017MNRAS.470.3206C}. If dynamical instabilities and giant impacts after the dispersal of the gas disc have played an important role in the final assembly of the super-Earths and mini-Neptunes observed by Kepler, then the envelopes we observe today must have survived the giant impacts.  Previous studies have investigated the conditions under which giant impacts can lead to ejection of an envelope, both by the shock that is driven through the envelope during the impact \citep{2003Icar..164..149G,2016ApJ...817L..13I,2019MNRAS.486.2780Y}, and because of the intense heating of the core and envelope that occurs when the impact energy is converted to thermal energy during the collision \citep{2019MNRAS.485.4454B}.

Following the discussion in \citet{2019MNRAS.485.4454B}, we make the simplifying assumption that the kinetic energy associated with an inelastic collision between two bodies is converted efficiently into thermal energy in the planetary core, and good thermal coupling between the core and envelope ensures that the base of the envelope achieves the same temperature as the core. If this temperature is such that the Bondi radius of the envelope is smaller than the core radius, then we assume that the atmosphere is lost, or is at least severely eroded. 

The impact energy, $E_{\textrm{imp}}=1/2\mu V_{\textrm{imp}}^{2}$, associated with each collision between two protoplanets is reported by our \texttt{SyMBA} simulations. Hence, in a post-processing step, we can determine the distribution of impact energies from our simulations and determine whether or not these are likely to be sufficient to erode any putative envelopes that the planets might possess.  The increase of the temperature, $\Delta T$, due to the impact event can be estimated by equating the impact energy to the change in thermal energy in the core that is present after the impact (which has mass $M_{\textrm{lr}}$):
\begin{equation}
E_{\textrm{imp}} = \eta  c_{\textrm{v}} M_{\textrm{lr}}  \Delta T,
\end{equation}
giving 
\begin{equation}\label{eq:delT}
\Delta T=\eta\frac{1}{2}\frac{\mu}{M_{\textrm{lr}}}\frac{V^{2}_{\textrm{imp}}}{c_{\textrm{v}}}.
\end{equation}
Here, $c_{\textrm{v}}$ is the specific heat capacity of the core, and $\eta$ is an energy conversion efficiency factor. The impact should lead to an increase of the final temperature of the core after the impact, $T_{\textrm{c,final}}=T_{\textrm{c,initial}}+\Delta T$, and here we take a conservative approach and assume the initial core temperature is negligible compared to the final value (i.e. $T_{\textrm{c,initial}} \simeq 0$). Assuming the base of the envelope has the same temperature as the core, the associated Bondi radius becomes
\begin{equation}\label{eq:RB}
R_{\textrm{B}}=\frac{2GM_{\textrm{lr}}}{c^{2}_{s}}=\frac{2GM_{\textrm{lr}}\mu_{m}}{\gamma k_{\textrm{B}}T_{\textrm{c,final}}},
\end{equation}
where $c_{s}$ is the sound speed, $\mu_{m}$ is the mean molecular weight, $\gamma$ is the adiabatic index and $k_{\textrm{B}}$ is the Boltzmann constant. We assume the atmosphere is likely lost due to an impact if $R_{\textrm{B}} \le R_{\rm core}$.

When calculating the value of $\Delta T$ and $R_{\textrm{B}}$, we assume the envelopes are a mixture of molecular hydrogen and atomic helium with $\mu_{m}=2.3\textrm{u}$ and $\gamma=7/5$. Previous studies have taken values of the specific heat capacity of the cores of super-Earths and mini-Neptunes in the interval $c_{\textrm{v}}=500-1000$ $\textrm{J}\textrm{kg}^{-1}\textrm{K}^{-1}$ \citep{2001PhRvB..64d5123A,2010A&A...516A..20V, 2011ApJ...733....2N,2012ApJ...761...59L}. In this study, we adopt the middle value within this range $c_{\textrm{v}}=750$ $\textrm{J}\textrm{kg}^{-1}\textrm{K}^{-1}$ as in \citet{2019MNRAS.485.4454B}. 

\begin{figure}
\centering
\includegraphics[width=8.4cm]{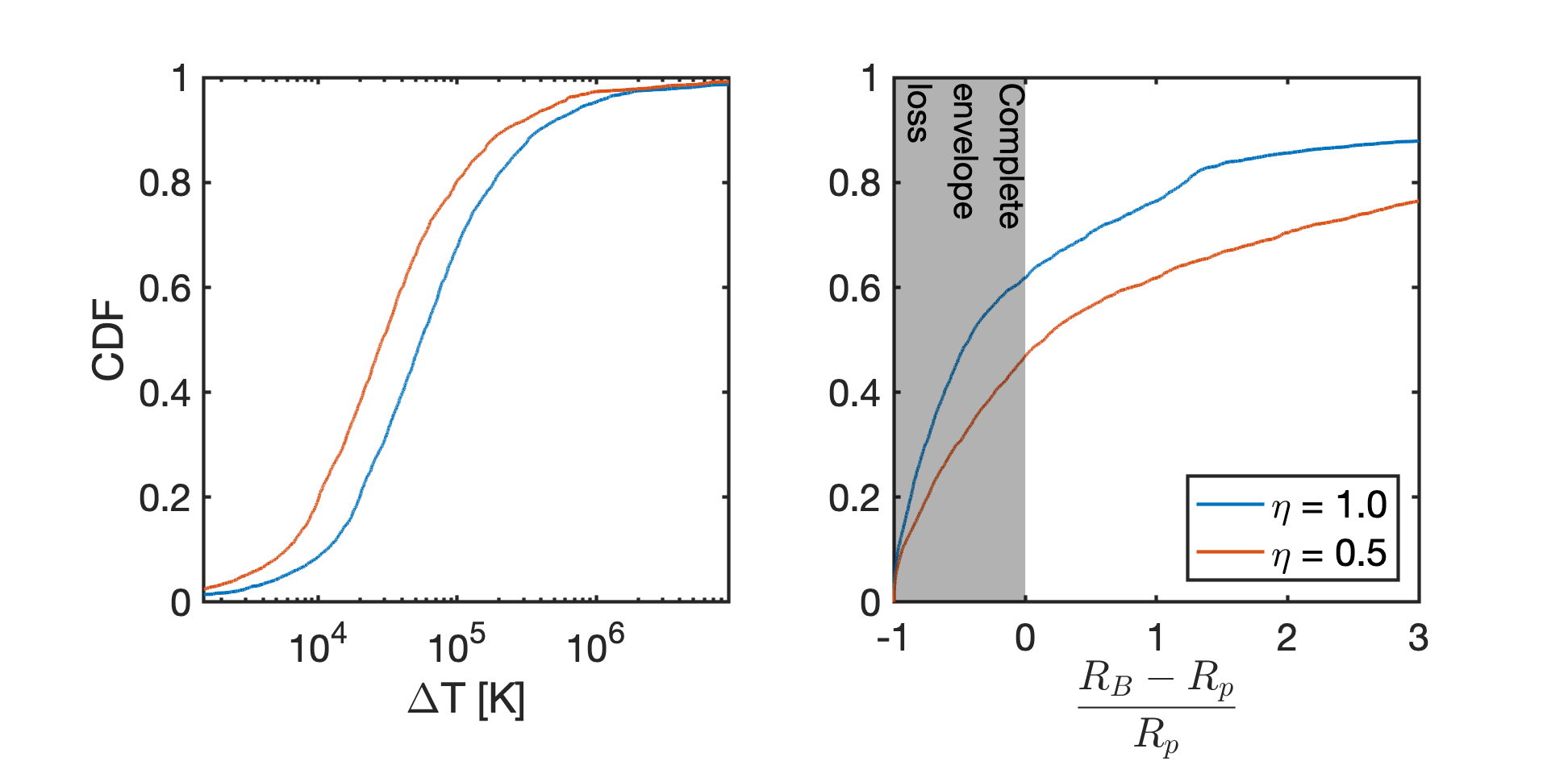}
\caption[Cumulative distributions of $\Delta T$ and $(R_{\textrm{B}}-R_{p})/R_{p}$]{Cumulative distributions of (left panel) $\Delta T$; (right panel) $(R_{\textrm{B}}-R_{p})/R_{p}$ calculated by all the giant impact events in our imperfect collision model simulations. Where $(R_{\textrm{B}}-R_{p})/R_{p}\le 0$ imply that the Bondi radius is equal to or smaller than the radius of the planet, i.e. complete H/He envelope loss (the grey area). The blue line denotes the value calculated with an energy conversion efficiency of 100\% ($\eta=1$), and red line denotes the value calculated with the energy conversion efficiency of 50\% ($\eta=0.5$).}\label{fig:r_diff}
\end{figure}

The left panel in Figure \ref{fig:r_diff} shows the cumulative distribution of the post-impact changes in core temperature, $\Delta T$, and the right panel shows the distribution of the quantity $(R_{\textrm{B}}-R_{\rm core})/R_{\rm core}$, such that a negative value implies substantial erosion of the envelope. From the figure, we can see more than 60\% of the collisions in our simulations could lead to envelope loss. These values were obtained by adopting $\eta=1$ in equation \ref{eq:delT}, corresponding to 100\% efficiency in converting impact kinetic energy into heat. Given that not all collisions are head-on, this is clearly an overestimate, as some of the energy can be converted into rotational energy or be taken away by post-impact debris \citep{2004ApJ...613L.157A}. 

\citet{2018LPI....49.2731C} investigated the Moon-forming impact and showed that around half of the impact kinetic energy is converted to internal energy, such that a more realistic figure would be $0.4<\eta<0.6$. It is worth noting that in practice, however, in a 5 planet system that was initially composed of 20 protoplanets, each remaining planet after the final assembly stage would have experienced 3 collisions on average, suggesting that significant atmospheric erosion should occur in super-Earth systems whose final assembly involves giant impacts. 

Clearly a more sophisticated approach is required to give a better quantitative estimate of the population of planets that are left with significant H/He envelopes after such a period of evolution. Such a calculation would provide one means of determining whether or not the observed population of super-Earths did indeed form via giant impacts. Even if the impacts themselves are unable to completely erode the envelopes, the remnant envelopes will be left in a bloated state and would therefore be more susceptible to photoevaporation by high energy radiation from the central star, as considered in the models of \citet{2017ApJ...847...29O}  and \citet{2018ApJ...853..163J}, for example. Hence, in the future it will be important to consider the evolution of envelopes during and after the giant impact phase to determine whether the resulting population of planets agrees with the observations.

\section{Synthetic observation of the final planetary systems}\label{sec:synobser}
The ability of any photometric observation of a planetary system to detect transits of all system members depends on the mutual inclinations of the planets. In addition, for any given planet with semi-major axis $a$, orbiting around a star with radius $R_*$, the probability of detecting a transit from a random viewing position scales as $R_*/a$, such that more distant planets around smaller stars are more difficult to detect. Based on these considerations, a meaningful comparison between the outcomes of planetary formation simulations and transit surveys, such as the one carried out by Kepler, must involve synthetic observation of the simulated planetary systems. 

Broadly speaking, the masses, orbital period ratios and planetary separations (as measured by the $K$-values) resulting from the $N$-body simulations show reasonable agreement with the inferred properties of the Kepler systems we have used as templates when setting up the initial conditions of the simulations. Here, we are interested in whether or not the distributions of the multiplicities of the simulated planetary systems, and the period ratios between neighbouring planets, when synthetically observed, agree with an appropriate sub-set of the Kepler systems. If such agreement was obtained, then it would support the hypothesis that the observed Kepler systems are all intrinsically high multiplicity systems with mutual inclinations similar to those that arise in the $N$-body simulations, which in turn would imply that the final assembly of the Kepler systems likely arose from a population of protoplanets that underwent dynamical instabilities and giant impacts, as considered in our $N$-body simulations. In addition, recent analyses have indicated that planets which are members of multiple systems have a statistically significant different eccentricity distribution compared to planets that are observed to be single \citep[e.g.][]{2016PNAS..11311431X, 2019AJ....157..198M}. We test whether this difference is matched by our simulations when synthetically observed.

\subsection{Observed multiplicities}\label{subsec:obsmulti}
Following the approach of \citet{2012ApJ...758...39J}, we consider the relative numbers of 1-planet, 2-planet, 3-planet, up to  7-planet systems that are detected when the simulation outcomes are synthetically observed from 100,000 randomly chosen viewing locations, isotropically distributed with respect to each host star. Using the observed numbers of 1-planet, 2-planet, 3-planet, etc. systems, we then define a Transit  Multiplicity Ratio (abbreviated to TMR hereafter) as follows:
\begin{equation}\label{eq:transitratio}
\textrm{TMR}(i,j)=\frac{\textrm{Number of } i\textrm{-planet systems}}{\textrm{Number of }j\textrm{-planet systems}},
\end{equation}
where $i$ and $j$ represent the numbers of planets detected during each of the synthetic observations. 

For comparison with the TMR values obtained from the $N$-body simulations, we take a sub-set of the Kepler Planet Candidates with the following cuts applied to the orbital periods, $P$, and planetary radii, $R_{\rm p}$, so that the Kepler sample roughly matches the simulation outcomes: $3 \textrm{ days} \le P \le 100 \textrm{ days}$ and $1 \textrm{ R}_{\oplus} \le R_{\rm p} \le 4 \textrm{ R}_{\oplus}$. In addition, to crudely account for the fact that the detection efficiency of Kepler decreases for small planets with longer orbital periods, we also required the planet to have a radius greater than the value given by $R_{\textrm{min}}=0.60(P/1 \textrm{ day})^{0.111}$ $R_{\oplus}$ when undertaking the synthetic observations. Incorporating this limit on the planet radius excluded around 20\%  of our final planets, but made essentially no difference to the TMRs obtained from the synthetic transit observations.
 
\begin{figure*}
\centering
\includegraphics[width=17.7cm]{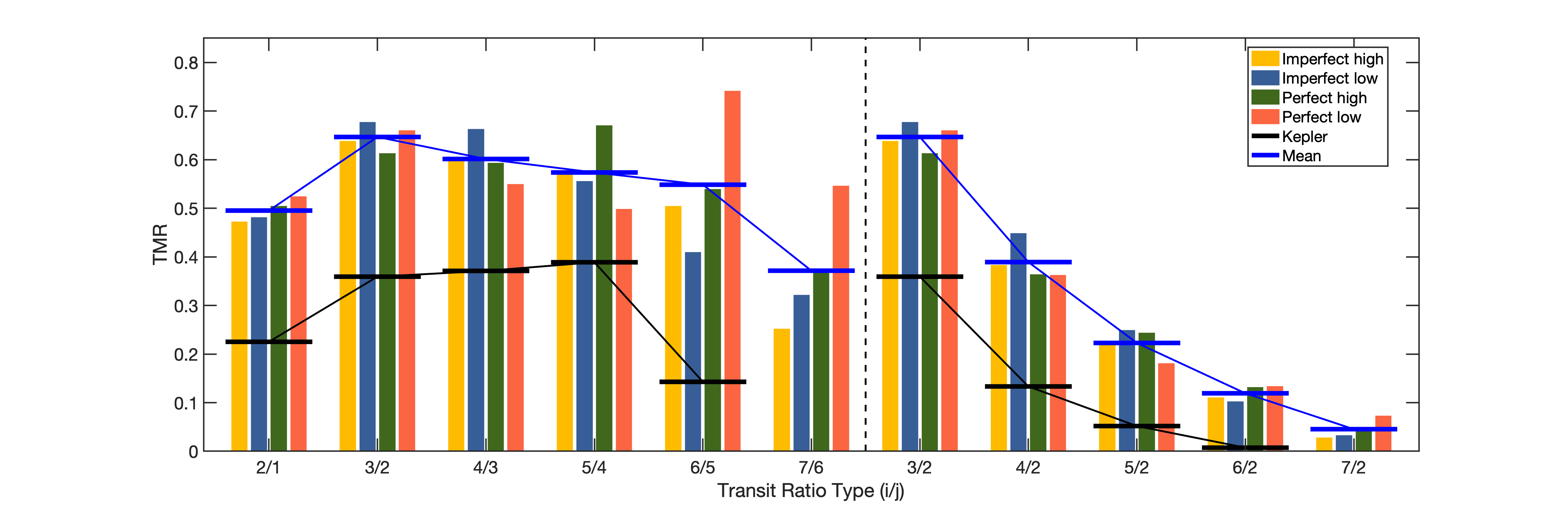}
\caption[Synthetic transit ratio by the simulation]{Synthetic transit multiplicity ratio from all four sets of our simulations. Yellow and blue bars denote the high and low initial eccentricity value simulations with imperfect accretion, respectively; green and red bars denote the high and low initial eccentricity simulations with perfect accretion, respectively. Black horizontal lines show the observed Kepler TMRs (as of 17/10/2018) and the blue horizontal lines show the mean values of the simulated TMRs.}\label{fig:transitbar}
\end{figure*}

The TMR values obtained are shown in figure~\ref{fig:transitbar}, where the coloured histograms show the values obtained from each of the different sets of $N$-body simulations, the blue horizontal bars show the mean values averaged over the different simulation sets, and the black horizontal bars show the TMRs from the Kepler data. The left panel shows TMRs for 2-planet:1-planet systems, 3-planet:2:-planet systems, 4-planet:3-planet systems, etc., and the right panel shows TMRs for n-planet systems relative to 2-planet systems, where n is an integer running between 3 and 7.  The results are very clear: the simulations consistently over produce high multiplicity systems relative to low multiplicity systems by a factor of between 1.5 and 2 compared to the Kepler systems. One reason for this is that the planet-planet scattering, leading to increases in the mutual inclinations of planetary orbits during the $N$-body simulations, does not increase the mutual inclinations sufficiently for agreement to be reached. The $N$-body simulation outcomes have mutual inclinations that are too low, with the RMS value obtained from the inclinations plotted in figure~\ref{fig:compareMandS_eim} being  $\langle I \rangle = 2.05^{\circ}$.

In their earlier studies of multiplicity ratios, \citet{2012ApJ...758...39J} and \citet{2012AJ....143...94T} suggest that mean mutual inclinations of $\langle I \rangle \simeq 5^{\circ}$ would be sufficient to provide agreement between their models and the Kepler data when comparing the relative numbers of 3-planet and 2-planet systems. This indicates that a factor of two increase in inclinations in our model systems would likely lead to much better agreement with the Kepler systems, given the factor of $\sim 2$ discrepancy shown between the TMRs shown in figure~\ref{fig:transitbar}. Even more recently, \citet{2019arXiv190208772I} and \citet{2019arXiv190302004C} have presented $N$-body simulations that provide much better agreement with the Kepler TMRs than our results do. In the case of the \citet{2019arXiv190208772I} study this improved agreement arises in part because they simulate the formation of more massive planetary systems than we do, leading to more effective gravitational scattering,  but in addition their simulations result in a number of systems with lower intrinsic multiplicities compared to our simulations.

\subsection{Period ratios}
\label{sec:PeriodRatios}
\begin{figure}
\centering
\includegraphics[width=8.4cm]{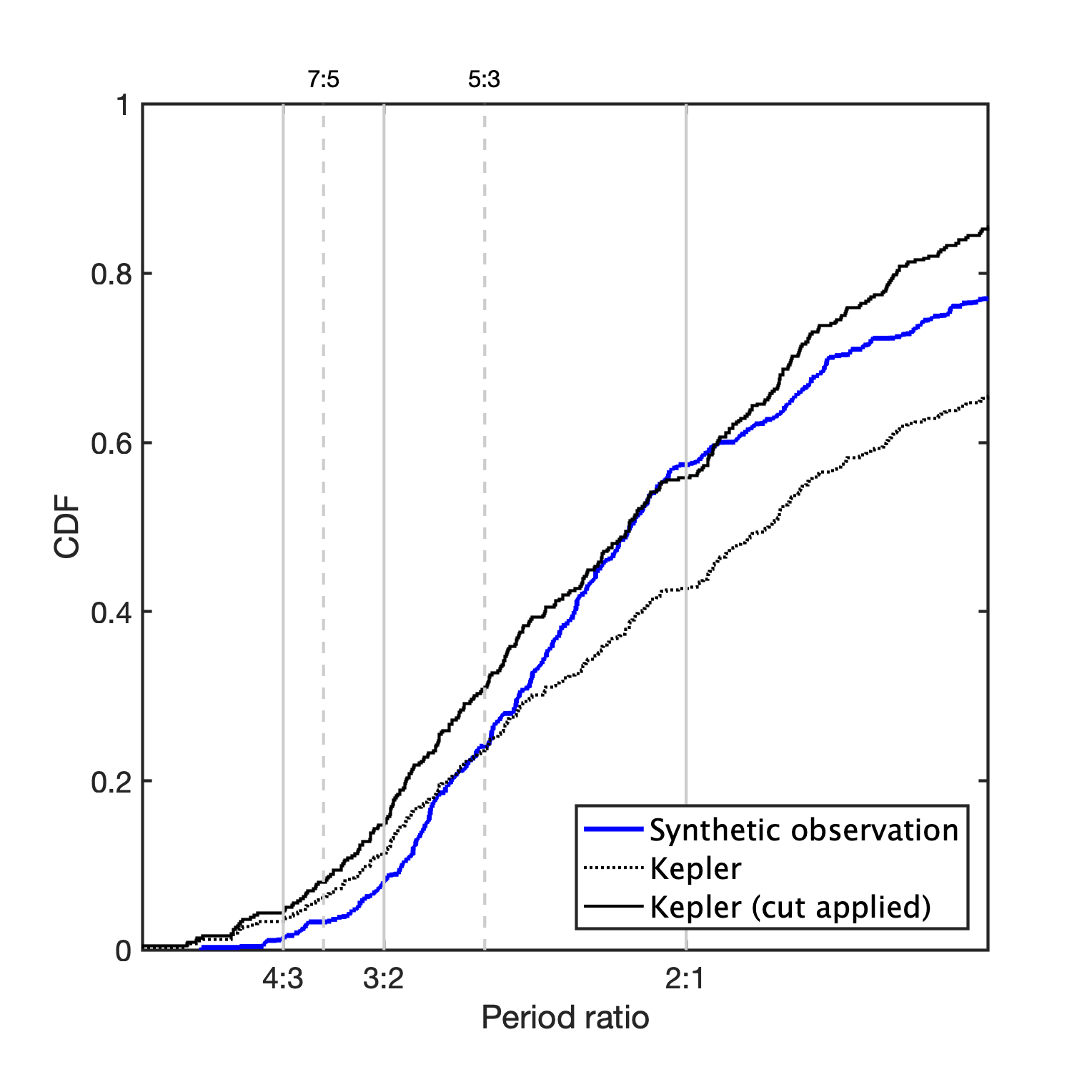}
\caption{CDFs of the period ratios obtained from the synthetic observations of the simulations, and for comparison the CDFs of period ratios obtained from our comparison sample of the Kepler data.}\label{fig:PeriodRatio-Synthetic}
\end{figure}
The CDF of the period ratios between neighbouring planets obtained from the synthetic observation of the simulation outcomes is shown in figure~\ref{fig:PeriodRatio-Synthetic}, along with that for the full set of Kepler planets that make up our comparison sample, and a subset of that sample for which the maximum period ratio is 3:1. This figure should be compared with figure~\ref{fig:periodratio_cumulative}, which shows the CDF of the intrinsic period ratios obtained from the simulations before being synthetically observed. This comparison demonstrates the importance of undertaking synthetic observations to mimic transit surveys, as the two distributions of period ratios are quite different from one another. We see from figure~\ref{fig:PeriodRatio-Synthetic} that we obtain a significant increase in the frequency of period ratios $>2$ when undertaking the synthetic observations, as planet pairs on mutually inclined orbits are not observed to simultaneously transit. However, it is also clear that the Kepler data still show a significant excess of large period ratios compared to the simulations, and in general the Kepler systems are more separated than the simulated systems. We also note that we recover the fact that the Kepler systems also show a significant excess of small period ratios compared to the simulations, discussed already in section~\ref{subsubsec:PeriodandK}. 

One curious feature of the CDF shown in figure~\ref{fig:PeriodRatio-Synthetic} is the flattening observed close to the location of the 2:1 resonance, which is reminiscent of the similar feature seen in the Kepler data due to there being a small deficit of planets at the 2:1 resonance location. This is not observed so strongly in the CDF of the intrinsic period ratios shown in figure~\ref{fig:periodratio_cumulative}, so we have examined the possibility that it arises here because mutual inclinations of planet pairs are increased near to this resonance. Plotting mutual inclinations against period ratios, however, showed no significant feature close to the 2:1 resonance, so for now this feature remains unexplained.

\subsection{Eccentricity distributions}
\begin{figure}
\centering
\includegraphics[width=8.4cm]{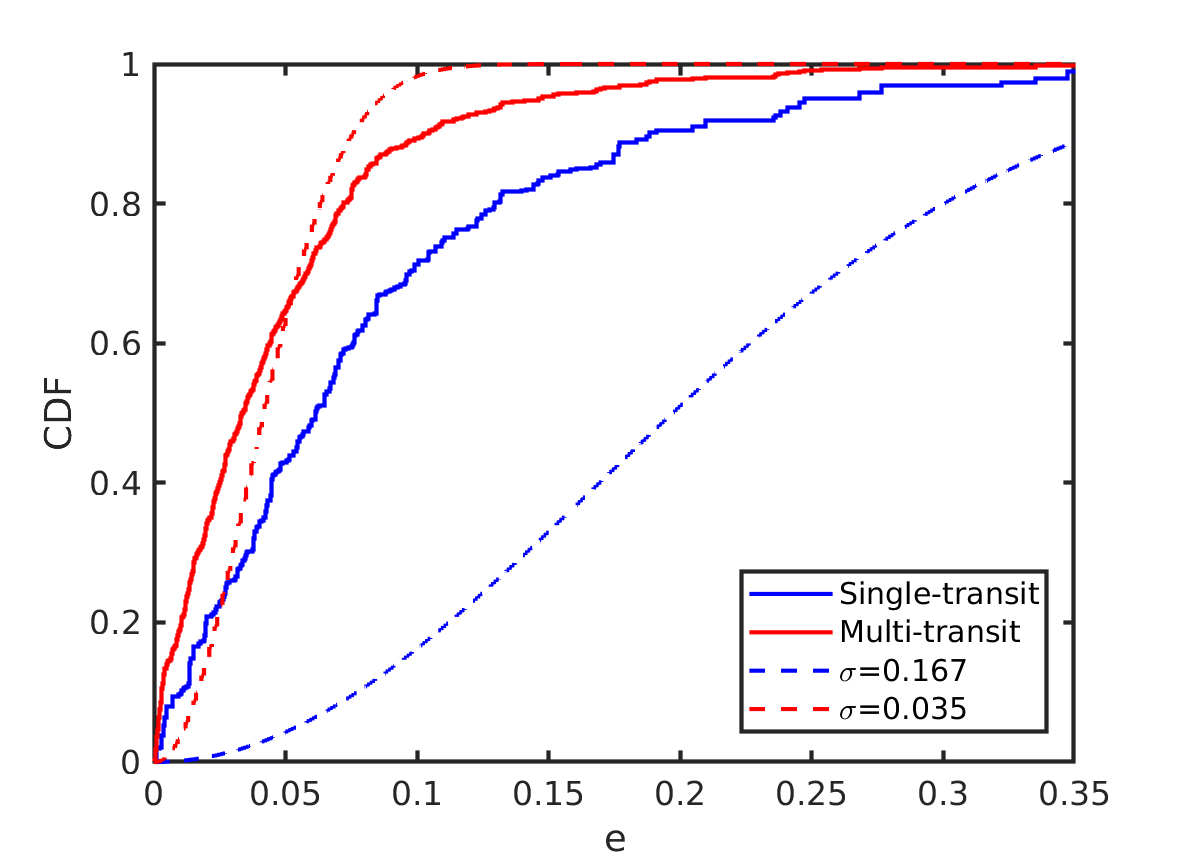}
\caption{CDFs of the eccentricities obtained from the synthetic observations of the simulations, and for comparison the CDFs of eccentricities drawn from Rayleigh distributions with eccentricity parameters 
$\sigma_e=0.035$ and $\sigma_e=0.167$.}\label{fig:e_obs}
\end{figure}
The CDFs of the eccentricities of the synthetically observed planets are shown in figure~\ref{fig:e_obs}, where the solid lines represent either systems observed to be singles or those observed to be multiples. The dashed lines show the CDFs for eccentricities drawn from a Rayleigh distribution with eccentricity parameters $\sigma_e=0.035$ and $\sigma_e=0.167$, which are the distributions and values for the observed Kepler multiple and single systems from \citet{2019AJ....157..198M}. While it is clear that the simulations produce single planets with systematically larger eccentricities than the planets in multi-planet systems because the singles are from systems that have undergone stronger scattering than the multiples, it is also clear that the simulations do not provide a good match to the observationally inferred distributions of eccentricities from \citet{2019AJ....157..198M}. In particular, the singles would need to be much more eccentric to match the observationally inferred distribution, and it is not at all clear that $N$-body simulations of the type presented here could fit the appropriate distribution of eccentricities, while also adopting planetary masses in line with those thought to make up the Kepler compact systems of super-Earths (such as shown in figure~\ref{fig:mass_cdf}). On the other hand, although the distribution from multiple systems is not particularly well fitted by the Rayleigh distribution, the range of eccentricities obtained is in much better agreement compared to those obtained for single transiting systems. Using a maximum likelihood estimation, and scanning through different values of $\sigma_e$, we find that an {\it assumed} Rayleigh distribution with parameter $\sigma_e=0.049$ provides the best fit to the multiple systems arising from the simulations. In future work, we will examine fitting the observed eccentricity distributions with the results of $N$-body simulations that consider different scenarios to those presented here.

Under conditions of strong scattering, the perturbed radial velocity of a planet relative to a circular Keplerian orbit, $v_r$, should correspond approximately to the escape velocity from the perturbing body. Assuming typical planet masses and radii ${\bar M_p}$ and ${\bar R_p}$, respectively, we have $v_r \sim \sqrt{2 G {\bar M_p} / {\bar R_p}}$. For small eccentricities $e\sim v_r / v_k$, where $v_k$ is the Keplerian velocity. From the CDF of simulated planet masses in figure~\ref{fig:mass_cdf}, the median planet mass ${\bar M_p} \sim 2.5$ M$_{\oplus}$ and the corresponding radius ${\bar R_p} \sim 1.56$ R$_{\oplus}$. For planets orbiting at $a_p \sim 0.1$ au, the eccentricity expected from strong scattering $e \sim 0.15$. The median eccentricity for single planets synthetically observed in the simulations is $e\sim 0.07$, and approximately 20\% of planets have eccentricities above the strong scattering value of $e \sim 0.15$. Hence, strong scattering contributes significantly to the eccentricity distribution, but weaker scattering events and collisional damping result in the majority of planets having smaller eccentricities. Finally, it is expected that the mean inclination, ${\bar i} \sim {\bar e}/2$ after dynamical relaxation \citep{2005dpps.conf...41K}. For ${\bar e} \sim 0.07$ the expected mean inclination is ${\bar i} \sim 2^{\circ}$, very similar to the mean value observed in the simulations, as discussed in section~\ref{subsubsec:EIM}.

Finally, we comment that for strong scattering it is expected that the resulting eccentricities will scale as $e \sim \sqrt{M_p}$. The discrepancy between the median eccentricity of the simulated single planets and the Rayleigh distribution with $\sigma_e=0.167$, shown in figure~\ref{fig:e_obs}, is about a factor of 3 ($\sim 0.07$ versus $\sim 0.2$). Hence, to generate this shift would require an increase in the masses of the planets by a factor of $\sim 9$. The CDFs for the planet masses from the simulations, and those inferred from the Kepler data, are shown in figure~\ref{fig:mass_cdf}, and there we observe about a factor of 3 discrepancy between the simulated planets and the Kepler planets. Hence, it is not clear at present whether or not strong scattering of planets that appear to be singles in the Kepler data can account for the inferred eccentricity distribution of these planets as derived by \citet{2019AJ....157..198M}. 

\section{Discussion and conclusions}\label{sec:discuss}
\subsection{Recap of simulation set-up}
We have presented the results of $N$-body simulations of in situ planetary system formation. These examine whether or not the final assembly of the Kepler compact multi-planet systems, and perhaps the wider population of Kepler planets, can be explained by a scenario in which a large number of orbiting protoplanets experience dynamical instability after the gas disc has dispersed, and accrete through giant impacts, until long term stable systems emerge. Our approach to creating initial conditions was to take eight of the known Kepler 5-planet systems, and to use these as templates for producing systems of 20 protoplanets, whose total mass was the same as the original Kepler systems (under the assumption of a particular mass-radius relation). For each Kepler template, we considered two different distributions of initial  eccentricities and inclinations, a `high set' for which the maximum values $\left[e_{\rm max},I_{\rm max}\right]=\left[0.02, 0.01\right]$, and a `low set' for which $\left[e_{\rm max}, I_{\rm max}\right]= \left[0.002, 0.001\right]$. For each of these sets we also adopted two different routines for handling collisional growth: a traditional, perfect accretion model that assumes hit-and-stick collisions; an imperfect accretion model which allows for a range of collision outcomes based on the prescriptions of \citet{2012ApJ...745...79L}. One of our main results is that the simulation outcomes had almost no detectable dependence on the collision model adopted, and this is because the systems we explored did not dynamically excite themselves sufficiently for collisions to be highly disruptive. This is in agreement with the recent study by \citet{2018MNRAS.478.2896M}. Consideration of more massive planetary systems, however, may lead to outcomes that depend on the collision model as the enhanced gravitational scattering may lead to higher collision velocities.

\subsection{Recap of main results}
All of the $N$-body simulations resulted in dynamical instability and collisions between protoplanets. The mean time scale for collisions to occur was approximately $3 \times 10^5$ orbital periods, measured at the centre of the annulus containing the protoplanets, indicating that the systems had time for dynamical relaxation to occur during the process of collisional growth. 90\% of collisions occurred within 1 Myr, and we ran the simulations for a total of 10 Myr. 

\begin{figure}
\centering
\includegraphics[width=8.4cm]{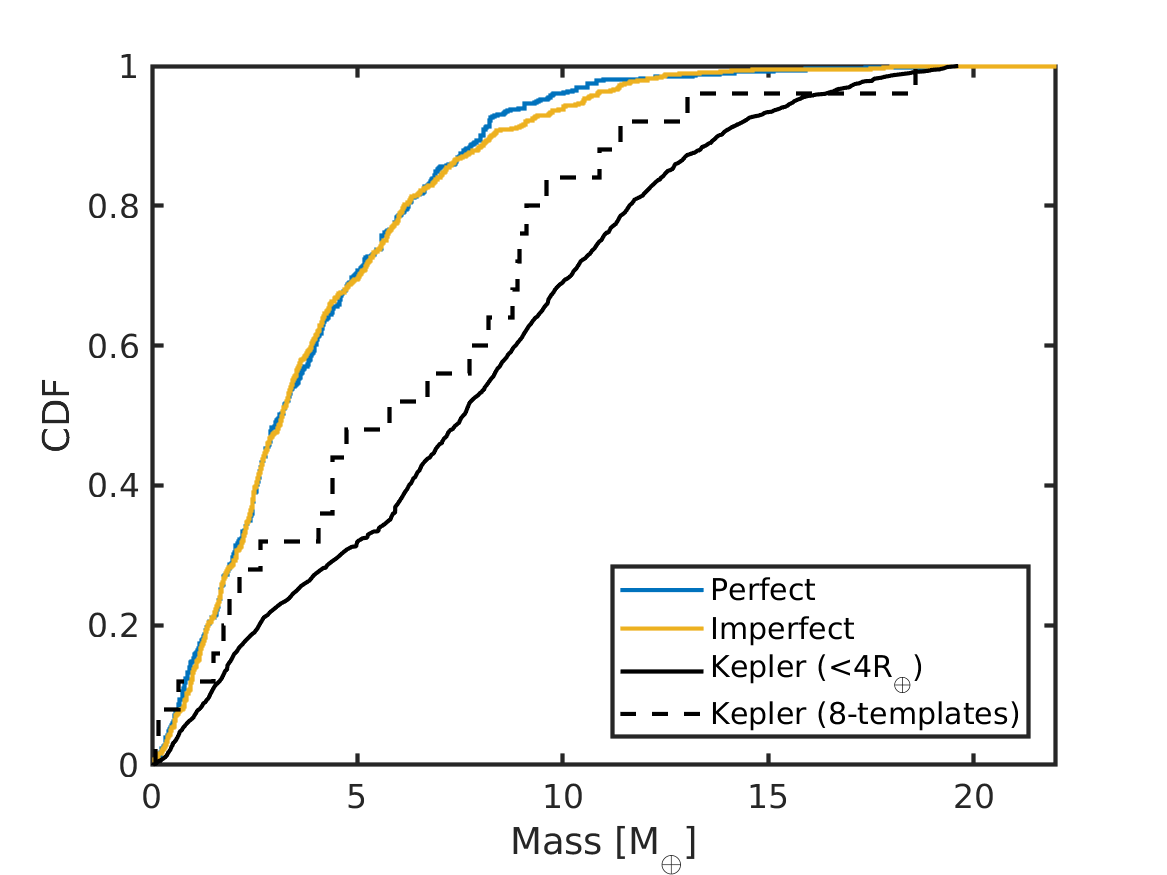}
\caption{Cumulative distributions of the planet masses obtained from the simulations and inferred from the Kepler data using the mass-radius relation described in the text.}\label{fig:mass_cdf}
\end{figure}

The final outcomes of the simulations generally showed good agreement with the Kepler systems we used as templates, indicating that the procedure adopted for setting up the simulations gave rise to plausible initial conditions. In particular, the final distributions of planet masses, orbital period ratios, separations between neighbouring planets, and intrinsic multiplicities showed good agreement with the templates on average. Notably, our simulations failed to produce any 1-planet or 2-planet systems. This suggests that if single or double planet systems are intrinsically common among the Kepler systems, as has been suggested \citep{2012ApJ...758...39J}, then the formation scenario presented here cannot explain them. If the single transiting planets are instead members of multi-planet systems whose mutual inclinations prevent all planets being observed, however, then final assembly through planet-planet scattering and giant impacts remains plausible \citep{2019arXiv190208772I, 2019arXiv190302004C}. Similarly, widely-spaced pairs of neighbouring planets with large period ratios are very difficult to explain in a model where the initial distribution of protoplanets is smooth and continuous. For these latter systems, it would appear necessary for the initial distributions of planetary building blocks to contain localised concentrations of protoplanets in order to produce the large period ratios seen in the Kepler data. Alternatively, some other process, such as migration, that can cause planets to move relative to one another, would need to be included in the models to explain the well-separated planet pairs that have been observed.

We undertook synthetic transit observations of the final planetary systems formed in the simulations. We counted the relative numbers of 1-planet, 2-planet, 3-planet, ..., 7-planet systems detected by the synthetic observations, and compared this with an appropriate sub-set of the Kepler data that matched the parameters of the model planetary systems. We found that the simulated systems over-produce, by about a factor of 2, the numbers of high multiplicity versus low multiplicity systems compared to the Kepler systems. This arises in part because the excitation of mutual inclinations in our simulations is too small by about a factor of 2, and also because our planetary systems produce too few low-multiplicity systems. We have examined the distributions of the masses of the planets obtained in the simulations, and find that these are smaller than the typical inferred masses of the planets in the Kepler data set (considering planets with radii $< 4 \, R_{\oplus}$). Figure~\ref{fig:mass_cdf} shows the CDFs of the planet masses from the simulations and the Kepler planets, and it shows: (i) the Kepler systems we chose as templates have moderately smaller masses than the Kepler data set as a whole we are comparing against; (ii) the simulations produce too many low mass planets compared to both the templates and the Kepler systems as a whole, because systems with higher multiplicity than 5 are formed. It is likely that by choosing more massive discs of protoplanets as initial conditions, the resulting planet masses, and the enhanced scattering they would experience, would lead to final systems more in agreement with the Kepler data, both in terms of inferred planet masses and in terms of the distribution of multiplicities, because of the larger mutual inclinations that would have been excited.  This may also result in eccentricity distributions that are in better agreement with the observations than we obtained in the simulations presented here. Alternatively, if final period of dynamical instability in multiplanet systems is initiated when the planets are essentially fully formed, rather than when the system consists of numerous low mass protoplanets, then it may be possible to achieve higher eccentricities and mutual inclinations because the scattering may be stronger and collisions may occur less frequently. 

\subsection{Recent relevant planet formation studies}
In their recent study, \citet{2019arXiv190208772I} were able to construct a population of planetary systems, that when synthetically observed, provided good agreement with the Kepler multiplicity distribution. This was achieved by combining simulations that resulted in resonant chains of planets that became dynamically unstable with simulations in which the resonant chains remained stable. It is noteworthy that only 5\% of the included planetary systems were intrinsically 1-planet systems, with most of the rest being multiple systems in which the mutual inclinations typically exceeded 4$^{\circ}$. In an earlier study, \citet{2016ApJ...832...34M} undertook a study of the multiplicity distributions arising from $N$-body simulations of planet formation, adopting a range of surface density profiles and masses in their initial discs of protoplanets and planetesimals. As with our simulations, theirs also formed planetary systems with intrinsic multiplicities $N_{\rm p} \ge 3$, and by suitably combining their different simulation results they were able to produce a population of planets that agreed with the Kepler distribution of multiplicities when their simulated systems were synthetically observed. Hence, it appears that combining a range of initial conditions for planet formation simulations, that ultimately result in dynamical instabilities and giant impacts, can lead to systems that collectively provide mutual inclination and intrinsic multiplicities that agree with observations when their transits are simulated. 

\subsection{Intrinsic multiplicities from RV studies}
The agreement between these simulations and the observations raises an important question: are essentially all planetary systems intrinsically multiple systems, even when observed to be singles by transit surveys? And if so, what is the underlying multiplicity distribution? Transit surveys cannot directly answer this. The detection of transit timing variations in apparently single planet systems discovered by Kepler, however, shows that a number of these planets have neighbours close to mean motion resonances \citep{2019AJ....157..171K}. In addition, the fact that the eccentricities of single planets appears to be systematically higher than in multiple planet systems \citep{2019AJ....157..198M}, indicates that a number of apparently single planets have likely been subject to gravitational scattering, and hence are members of multi-planet systems. Radial velocity surveys can in principle detect nearby companions in compact multi-planet systems, assuming modest mutual inclinations, although they are constrained by limits imposed by spectral resolution, instrument stability and stellar variability, and numerous compact multi-planet systems have been discovered by this method \citep[e.g.][]{2011arXiv1109.2497M}. Nonetheless, there are hints in the data that super-Earths do not always come as members of compact multiple systems. For example, the recently discovered super-Earth orbiting with a period of 233 days around Barnard's star indicates a lack of close orbiting planets with similar masses in that system \citep{2018Natur.563..365R}, and certainly none that became anchored at the inner edge of the protoplanetary disc during their formation, as often occurs in $N$-body simulations of planet formation that involve pebble drift or planet migration. Similarly, Proxima b, orbiting with a period of 11 days does not appear to have closely neighbouring planets of similar masses \citep{2016Natur.536..437A}. While these are only individual examples, they indicate that not all planetary systems are compact multiples. Future high precision RV surveys targeted at characterising the multiplicities of short period super-Earth systems will have the power to determine the whether or not final assembly of planetary systems via dynamical instability is the dominant mode of planet formation, or if instead a substantial population of relatively isolated planets exists that cannot be explained by the giant impact formation route. 

\subsection{Compact non-resonant systems}
Furthermore, compact systems of super-Earths, such as Kepler-11 \citep{2014ApJ...795...32M}, that contain planets on low eccentricity orbits, and which are apparently close to instability, are also difficult to assemble via dynamical instabilities and giant impacts. Kepler-11 appears to have been assembled in a highly dissipative environment, presumably in a gaseous protoplanetary disc -- which is supported the low densities of some of the planets \citep{2013ApJ...770..131L}, but none of the planet pairs are in mean motion resonance. Hence, while dynamically quiet formation in a disc seems necessary, the lack of resonances indicates that disc-driven migration may not have played an important role in this system. We note, however, that recent analyses of single and multiple planet migration in inviscid protoplanetary discs by \citet{2019MNRAS.484..728M} and \citet{2019MNRAS.tmpL.120M}, lead to more complex migration behaviour of planets than has been found to traditionally occur in viscous discs. Hence, the formation of resonant chains is not a foregone conclusion in inviscid discs, and such an environment may provide a way of forming systems such as Kepler-11. A reasonable conclusion is that the observational evidence appears to indicate that a number of different pathways are required for the final assembly of planetary systems.

\subsection{Co-orbital systems}
Approximately 1\% of our simulations gave rise to pairs of planets in apparently long-term stable 1:1 co-orbital resonances, occupying both tadpole and horseshoe orbits. These normally form early in the simulations, when the numbers of protoplanets are high and the planetary systems are undergoing strong planet-planet interactions. The co-orbital pairs arise as a result of three body encounters removing the requisite energy and angular momentum from a pair of planets such that the co-orbital configuration can form. These co-orbital planet pairs occur with equal frequency in the perfect and imperfect collision model simulations, indicating that the treatment of collisions has no effect on their formation. In spite of intensive searches through the Kepler data, no co-orbital planet systems have been found. We note, however, that there are just over one hundred Kepler systems with known multiplicity $\ge 3$ that fall within the parameter ranges covered by our simulations, so a 1\% occurrence rate, that would agree with the simulations results, leads to an expectation that just one co-orbital system would have been found. Hence, the current data set is too small to determine if the observed planets indeed formed from a large number of protoplanets undergoing dynamical relaxation and collisions, leading to co-orbital pairs forming with an efficiency of $\sim 1\%$ per system. Future missions, such as PLATO, will monitor many more stars than Kepler \citep{2014ExA....38..249R}, and hence will place more stringent constraints on the formation histories of compact multiplanet systems.
 
 \subsection{Composition changes and envelope loss}
Recent observations have indicated the presence of a valley in the distribution of planet radii for short period Kepler planets \citep{2017AJ....154..109F, 2018AJ....156..264F, 2018MNRAS.479.4786V}, and models of envelope photoevaporation suggest that the position of the valley is most easily explained if the cores of super-Earths that are subject to photoevaporation are rocky rather than volatile rich \citep{2017ApJ...847...29O,2018ApJ...853..163J}. 
In this context, we examined whether or not high energy collisions during our simulations could significantly modify the compositions of the final planets relative to the initial protoplanets. In post-processing, we used the scaling relations between collision energies and compositional changes presented by \citet{2010ApJ...719L..45M}, and examined whether or not high energy collisions occurred frequently enough that they could remove a large fraction of  water-rich mantles of colliding, differentiated protoplanets whose initial compositions were 50\% rock and 50\% water-ice. The results of this analysis suggest that collisional stripping of water-rich mantles cannot explain the fact that the apparently naked cores observed by Kepler have an Earth-like composition, instead of a mixture of rock and water-ice. It seems unlikely, therefore, that these now naked cores formed exterior to the ice line and migrated to their current locations. Using a similar analysis, we also examined whether or not the impact energies of collisions between protoplanets could be sufficient to remove any H/He envelopes they might possess. Our simple analysis suggests that $\sim 30\%$ of collisions occurring in the simulations could remove gaseous envelopes, such that planets experiencing multiple collisions would have a high probability of losing their envelopes completely. Further modelling of this process could place significant constraints on the collisional history of the observed population of super-Earths and mini-Neptunes. Compared to the scenario examined here in which a large number of protoplanets undergo collisional accretion during the final assembly of exoplanet systems, the exoplanet data likely point to an origin in which systems of fewer, essentially fully formed planets undergo dynamical relaxation after dispersal of the gas disc in order to produce the observed orbital architectures, while undergoing fewer collisions in order to maintain the gaseous envelopes possessed by a large number of the observed exoplanets.

In future work we will present simulations with a significantly broader range of initial conditions, to assess the conditions under which dynamical instabilities in multi-planet systems may have contributed to the final stage assembly of the observed short-period super-Earths and mini-Neptunes.

\section*{Acknowledgements}
This research utilised Queen Mary's Apocrita HPC facility, supported by QMUL Research-IT \footnote{http://doi.org/10.5281/zenodo.438045}. 




\bibliographystyle{mnras}
\bibliography{Sanson} 




\appendix
\section{Planetesimal ring fragmentation}\label{app:pfr}
For the imperfect collision simulations, which adopted the collision model from \citet{2012ApJ...745...79L}, approximately 10\% of our simulations experienced at least one super-catastrophic collision, leading to the formation of a ring composed of collision debris in the form of planetesimals. This ring was often confined to the inner most regions close to the star, where collision velocities can reach their highest values, and this led to large numbers of particles needing to be integrated using small time step sizes. In order to overcome this problem, we developed a scheme for reducing the masses of the ring particles on a time scale corresponding to their collision time scales. Then, once the mass of the ring reaches negligible values, the ring particles could be removed from the simulations, since the planetesimals would then be ground down to dust, which would be removed in reality by the Poynting-Robertson effect for grain sizes between 1 mm to 1 $\mu$m  \citep{1903MNRAS..64A...1P,1937MNRAS..97..423R,1962ApJ...135..855G} or radiation pressure for grains smaller than 1 $\mu$m \citep{1979Icar...40....1B}. 
When there is a fragmentation ring detected between the inner-most big body and the host star, we apply the following step during the simulation. 

The total number of bodies in the ring, $N$, the mean semi-major axis, $\left \langle a \right \rangle$, eccentricity, $\left \langle e \right \rangle$, inclination, $\left \langle I \right \rangle$, and semi-major axis of the inner- and outer-most bodies in the ring ($a_{out}$ and $a_{in}$) can be found from the simulations. These can be used to calculate the collision time in the ring, $\tau_{\textrm{coll}}$, according to
\begin{equation} \label{eq:mft}
\tau_{\textrm{coll}}=\frac{1}{n\sigma \left \langle v \right \rangle},
\end{equation}
where $n$ is the number density given by,
\begin{equation} \label{eq:nd}
n\approx \frac{N}{2 \pi (a^{2}_{out}-a^{2}_{in})  \langle a \rangle  \langle I  \rangle}.
\end{equation}
The velocity dispersion  $\left \langle v \right \rangle$ is given by,
\begin{equation} \label{eq:meanv}
\left \langle v \right \rangle\approx \left \langle v_{k} \right \rangle \sqrt{ \left \langle e^{2} \right \rangle + \left \langle I^{2} \right \rangle},
\end{equation}
where $\left \langle v_{k} \right \rangle$ is the mean Keplerian velocity of the objects in the ring. The collision cross-section is simply,
\begin{equation} \label{eq:crosssection}
\sigma=\pi R^{2}_{p}.
\end{equation}
Having obtained the collision time, we then decrease the mass of the ring particles according to
\begin{equation} \label{eq:innerringmass}
m_{p}(t)=m_{p}(t_{0})e^{-(t-t_0)/\tau_{\rm coll})},
\end{equation}
where $t_0$ is the time of ring formation,
and after 15 e-folding times we remove the ring particles, since their masses are then negligible.

\begin{figure}
\centering
\includegraphics[width=8.4cm]{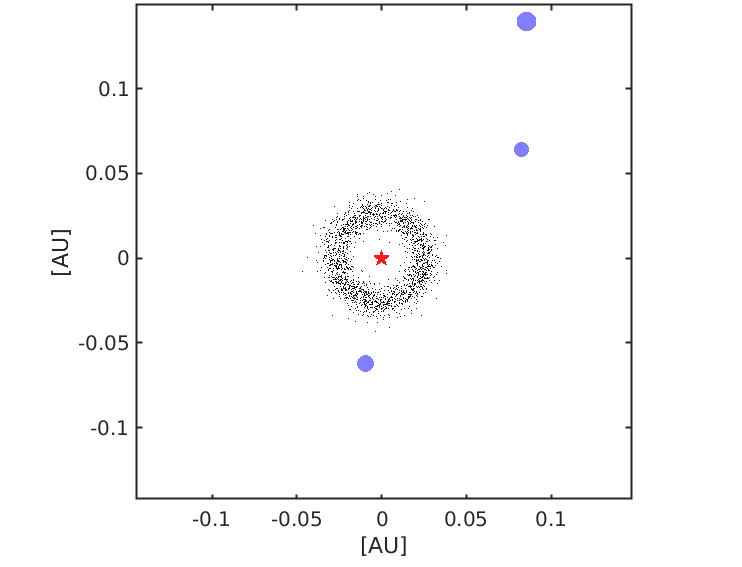}
\caption[All inner ring object position]{An example of an inner ring from one of our simulations, where the small black dots are the position of the inner ring objects. The blue circles denote the planets with their size is relative in term of their mass. The centre red pentagram is the position of the host star. Viewed from the top of the system all bodies are orbiting the host star in an anti-clockwise direction. Parameter information about this inner ring is listed in figure \ref{fig:innerring_para} and table \ref{tab:innerring}.}\label{fig:innerring}
\end{figure}

\begin{figure}
\centering
\includegraphics[width=8.4cm]{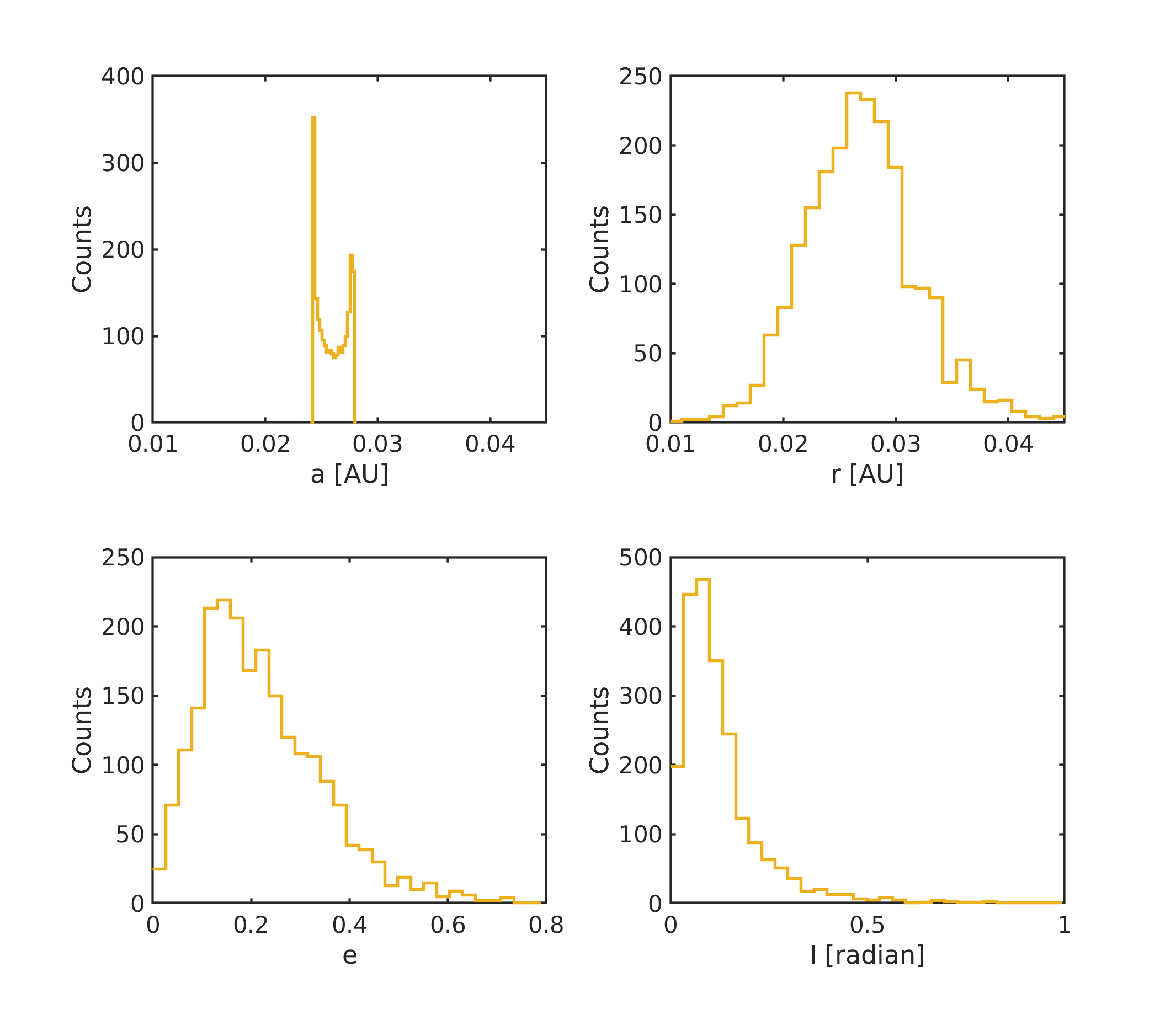}
\caption[All inner ring object parameters]{Counting distribution of all inner ring object parameters. Where it shows the distribution of the (top left) semi-major axis, (top right) distance, $r$, from the host star, (bottom left) eccentricities, and (bottom right) inclinations.}\label{fig:innerring_para}
\end{figure}

Figure \ref{fig:innerring} shows the ring formed in one of our simulations. Figure \ref{fig:innerring_para} shows the parameters of the ring shown in figure \ref{fig:innerring}, and table \ref{tab:innerring} lists the ring parameters. We can see 2177  bodies with masses $\sim$0.0015 $M_{\oplus}$ concentrated between 0.01 AU and 0.05 AU. Applying the parameters of the ring listed in table \ref{tab:innerring} to equations \ref{eq:mft}, we get $\tau_{\textrm{coll}}= 2.12$ years, leading to rapid removal of the ring.

\begin{figure}
\centering
\includegraphics[width=8.4cm]{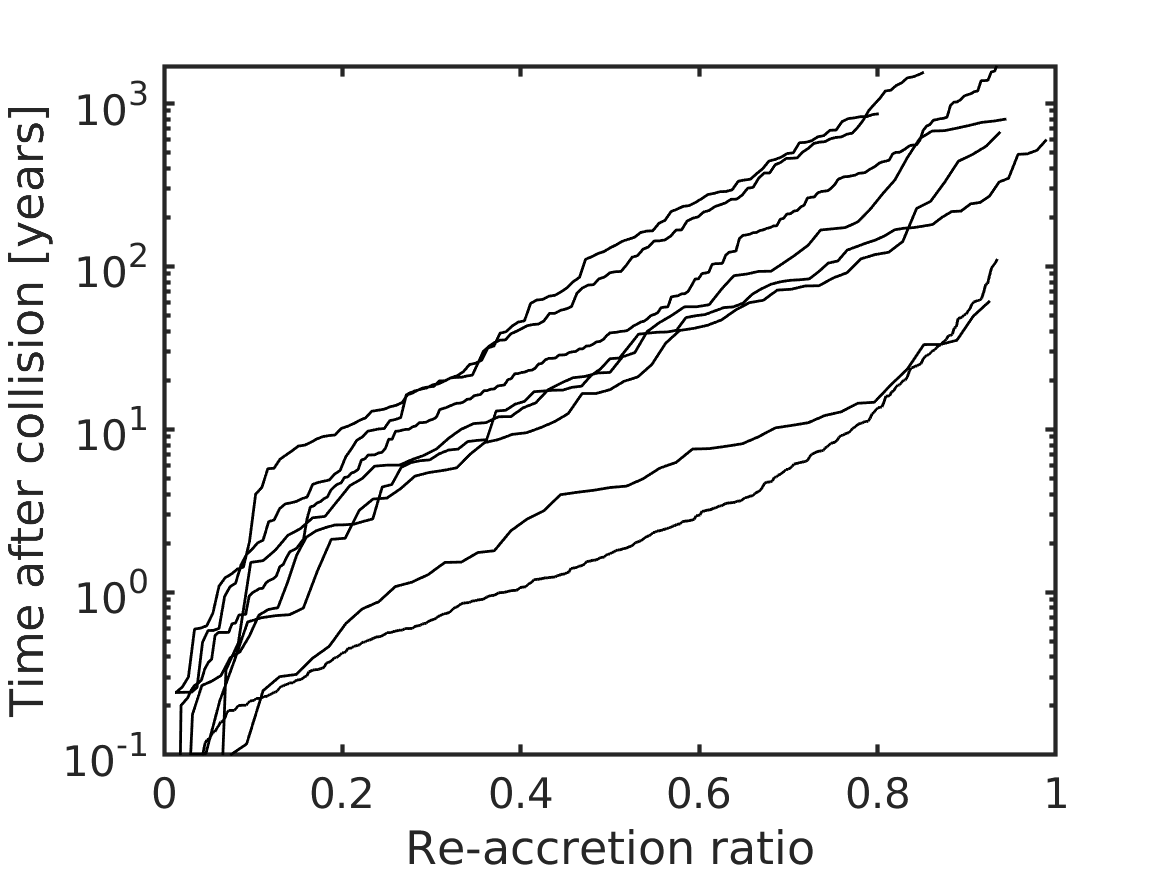}
\caption[]{Time since collision versus the fraction of debris particles remaining since the collision for 8 collision events selected from the simulations.} 
\label{fig:reacc}
\end{figure}

We note \citet{2018MNRAS.478.2896M} also include an imperfect accretion model in their simulations, and immediately remove all small debris particles after they are formed because of the small time scale for collisions and collisional grinding versus the re-accretion time scale of the debris onto nearby protoplanets. We have examined the re-accretion time scale for debris particles in our simulations, and find that after a debris cloud is generated the time scale for half of the debris particles generated to be re-accreted is between $\sim 2$ years up to $\sim 100$ years, as shown in Figure~\ref{fig:reacc}. This is considerably shorter than the re-accretion time scale of 30,000 years estimated by \citet{2018MNRAS.478.2896M}, but perhaps comparable to or longer than the typical grinding time scales for debris clouds generated by during collisions. This suggests that a realistic model for the evolution of post-collision debris should allow a fraction of it to re-accrete while the other fraction is ground down and removed by radiation pressure. Incorporating such a model, however, goes beyond the scope of this paper. 

\begin{table}
	\centering
	\caption{Fragmentation ring parameters for \texttt{Kepler55\_low\_06}, displayed in figure \ref{fig:innerring}.}
	\label{tab:innerring}
	\begin{tabular}{cccccc} 
		\hline
		\hline
		$N$ & $a_{in}$ (AU) & $a_{out}$ (AU) & $\left\langle a \right\rangle$ (AU) & $\left\langle e \right\rangle$ & $\left\langle I \right\rangle$ (rad) \\
		\hline 
		2177  & 0.0241 & 0.0283 & 0.0260 & 0.2213 & 0.1272 \\ 
		\hline
		\hline
	\end{tabular}
\end{table}

\section{Surface density fitting model}\label{app:Sigmafit}
\begin{table*}
	\centering
	\caption{Surface density fitting model for the eight Kepler templates. For model details see equation \ref{eq:Sigmafit}.}
	\label{tab:fittype}
	\begin{tabular}{lccccc}
		\hline
		\hline
		System template & Fitting model & $c_1$ & $c_2$ & $c_3$ & $c_4$\\
		\hline
		\texttt{Kepler55} & Power law & $1.047\times10^{-1}$ & -2.488 & 73.80 & /\\
		\texttt{Kepler80} & Polynomial & $-1.687\times10^{7}$ & $2.597\times10^{6}$ & $-1.097\times10^{5}$ & $1.917\times10^{3}$\\
		\texttt{Kepler84} & Polynomial & $2.062\times10^{5}$ & $-9.379\times10^{4}$ & $1.142\times10^{4}$ & $-1.507\times10^{2}$\\
		\texttt{Kepler102} & Fourier series & $1.042\times10^{2}$ & 57.58 & 37.81 & 60.55\\
		\texttt{Kepler154} & Polynomial & $6.073\times10^{4}$ & $-3.512\times10^{4}$ & $5.303\times10^{3}$ & -25.78\\
		\texttt{Kepler169} & Exponential & $5.233\times10^{2}$ & -17.72 & 0 & /\\
		\texttt{Kepler292} & Power law & 29.21 & $9.643\times10^{-1}$ & 0 & /\\
		\texttt{Kepler296} & Power law & $9.924\times10^{-1}$ & -2.082 & 0 & /\\
		\hline
		\hline
	\end{tabular}
\end{table*}

Table \ref{tab:fittype} shows all the coefficients adopted to fit the surface density, $\Sigma_{\rm fit}$, together with the type of fitting model chosen from the models described in section m\ref{subsec:surfdensity}. The choice of fitting model was made by selecting the one that gave the best least-squares fit.

\section{Simulation outputs for individual templates}\label{app:remain}
This section provides the simulation outcomes from all system templates that were not shown in the main body of the paper: \texttt{Kepler55}, \texttt{84}, \texttt{154}, \texttt{296} (imperfect collision simulations) and all 8 templates for the perfect collision simulations.

\newpage
\begin{figure*}
\centering
\includegraphics[width=16cm]{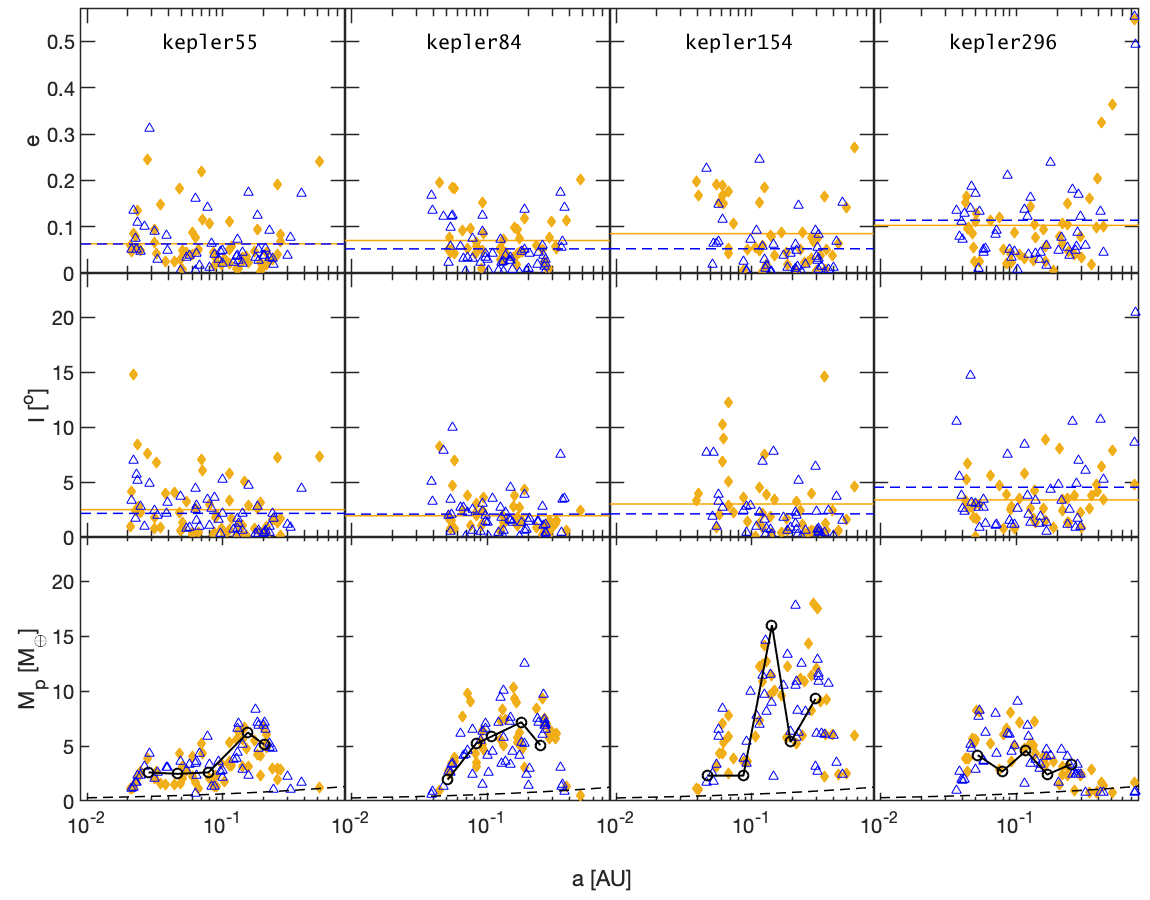}
\caption[All imperfect collision results of Kepler-55, 84, 154, and 296]{Same plot as figure \ref{fig:allseim} but for the imperfect Kepler-55, 84, 154, and 296 templates. }\label{fig:appseim}
\end{figure*}

\begin{figure*}
\centering
\includegraphics[width=16cm]{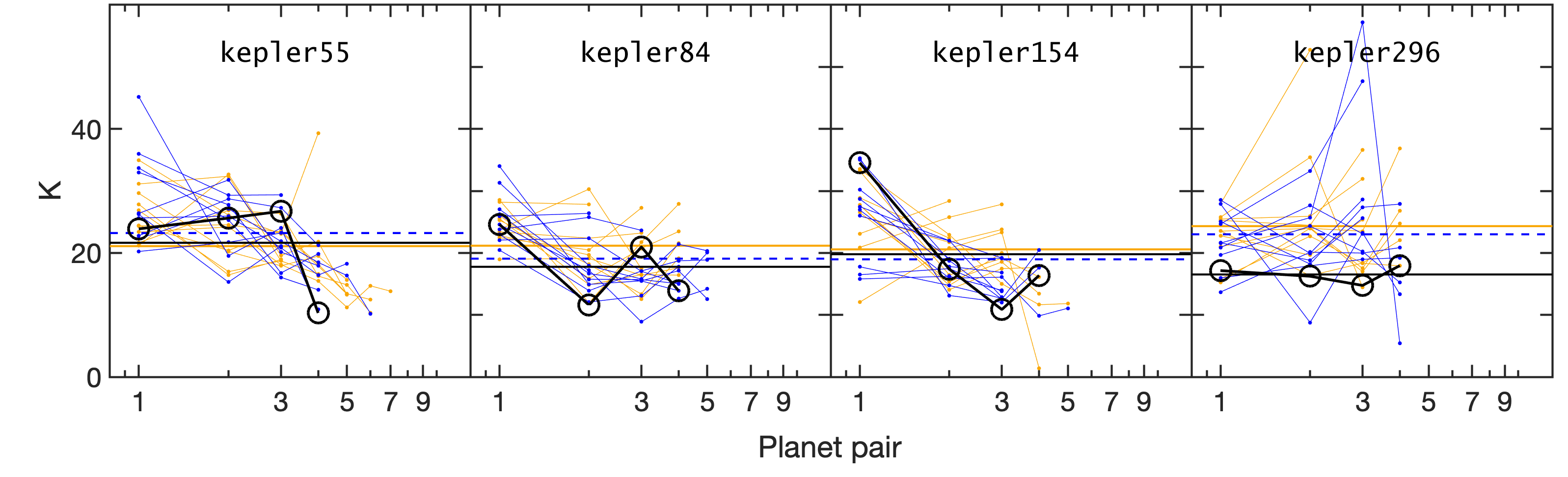}
\caption[All $K$-vlaue from imperfect collision of Kepler-55, 84, 154, and 296]{Same plot as figure \ref{fig:allsK} but for the imperfect \texttt{Kepler55}, \texttt{84}, \texttt{154}, and \texttt{296} templates.}\label{fig:appsK}
\end{figure*}

\begin{figure*}
\centering
\includegraphics[width=16cm]{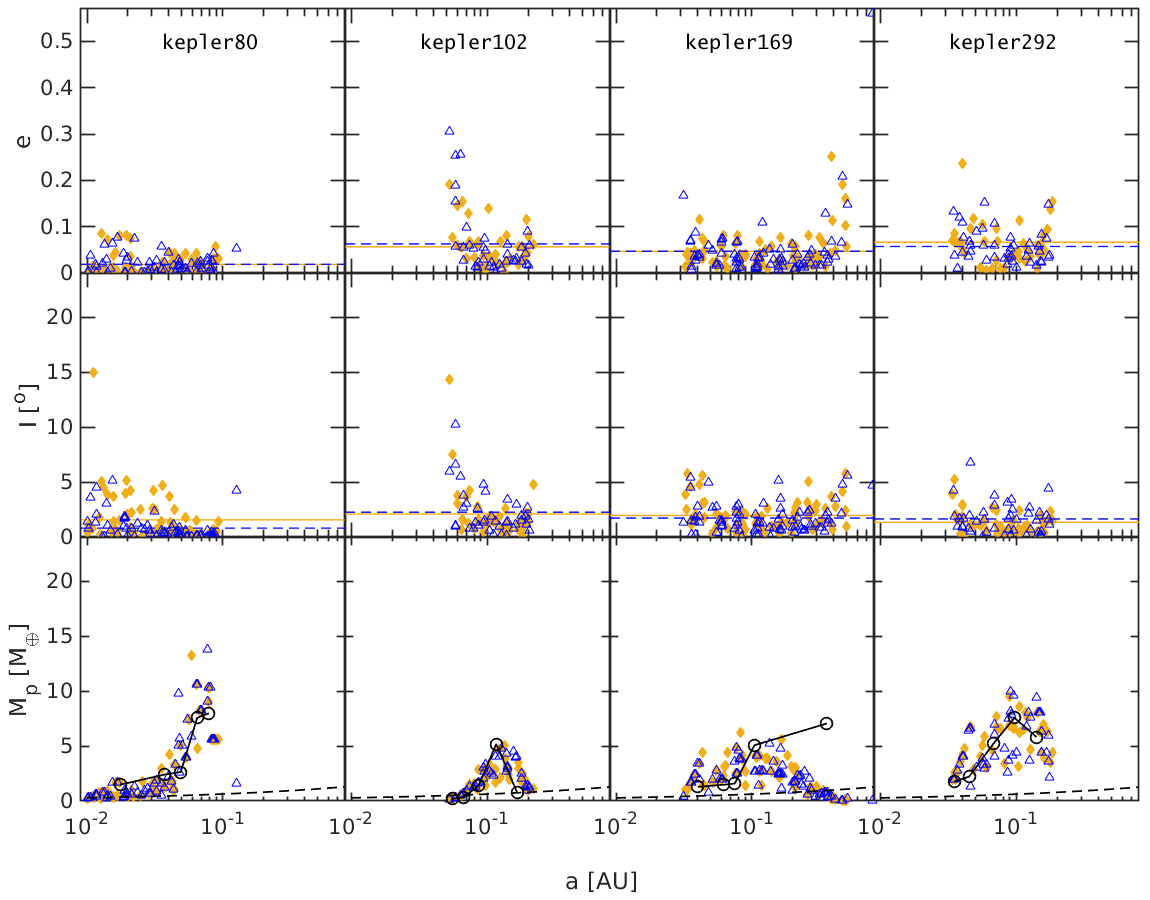}
\caption[All perfect collision results]{Same plot as figure \ref{fig:allseim} but for the perfect \texttt{Kepler80}, \texttt{102}, \texttt{169}, and \texttt{292} templates. }\label{fig:allmeim}
\end{figure*}

\begin{figure*}
\centering
\includegraphics[width=16cm]{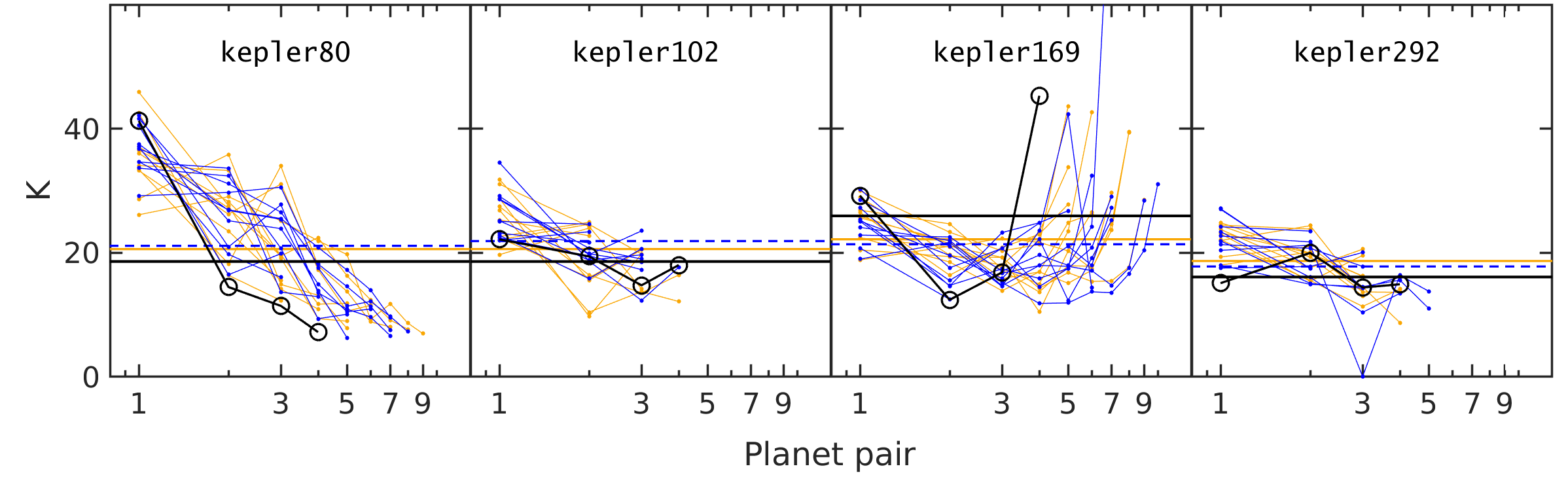}
\caption[All $K$-vlaue from perfect collision]{Same plot as figure \ref{fig:allsK} but for the perfect \texttt{Kepler80}, \texttt{102}, \texttt{169}, and \texttt{292} templates.}\label{fig:allmK}
\end{figure*}

\begin{figure*}
\centering
\includegraphics[width=16cm]{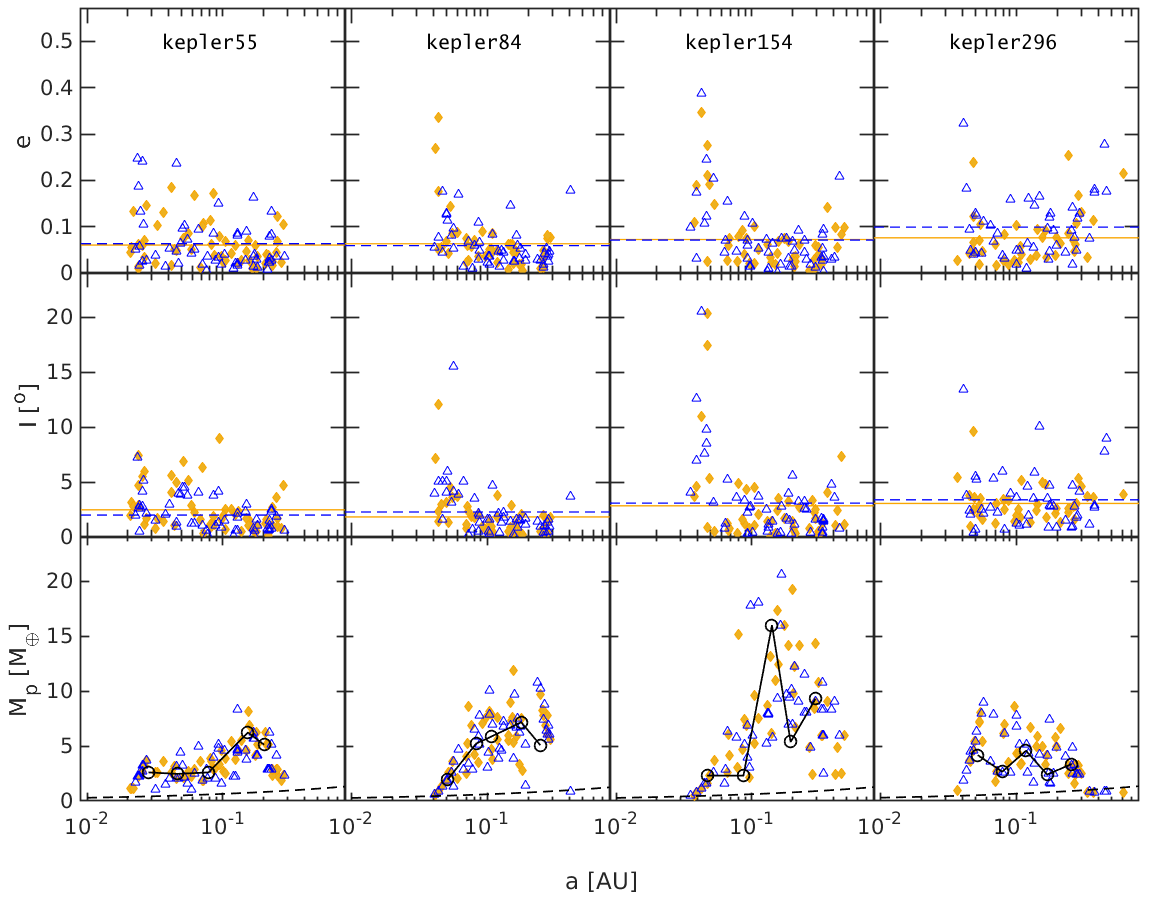}
\caption[All perfect collision results of \texttt{Kepler55}, \texttt{84}, \texttt{154}, and \texttt{296}]{Same plot as figure \ref{fig:allseim} but for the perfect \texttt{Kepler55}, \texttt{84}, \texttt{154}, and \texttt{296} templates. }\label{fig:appmeim}
\end{figure*}

\begin{figure*}
\centering
\includegraphics[width=16cm]{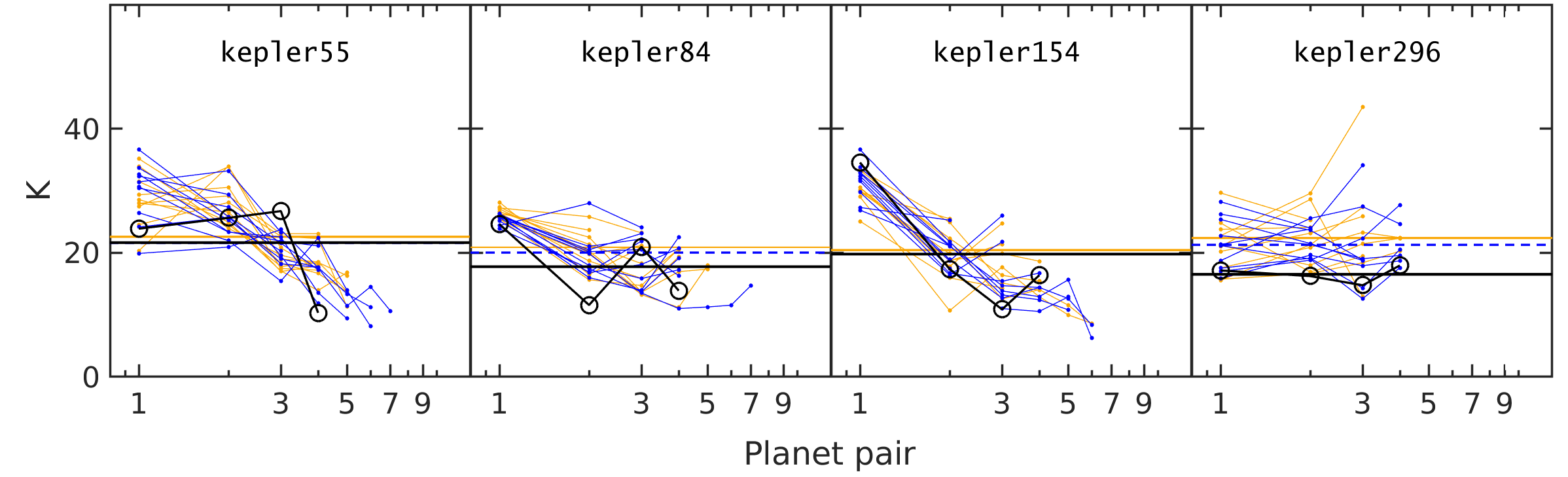}
\caption[All $K$-vlaue from perfect collision of \texttt{Kepler55}, \texttt{84}, \texttt{154}, and \texttt{296}]{Same plot as figure \ref{fig:allsK} but for the perfect \texttt{Kepler55}, \texttt{84}, \texttt{154}, and \texttt{296} templates.}\label{fig:appmK}
\end{figure*}

\begin{figure}
\centering
\includegraphics[width=8.2cm]{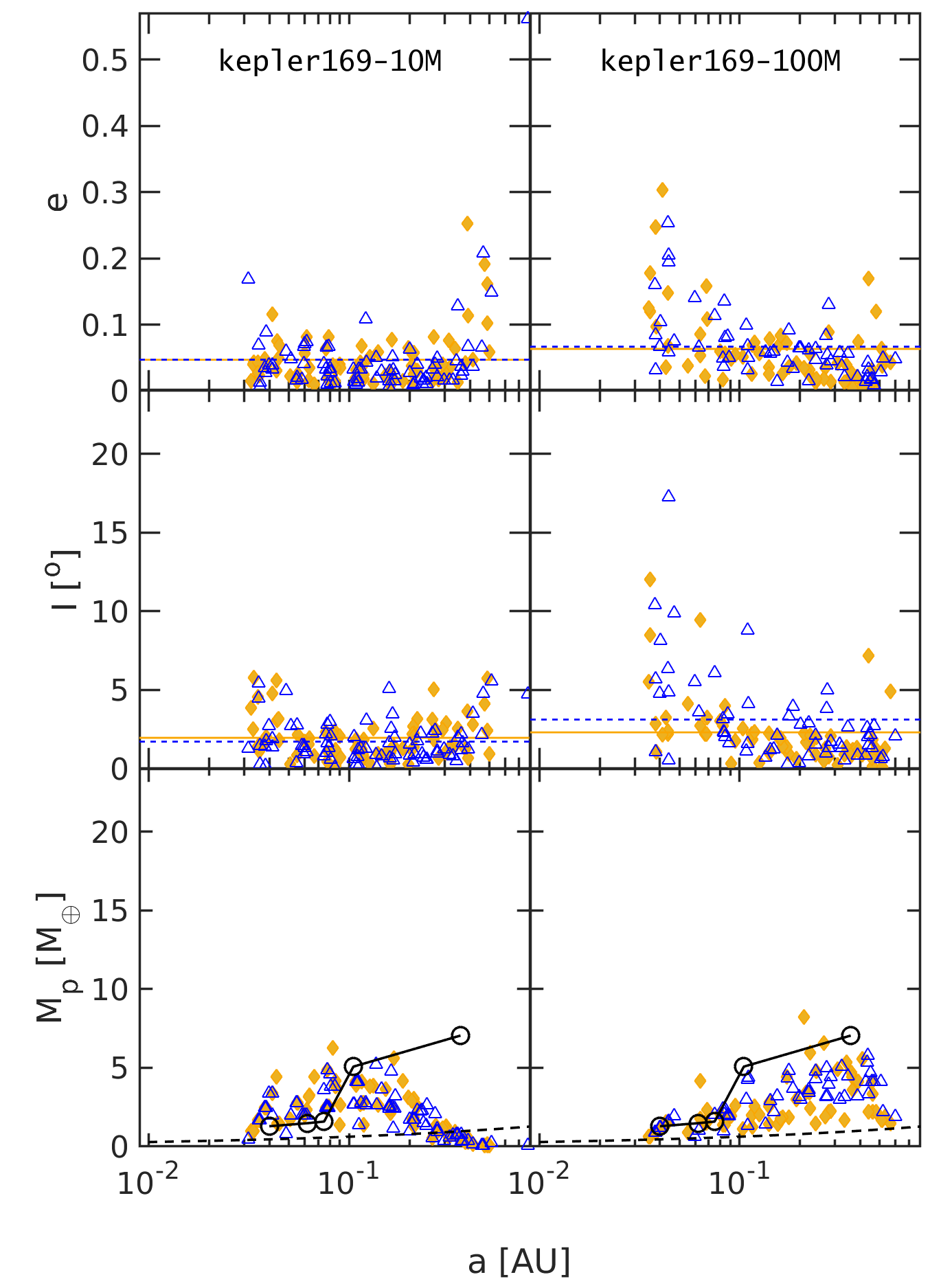}
\caption[\texttt{Kepler169} 10 Myr vs. 100 Myr]{Similar to figure \ref{fig:allseim} but for the comparison of template \texttt{Kepler169} (perfect routine) after running 10 Myr and 100 Myr.}\label{fig:169_10100_eim}
\end{figure}

\begin{figure}
\centering
\includegraphics[width=8.2cm]{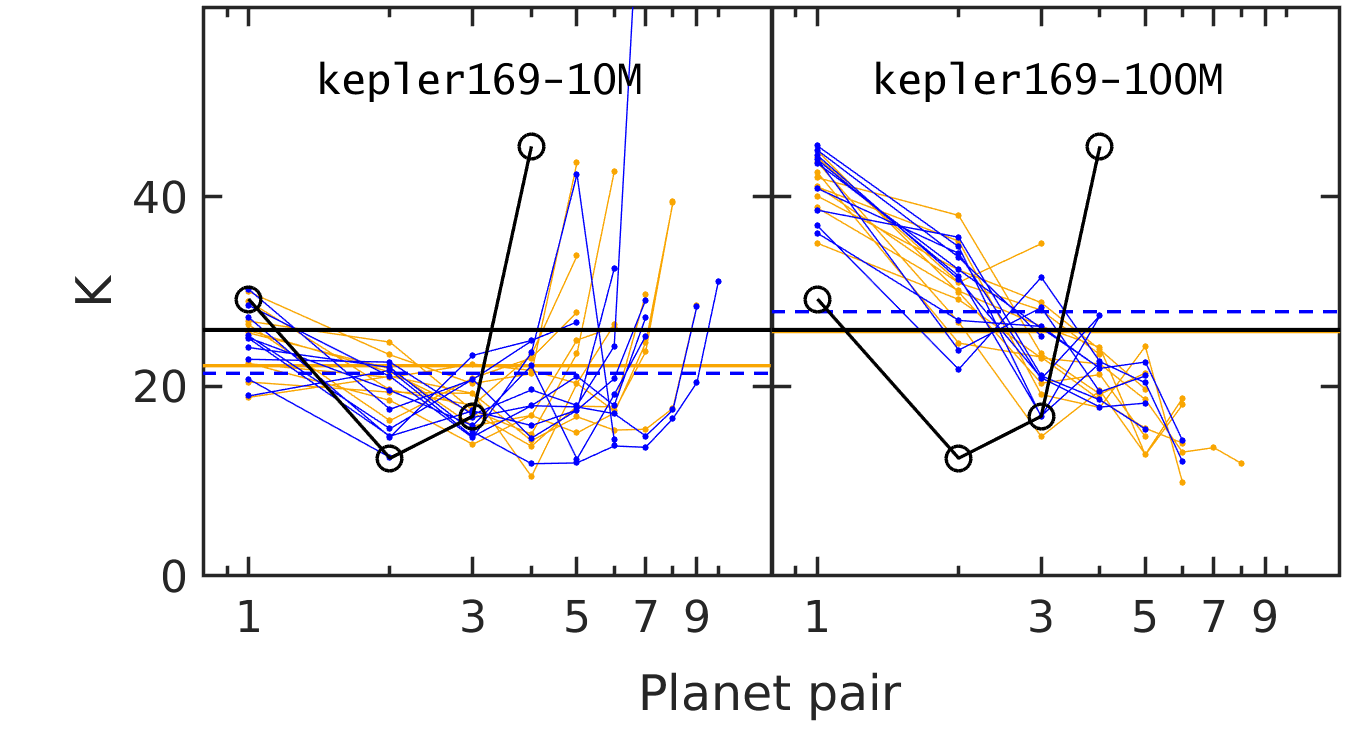}
\caption[\texttt{Kepler169} 10 Myr vs. 100 Myr]{Similar to figure \ref{fig:allK} but for the comparison of template \texttt{Kepler169} (perfect routine) after running 10 Myr and 100 Myr.}\label{fig:169_10100_K}
\end{figure}

\bsp	
\label{lastpage}
\end{document}